\documentclass[preprint,3p,10pt]{elsarticle}

\journal{CR Physique}

\usepackage{xparse}
\usepackage{xspace}
\usepackage{xargs}
\usepackage{ifthen}
\usepackage{bm}

\usepackage[x11names,svgnames]{xcolor}

\usepackage[T1]{fontenc}
\usepackage{csquotes}
\usepackage[kerning=true,babel=true]{microtype} 

\usepackage[bitstream-charter]{mathdesign}

\usepackage{amsmath}
\usepackage{stmaryrd}
\usepackage{mathtools}
\usepackage{dsfont}
\makeatletter
\DeclareFontFamily{OMX}{MnSymbolE}{}
\DeclareSymbolFont{MnLargeSymbols}{OMX}{MnSymbolE}{m}{n}
\SetSymbolFont{MnLargeSymbols}{bold}{OMX}{MnSymbolE}{b}{n}
\DeclareFontShape{OMX}{MnSymbolE}{m}{n}{
    <-6>  MnSymbolE5
   <6-7>  MnSymbolE6
   <7-8>  MnSymbolE7
   <8-9>  MnSymbolE8
   <9-10> MnSymbolE9
  <10-12> MnSymbolE10
  <12->   MnSymbolE12
}{}
\DeclareFontShape{OMX}{MnSymbolE}{b}{n}{
    <-6>  MnSymbolE-Bold5
   <6-7>  MnSymbolE-Bold6
   <7-8>  MnSymbolE-Bold7
   <8-9>  MnSymbolE-Bold8
   <9-10> MnSymbolE-Bold9
  <10-12> MnSymbolE-Bold10
  <12->   MnSymbolE-Bold12
}{}

\let\llangle\@undefined
\let\rrangle\@undefined
\DeclareMathDelimiter{\llangle}{\mathopen}%
                     {MnLargeSymbols}{'164}{MnLargeSymbols}{'164}
\DeclareMathDelimiter{\rrangle}{\mathclose}%
                     {MnLargeSymbols}{'171}{MnLargeSymbols}{'171}
\makeatother

\usepackage{braket}
\usepackage{mathtools}
\usepackage{eucal}
\usepackage{upgreek} 

\usepackage[version=3]{mhchem}

\usepackage{graphicx}

\usepackage[unicode]{hyperref}
\hypersetup{
    colorlinks=true,
}

\newcommandx{\Iverson}[1]{\ensuremath{\left[ #1 \right] }}

\newcommandx\PermutationGroup[1]{\ensuremath{\mathfrak{S}_{#1}}}

\newcommandx\GeneralLinearGroup[2][2={}]{
\ifthenelse{\equal{#2}{}}{
\ensuremath{\text{GL}(#1)}
}{
\ensuremath{\text{GL}(#1,#2)}
}\xspace
}
\newcommandx\SpecialLinearGroup[2][2={}]{
\ifthenelse{\equal{#2}{}}{
\ensuremath{\text{SL}(#1)}
}{
\ensuremath{\text{SL}(#1,#2)}
}\xspace
}

\newcommandx\ContinuityClass[3][2={},3={}]{
\ifthenelse{\equal{#2}{}}{
    \ensuremath{\mathcal{C}^{#1}}
  }{
    \ifthenelse{\equal{#3}{}}{
      \ensuremath{\mathcal{C}^{#1}(#2)}
    }{
      \ensuremath{\mathcal{C}^{#1}(#2,#3)}
  }
}\xspace
}

\newcommandx\IntegerPart[1]{\ensuremath{\text{E}\left[ #1 \right]}}

\newcommandx\PauliMatrix[2][1={\sigma}]{\xspace\ensuremath{
\ifthenelse{\equal{#2}{}}{
#1
}{
#1_{#2}
}
}\xspace}
\newcommandx\ExteriorAlgebra[2][1={}]{\xspace\ensuremath{
\ifthenelse{\equal{#1}{}}{
\Lambda #2
}{
\Lambda^{#1} #2
}
}\xspace}

\newcommandx\Norm[1]{\ensuremath{\left\lVert #1 \right\rVert}}
\newcommandx\norm[1]{\ensuremath{\lVert #1 \rVert}}

\renewcommand{\Re}{\operatorname{Re}}
\renewcommand{\Im}{\operatorname{Im}}

\NewDocumentCommand{\GammaMatrix}{mg}{
\IfNoValueTF{#2}{
\Gamma_{#1}
}{
\Gamma_{#1 #2}
}
}

\NewDocumentCommand{\TRIClosedCurve}{O{} m}{\ensuremath{\mathcal{C}_{#2}^{#1}}}

\begin{document}
\hypersetup{
    citecolor=PaleGreen4!80!black,
    linkcolor=DarkRed, 
    urlcolor=DarkSeaGreen4!90!black}

\centerline{}
\begin{frontmatter}


\title{Driven dissipative dynamics and topology of quantum impurity systems}\footnote{French Title: Dynamique Dissipative et Topologie dans des Syst\` emes d'impuret\' es quantiques}

\author{Karyn Le Hur$^{1}$, Lo\" ic Henriet$^{2}$, Lo\" ic Herviou$^{1,3}$, Kirill Plekhanov$^{4,1}$, Alexandru Petrescu$^{5}$,
\\ Tal Goren$^{1}$,  Marco Schiro$^{6}$, Christophe Mora$^{3}$, Peter P. Orth$^{7}$}
\address{$^{1}$ Centre de Physique Th\'{e}orique, Ecole Polytechnique, CNRS, Universit\' e Paris-Saclay, 91128 Palaiseau Cedex France
\\ $^{2}$ ICFO-Institut de Ciencies Fotoniques, The Barcelona Institute of Science and Technology, 08860 Castelldefels (Barcelona), Spain 
\\ $^{3}$ Laboratoire Pierre Aigrin, Ecole Normale  Sup\' erieure-PSL Research  University, CNRS, Universit\' e Pierre et Marie Curie Sorbonne Universit\' es, Universit\' e Paris-Sorbonne Paris Cit\' e, 24 rue Lhomond 75231 Paris Cedex 05 France  \\
$^{4}$ LPTMS, CNRS, Univ. Paris-Sud, Universit\' e Paris-Saclay, 91405 Orsay, France
\\ $^{5}$ Department of Electrical Engineering, Princeton University, Princeton, New Jersey, 08544 
\\ $^{6}$ Institut de Physique Th\' eorique, Universit\' e Paris Saclay, CNRS, CEA, F-91191 Gif-sur-Yvette, France \\
$^{7}$ Department of Physics and Astronomy, Iowa State University, Ames, Iowa 50011, USA}

\begin{abstract}
In this review, we provide an introduction and overview to some more recent advances in real-time dynamics of quantum impurity models and their realizations in quantum devices. We focus on the Ohmic spin-boson and related models, which describes a single spin-1/2 coupled to an infinite collection of harmonic oscillators. The topics are largely drawn from our efforts over the past years, but we also present a few novel results. In the first part of this review, we begin with a pedagogical introduction to the real-time dynamics of a dissipative spin at both high and low temperatures. We then focus on the driven dynamics in the quantum regime beyond the limit of weak spin-bath coupling. In these situations, the non-perturbative stochastic Schroedinger equation method is ideally suited to numerically obtain the spin dynamics as it can incorporate bias fields $h_z(t)$ of arbitrary time-dependence in the Hamiltonian. We present different recent applications of this method: (i) how topological properties of the spin such as the Berry curvature and the Chern number can be measured dynamically, and how dissipation affects the topology and the measurement protocol, (ii) how quantum spin chains can experience synchronization dynamics via coupling to a common bath. In the second part of this review, we discuss quantum engineering of spin-boson and related models in circuit quantum electrodynamics (cQED), quantum electrical circuits and cold-atoms architectures. In different realizations, the Ohmic environment can be represented by a long (microwave) transmission line, a Luttinger liquid, a one-dimensional Bose-Einstein condensate, a chain of superconducting Josephson junctions. We show that the quantum impurity can be used as a quantum sensor to detect properties of a bath at minimal coupling, and how dissipative spin dynamics can lead to new insight in the Mott-Superfluid transition.

\vskip 1\baselineskip


\end{abstract}

\begin{keyword}
Spin-Boson, Ising and Kondo models, Berezinskii-Kosterlitz-Thouless transition, Stochastic Spin Dynamics and Bloch Equations, Disorder in Time, Boltzmann-Gibbs Description and Thermalization, Stochastic Mean-Field Theory, Topology and Floquet Systems, Quantum Brownian Motion, Mott-Superfluid Transition, Light-Matter and Hybrid Systems, Majorana quantum impurity models, Quantum Materials and Quantum State Engineering
\vskip 0.5\baselineskip


\end{keyword}

\end{frontmatter}


\section{General introduction and motivation}

The study of quantum impurities is a large field of research whose motivation is at least twofold. First, impurities are abundant in various situations of physics. They take different roles that can be classified as follows: (i) the impurity can be the quantum system of interest itself, (ii) it can be a component of a quantum sensing device, or (iii) it acts as a scatterer in a host material. Let us name just a few well-known examples, some of which we will discuss in more detail later in this review. Impurities are quantum systems of interest as qubits in quantum computing~\cite{NielsenChuang-QC} and quantum simulations~\cite{Feynman-QuantumSimulators}. Here, major research goals are the dynamical and coherent control and read-out of their quantum state and the minimization of decoherence effects due to the coupling to the environment. As quantum simulators one often strives to reach a rather strong coupling between the impurity and its environment in order to simulate in the non-perturbative regime. Impurities are employed as quantum sensors to read out properties of nearby (quantum) systems with high sensitivity and minimal coupling~\cite{Cappallaro-RMP}. Examples are quantum point contacts that detect nearby charge states, ancilla qubits that non-destructively read-out the state of other qubits, and nitrogen vacancy centers that can sense magnetic fields with very high precision. Finally, quantum impurities appear as (magnetic) scatterers in electronic host materials, where they crucially affect transport properties, for example, via the famous Kondo effect~\cite{Kondo, Hewson}. This describes the hybridization of a local quantum spin with the host electrons. The resulting entanglement of the quantum impurity with the host electrons can give rise to fascinating non-Fermi liquid behavior as seen in various heavy-fermion materials~\cite{Hewson} and multi-channel Kondo models~\cite{Schofield-NFL}. In this review, we focus on recent experimental realizations of quantum impurities in circuit electrodynamics (cQED)~\cite{cavity1, cavity2, Yale} and cold-atom setups~\cite{DIW}. These platforms offer unique control capabilities that promise, for example, quantum simulations in the non-perturbative regime, where the impurity is strongly entangled with its environment. 

A second, independent motivation to investigate quantum impurity models is that they are a perfect testbed for methodological advances of both analytical and numerical techniques in quantum many-body physics. It is extremely challenging to (quantitatively) describe a system that is composed of many interacting particles. Zero-dimensional, few level quantum impurity systems coupled to infinitely many environmental degrees of freedom are often suitable starting points to develop, test and compare different methods. An important step made by Feynman and Vernon in 1963 was to suggest such a Hamiltonian description of an environment in terms of a bath of harmonic oscillators~\cite{FeynmanVernon}. Famous examples of impurity models include a heavy particle undergoing Brownian motion in a thermally equilibrated molecular medium~\cite{Hanggi} or the Caldeira-Leggett model~\cite{CaldeiraLeggett} describing a dissipative environment of a small quantum system such as a resistive element in a quantum electrical circuit. Other famous examples are the Kondo model, originally developed to model a magnetic impurity in a metallic host~\cite{Kondo,Hewson,AndersonRG,Wilson,Nozieres,IanCFT,NozieresBlandin}, and the spin-boson model~\cite{Blume,leggett,weiss}. These two models are the main focus of this review. They describe a quantum spin $\boldsymbol{S}$ (e.g. with $S=1/2$) that is coupled to infinitely many fermionic or bosonic environmental degrees of freedom as described by a Hamiltonian of generic form
\begin{align}
  \label{eq:10}
  H &= - \hat{\boldsymbol{h}} \cdot \boldsymbol{S} + H_0[\boldsymbol{S}] + H_{\text{env}}[\hat{\boldsymbol{h}}]\,. 
\end{align}
Here, $\hat{\boldsymbol{h}}$ describes the environmental degrees of freedom that are coupled to the spin, which could be bosonic, fermionic or spins themselves. As described in detail below, they act as fluctuating fields $\hat{\boldsymbol{h}}$ from the point of view of the spin, which randomize its state and induce decoherence and dissipation. 

The goal to understand the thermodynamics and dynamics of quantum impurity models has inspired researchers to develop various ground-breaking methods. On the analytical front, these are, for example, renormalization group (RG) approaches starting from Anderson's ``poor-man'' scaling solution of the Kondo model~\cite{AndersonRG, leggett, weiss, BrayMoore}, functional integral techniques that are applicable both in and out-of-equilibrium~\cite{FeynmanVernon, leggett, anderson, anderson2}, various functional RG methods~\cite{Schoeller-FRG, Schoeller-FRG-Review, Metzner-FRG}, exact and approximate master equations~\cite{Breuer-MasterEquations}, variational approaches~\cite{variational_1, variational_2,variational_3}, bosonization techniques~\cite{Giamarchi}, and exact solutions based on the Bethe ansatz~\cite{AlexeiPaul, NatanJohannesson}. On the numerical side, there is Wilson's numerical RG (NRG) solution of the Kondo model~\cite{Wilson} that spurred a lot of further development of NRG methods~\cite{NRG-Review}. These include, among others, extensions to bosonic baths and time-dependent situations~\cite{Hofstetter-NRGDynamics, Meirong1, Bulla,Andersschiller,subPhilippe,NRG2spins, CostiNRG}, relevant to describe spin-boson dynamics. Other numerical approaches are the density-matrix renormalization group (DMRG)~\cite{Schollwoeck, White-TDDMRG, Schmitteckert-TDDMRG}, path-integral Monte-Carlo methods~\cite{QMC2spins,MarcoMC,Thomas,WernerOM}, master equations~\cite{Breuer-MasterEquations,Maxim,BDA}, matrix product states (MPS) methods~\cite{Weichselbaum,TEMPO}, variational approaches and cluster mean-field approaches~\cite{Cirac,Lesanosky,AMRey,Rossini, Blunden-Codd}, and stochastic approaches~\cite{stoch1,stoch2,stoch3,Breuer, S1, S2, S3, S4, S5, S6, S7, PAK-1, PAK-2, Loic1, Loic2, Loic3, Loicthese}. In this review, we illustrate physical ideas behind the stochastic Schroedinger equation (SSE) method~\cite{PAK-1, PAK-2, Loic1, Loic2, Loic3, Loicthese}, on which several of the authors have worked over the past few years, making a connection to classical Bloch equations. The mathematical steps, starting from the real-time path integral ``sojourn-blip'' representation \cite{leggett,weiss}, have been summarized in Refs. \cite{PAK-1, PAK-2, Loic1, Loic2, Loic3, Loicthese}.

Finally, we note that an important fundamental link between quantum impurity physics and that of higher-dimensional lattice models, such as the Hubbard model, is provided by the powerful dynamical mean-field theory (DMFT) method~\cite{DMFT}. Within DMFT, one takes a local approach to interacting lattice models by mapping them to a quantum (Anderson) impurity model that is embedded in a self-consistent medium. Stochastic or Monte-Carlo approaches have also been developed further for quantum lattice systems at equilibrium~\cite{Werner,Werner2,Subir,Pankov,Olivier,Vojta,Michel,Walter} and for driven and open systems \cite{Jonathan,Cristiano}. 

Considering the vast existing literature on quantum impurity systems, let us explain what we aim to achieve in this review. We want to provide an intuitive and pedagogical introduction to some of the more recent advances in quantum impurity models. In Chapter~\ref{sec:dynamics-spins}, we discuss the dynamics and topology of the spin-boson model. We focus on the development of the stochastic Schroedinger equation (SSE) method and its use to investigate the influence of the environment on the topological properties of a spin such as its Berry curvature and topological Chern number~\cite{Berry}. We describe a dynamical measurement of the Chern number~\cite{PolkovnikovGritsev, Review} and its breakdown in the presence of sufficiently strong dissipation~\cite{Loic3}. We also discuss the phenomenon of dynamical synchronization in spin systems that are coupled via a common bath~\cite{NRG2spins,JeanNoel, synchronization_salomon}. In Chapter~\ref{sec:real-spin-boson}, we then consider various experimental realizations of spin-boson and Kondo models in cold-atom, circuit quantum electrodynamics, and cQED architectures~\cite{Yale,Markus, recati_fedichev, orth_stanic_lehur, Karyn2, Houck, CRAS2016,Goldstein,Clerk}. These realistic setups provide concrete examples of quantum impurities acting as the quantum system of interest that can simulate strong-coupling Kondo physics. In other setups, the impurity rather acts as a sensing device that probes its environment, or as scatterers that strongly affect the transport of electrons or photons via the emergence of a many-body entangled state. Finally, in Chapter~\ref{sec:diss-array-spin}, we consider the dynamics in arrays of dissipative spins and their potential realization in cold-atom and cQED setups. We briefly conclude in Chapter~\ref{sec:conclusion}. 

In the following, we will complement this general discussion by introducing the spin-boson Hamiltonian, its relation to the Kondo model and a qualitative description of some of its fundamental properties. We also briefly discuss some of the main results that are then described in more detail in the body of the paper. We deliberately point out various interesting cross-links in the literature and also mention a number of references to some of the works related to quantum impurity physics that will not be at the center of this review. We believe that this is nevertheless useful as it not only exemplifies the broad impact of impurity models, but also serves as a guide to the interested reader who wants to explore applications beyond the core topics of this review. 

\subsection{Spin-boson and Kondo models}
\label{sec:spin-boson-model}
In the following, we consider mostly the case of an ohmic dissipative bosonic environment where the number of modes at low frequency grows linearly. We start with the spin-1/2 impurity model whose Hamiltonian takes the (standard) spin-boson form~\cite{leggett,weiss}:
\begin{equation}
\label{eq:11}
\hat{H} = \frac{\Delta \sigma_x}{2}+ \frac{h_z \sigma_z}{2} + \frac{\sigma_z}{2}\sum_k \lambda_k(b^{\dagger}_k + b_k) + \sum_k \omega_k b^{\dagger}_k b_k\,.
\end{equation}
The Planck constant $\hbar$ is fixed to unity. This model describes a spin that is coupled via its $\sigma_z$ component to a bath of harmonic oscillators with creation operators $b^\dag_k$. The bath spectral function takes the form $J(\omega) = \pi \sum_k \lambda_k^2 \delta(\omega-\omega_k) = 2\pi \alpha \omega e^{-\omega/\omega_c}$, which we assume to be of Ohmic form and described by the dimensionless dissipation coefficient $\alpha$. Typical values of $\alpha$ range from $\alpha = 0$ (no dissipation) to $\alpha = 1$ corresponding to the strong-coupling limit with the environment. Here, $\Delta$ plays the role of the transverse field, which is affected by the presence of the bath. One can absorb the effect of the bath through the unitary transformation
$H' = U^\dag H U = \frac{\Delta}{2} \bigl( \sigma_+ e^{i \phi} + \sigma_- e^{-i\phi} \bigr) + \frac{h_z \sigma_z}{2} +  \sum_k \omega_k b^{\dagger}_k b_k$ with $U = \exp(i \phi \sigma_z/2)$, and $\phi = \sum_k \frac{\lambda_k}{\omega_k} (b^\dag_k - b_k)$~\cite{weiss}. The bath then renormalizes the transverse field as $\Delta_r = \Delta \langle e^{i\phi} \rangle$, which can then be calculated and identified to $\Delta_r = \Delta (\Delta/\omega_c)^{\alpha/1-\alpha}$~\cite{leggett,weiss}. Here, we have taken the average over the bath ground state of harmonic oscillators $\langle e^{i\phi} \rangle = e^{-\langle \phi^2 \rangle/2}$ and summed over modes from frequencies of the order of $\Delta$ to frequencies of the order of $\omega_c$~\cite{leggett}.

We will describe quantitatively the interplay between dissipation mediated by an environment and dynamical effects imposed by time-dependent signals, \emph{i.e.}, time-dependent parameters in the Hamiltonian. We put a particular focus on parameters whose time dependence is periodic, which are referred to as Floquet perturbations. The general class of problems we treat refers to dissipative, driven and open quantum systems~\cite{Frank,ZollerPeter,Torre}, which may also find applications to realize and probe topological phases~\cite{Bardyn,MobilTopo,cQEDWalk}. 

The spin-boson model can be mapped onto the anisotropic Kondo model and Ising model with long-range forces~\cite{anderson}. We consider the case where the cutoff frequency $\omega_c \gg (\Delta, h_z)$ is the largest frequency scale of the system. For ohmic dissipation,  the spin-boson model exhibits a quantum phase transition which is of Berezinskii-Kosterlitz-Thouless type at $\alpha\sim 1$~\cite{Berezinskii,KT} by reminiscence of the two-dimensional XY model, and is associated with the production of defects in the time domain (here corresponding to spin flips of the spin-1/2 particle). When increasing the coupling between the spin-1/2 impurity and the environment at zero temperature, this engenders a jump of the spin magnetization at zero temperature~\cite{anderson2,KLH,KLHannal}, by analogy to the jump of the superfluid stiffness at finite temperature in the two-dimensional classical XY model \cite{Reppy,Dalibard}, and the spin remains localized by the environment in one of the two wells or one of the two spin polarizations \cite{anderson2,Thouless,Meirong1}. This analogy can be understood from perturbative renormalization group \cite{AndersonRG}, Numerical Renormalization Group and Bethe Ansatz arguments \cite{KLH,KLHannal,Kopp}. It is important here to mention the current efforts to realize the spin-boson model \cite{Gross}; in particular, the (relatively) strong-coupling limit of such a ohmic spin-boson model has been realized recently \cite{Lupascu,Martinez}. Furthermore, dissipation effects at quantum phase transitions and dissipation-induced quantum phase transitions have been observed in other systems such as superconducting systems \cite{StanfordNadya}, cQED systems \cite{Fitzpatrick}, electronic quantum circuits \cite{LPN} and resonant level architectures \cite{Finkelstein}. Analogies and applications to the spin-boson model reach as far are to quantum biology~\cite{huelga_plenio}, and in the description of quantum fluids of light~\cite{SchmidtKoch, Hartmann, Angelakisreview,CRIA}.
 
\subsection{Stochastic dynamics and disorder in time}
\label{sec:stoch-dynam-disord}

Integrating out the degrees of freedom of an ohmic (Gaussian) type environment produces unusual memory effects in time in the low-temperature regime \cite{FeynmanVernon,anderson}.
In the context of quantum spin trajectories on the Poincar\' e-Bloch sphere, this has resulted in the representation of interacting quantum blips and sojourns in time \cite{leggett,weiss}, where the interaction with the environment produces long-range spin-spin interaction in time. This interaction can be viewed as the effect of a stochastic colored noise (with memory) on the system. This disorder in time induces novel many-body physics, related to Kondo, Ising physics and localization transitions~\cite{Blume,anderson}. Solving real-time dynamics is often difficult and requires the development of new methods, even in a mean-field picture.  Exactly solvable limits exist, for example related to mathematical mappings between bosons and fermions in one dimension, at and around the Toulouse limit (point)\cite{Toulouse,resonant,Schoeller}. Decoupling the interactions in time is possible through the introduction of Hubbard-Stratonovich variables in the Feynman path integral sense. Then, one reaches a quantum stochastic theory with Hubbard-Stratonovich fields as stochastic classical variables \cite{S1,S2,S3,S4,S5,S6,S7}. Averaging over these noisy fields, one can follow the reduced density matrix of the system in time. This way of thinking has led to the development of local type stochastic Schr\" odinger equation approaches \cite{S1,S2,S3,S4,S5,S6,S7}. Inspired by these earlier works, we have developed and applied the stochastic Schr\" odinger equation approach  to driven and dissipative Landau-Zener models \cite{PAK-1,PAK-2}, Rabi models and dissipative spin systems \cite{Loic1,Loic2,Loic3,Loicthese}, and addressed various classes of time-dependent situations. By analogy with disordered problems in real space~\cite{Zoran}, driven light-matter systems and coupled spin systems require more than one stochastic field to treat the real-time dynamics rigorously \cite{Loic1,Loic2,Loic3}. Below, we review pedagogically some theoretical aspects of quantum spin dynamics related to the stochastic Schr\" odinger equation approach in Sec. 2 and present novel results (in Secs. 2.6 and 2.7). The stochastic approach is directly related to current efforts in quantum circuits, where single trajectories can now be tracked on the Bloch sphere \cite{Berkeley,NicolascQED}, and in ultra-cold atoms \cite{atomChern2,Ian}. This approach could also serve as a rigorous mean-field starting point to capture the real-time dynamics in quantum materials. 

Furthermore, it is relevant to recall that in the case of a non-magnetic impurity (tunnel junction), the effect of such an ohmic dissipative environment (modeling the resistance from the surrounding circuit) \cite{Portier}, can deeply affect the I-V characteristics, and produce dynamical Coulomb blockade physics at large resistances where the current becomes progressively blockaded at small voltages \cite{LPN,IngoldNazarov}. This physics is also directly related to the Kane-Fisher tunnel barrier modelling an impurity in a one-dimensional Luttinger liquid \cite{Kane-Fisher,SafiSaleur}, and more generally to disorder physics produced by many impurities in Luttinger liquids \cite{GiamarchiSchulz}. Indeed, a DC resistance can be seen as a long LC transmission line \cite{Markus}, which is then related to the Luttinger liquid through bosonization \cite{Haldane,Giamarchi}. Recently, efforts in mesoscopic physics have been realized to achieve time-dependent potentials in tunnel junctions and quantum point contacts offering a very rich dynamics in time of electron wave packets \cite{Glattli,LPA,Julien}. Ramsey interferometry in tunnel junctions (with a continuum of states in the metallic leads) would allow to probe dynamical Coulomb blockade physics in time at weak resistances through current noise measurements by adjusting the time scale between the two time-dependent pulses \cite{Tal}. Ramsey interferometry of an atom with discrete energy levels has been generalized for ensembles of fields \cite{Eric} and spins \cite{Immanuel}, and applied in the context of topological phases in ultra-cold atoms \cite{Munich}. The behavior of two-time correlatiors of wave-packets has been studied theoretically in Luttinger liquids \cite{MarcoAditi1,MarcoAditi2}. One can then apply bosonization methods \cite{Haldane,Giamarchi}, that can be combined with Keldysh methods to address non-equilibrium transport problems \cite{Torre}. 

\subsection{Many-Body Quantum Realizations}
\label{sec:quant-simul-exper}
\begin{figure}[t]
\center
\includegraphics[width=\linewidth]{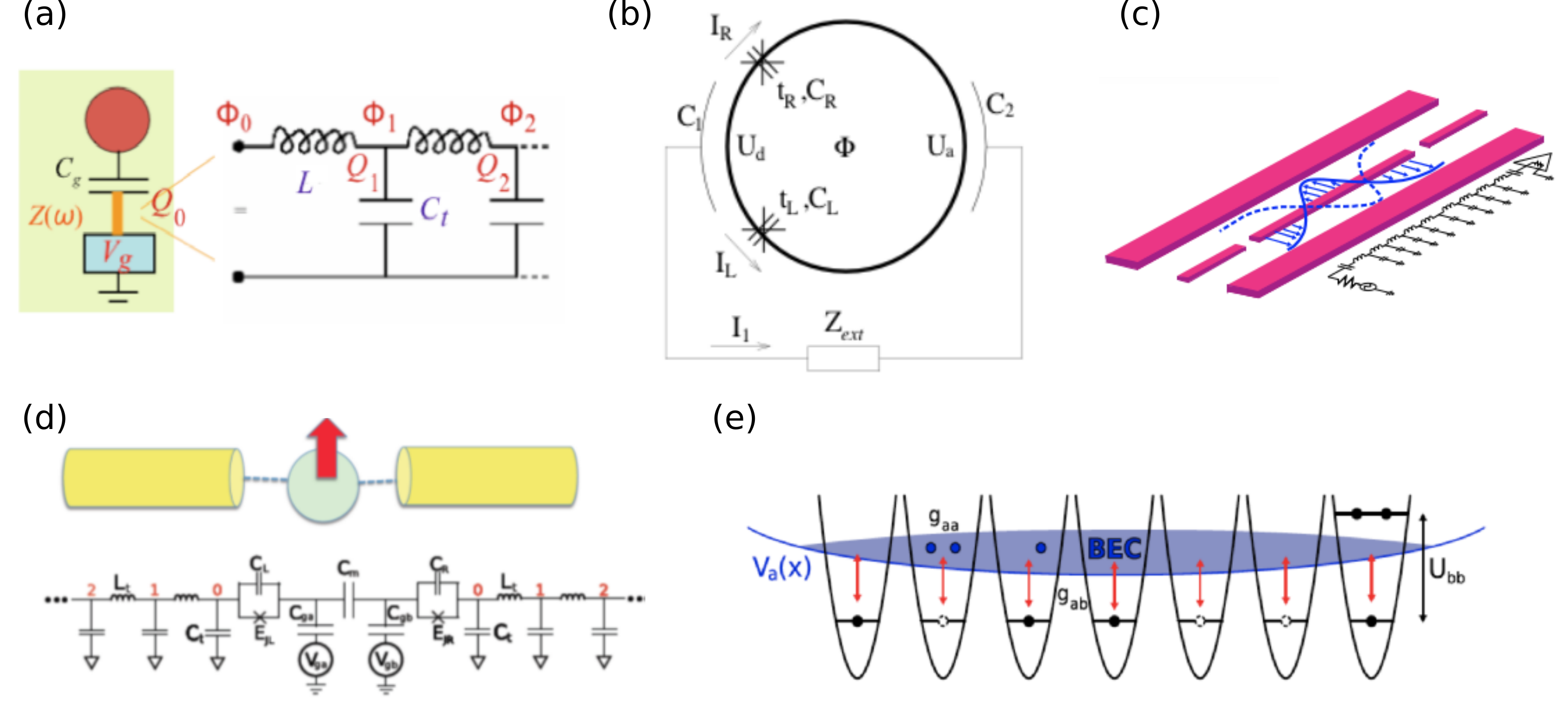} 
\caption{Spin-boson engineering. We summarize different geometries to be discussed below that realize the spin-boson model with ohmic dissipation in quantum electrical ciruits, circuit-QED, and cold-atoms. (a) A quantum dot that is capacitively coupled to a transmission line embodies a resonant level model~\cite{Makhlin}. The transmission line can be replaced by a Luttinger liquid or edges states of quantum Hall systems~\cite{MatveevFurusaki,LiHur}. A generalization to a spin-boson-fermion model has been proposed in Ref. \cite{Karyn2}. (b) In a ring geometry, the persistent current is strongly modified by the coupling to a dissipative impedance element $Z_{\text{ext}}$~\cite{Markus}. (c) Equivalently to the setup in (a) with the long transmission line, spin-boson models also arise in a circuit quantum electrodynamics (cQED) architecture~\cite{Yale,Fitzpatrick,Stevelecturenotes} or (d) in Josephson junction arrays~\cite{Karyn,Goldstein,Camalet}. This circuit is related to the photonic Kondo effect of light that is addressed in Sec.~\ref{sec:spin-boson-model-2} (e) Cold-atom setups can also realize spin-boson or Kondo physics. Shown is a one-dimensional array of tight traps~\cite{Optiqueatom} that can either host zero or one particle of some atomic species ``b'' (due to large on-site interactions $U_{bb}$). This array describes the spins. These are coupled (via scattering $g_{ab}$ and Raman coupling (red arrows)) to a Bose-Einstein condensate (BEC) realized by another atomic species ``a'' that experiences a shallow trapping potential $V_a(x)$. The low-energy sound modes of the BEC realize the bosonic bath~\cite{recati_fedichev}. We have generalized this setup to the case of an ensemble of atomic traps (deep in the Mott regime) \cite{orth_stanic_lehur}, realizing a dissipative quantum Ising model. Such a spin array will be discussed in Secs.~\ref{sec:solvable-mean-field},~\ref{sec:spin-boson-model-1} and~\ref{sec:diss-array-spin}. A related system of a mobile impurity in a one-dimensional Luttinger liquid realize similar physics~\cite{lamacraft}.}
\label{fig:1}
\end{figure}

In Sec.~\ref{sec:real-spin-boson}, we review specific simulators of the spin-boson system with ohmic dissipation. In Fig.~\ref{fig:1}, we show several implementations of the spin-boson model (in relation with our theoretical results and proposals). We also show novel results in Secs.~\ref{sec:spin-boson-model-1} and~\ref{sec:spin-boson-model-2}. 

An ohmic resistance can be engineered through a one-dimensional transmission line \cite{Markus,Karyn2} - an ensemble of harmonic oscillators with a plasma frequency - which offer new applications  in circuit-QED array \cite{Houck,CRAS2016}, Luttinger liquid physics, Carbon nanotubes or edge states of quantum Hall systems \cite{LPN,Finkelstein,MatveevFurusaki,LiHur}, Josephson junction arrays \cite{Neel,Pop,Fluxonium}, one-dimensional Bose-Einstein condensates (BECs)\cite{recati_fedichev,orth_stanic_lehur,Carlos_scientific_reports,lamacraft,Fischer,Anna}. A transmission line can also be realized in a long cavity multi-mode circuit (see Refs. \cite{Fitzpatrick,Stevelecturenotes} and Fig.~\ref{fig:1}(c)). It is important to stress that some efforts have been achieved recently to reach the strong-coupling limit of the spin-boson model with ohmic dissipation in solid-state devices with Josephson qubits coupled to a transmission line \cite{Lupascu,Martinez}. Efforts in measuring the renormalization of the qubit (spin) frequency or zero-point fluctuations corrections to the Lamb shift and the precise form of the damping rate have been performed \cite{Lupascu}. 

The dissipative spin-boson model can be realized with charge qubits built of mesoscopic quantum dots or Cooper pair superconducting boxes capacitively coupled to the transmission line  \cite{Makhlin} (see Fig.~\ref{fig:1}(b)). A charge qubit can be seen as a spin-1/2 object corresponding to the two degenerate charge states forming the low-energy sub-space \cite{Pashkin, Wiel}. Gate voltages can control the two charge states and allow to realize an effective magnetic field $h_z$ in Eq. (1). Here, $\Delta$ corresponds  to the tunneling amplitude between the (superconducting) box and the lead(s) or the Josephson coupling energy of the junction. If the gate source is placed in series with an external resistor, then this may describe the spin-boson model with ohmic dissipation \cite{Makhlin}. In ultra-cold atoms, the spin-1/2 can be built with one atom or zero atom in a tight optical trap limit \cite{Optiqueatom} as suggested in Refs. \cite{recati_fedichev,orth_stanic_lehur}. The bath here refers to the sound modes of a one-dimensional BEC and the coupling with the tight trap involves optical Raman transitions (close to resonance) and collisional interactions (see Fig.~\ref{fig:1}(e))). The spin-boson model can also be derived when coupling a quantum dot to a boson and fermion bath \cite{Meirong1,Karyn2,Pascal}.  Charge measurements provide generally the quantity $\langle \sigma_z\rangle$, which represents the occupation of the dot or island. In a ring geometry, the application of a magnetic flux generates a persistent current which is proportional to $\langle \sigma_x\rangle$ \cite{Markus} (see Fig.~\ref{fig:1}(b)).  The recent realizations \cite{Gross,Lupascu} of the model involve flux qubits;  transmon or Xmon qubits, could also offer a long lifetime \cite{Transmon,Xmon}.  These nano-systems also allow to address non-equilibrium transport at a given quantum phase transition, both theoretically and experimentally \cite{LPN,Finkelstein,Finkelstein2,Chung-Hou}. 

To be more precise regarding the geometries to be addressed in Sec. 3, following Ref. \cite{recati_fedichev} and Fig.~\ref{fig:1}(e), we present a novel application of the spin-boson model where the spin is used as a quantum microscope to probe the Mott-superfluid transition. This geometry can also be realized in Josephson junction arrays \cite{Neel}. Then, we discuss our implementation of the spin-boson model with Josephson circuits and microwave light (see Fig.~\ref{fig:1}(d) \cite{Karyn,Goldstein,Camalet} in relation with current technology \cite{Lupascu}. Then we shall generalize the discussion to spin-fermion systems which yield a mapping to the spin-boson model, and address hybrid systems of fermions and bosons. Pioneering circuits were introduced in Ref. \cite{Dousse}. 

We pursue the analysis with arrays of spin-boson models both in Sec. 2 and Sec. 4 by showing (new) results on dissipative spin chains. This study is directly motivated by our proposal in ultra-cold atoms in Ref. \cite{orth_stanic_lehur}.  In Sec. 4, we show that such dissipative spin chain models are related to the study of Meissner currents in ladder systems in the Mott regime. An artificial gauge field can then control the transverse magnetic field in the spin systems.

\section{Dynamics of spins}
\label{sec:dynamics-spins}
In this Section, we study the dynamics of the spin-boson model in Eq.~\eqref{eq:11}. We start with the well-known and intuitive Bloch equations for a spin in a magnetic field. We then show how the stochastic Schroedinger equation (SSE) approach~\cite{PAK-1,PAK-2} can be interpreted as a Bloch equation in a random magnetic field, where the randomness arises from the quantum and thermal fluctuations of the environment. We discuss that the long-time steady state can be characterized by an effective temperature $T^*$ (which can be negative~\cite{NegativeT}) capturing the entanglement of the spin with its environment~\cite{Williams, KLHannal,Kopp}. We then present two recent applications of the SSE method: (i) to investigate the topology of a dissipative quantum spin~\cite{Loic3}, and (ii) dynamics of spin chains that experience a common bath. 

The SSE allows one to identify the influence of the environment on the topological Chern number $C$. We discuss how a recently introduced dynamical measurement protocol~\cite{PolkovnikovGritsev,Review, Xu}, which was successfully implemented in cQED~\cite{Lehnert,Roushan,AndreasBerry}, and cold-atoms~\cite{atomChern,atomChern2,Esslinger}, breaks down at stronger spin-bath coupling via a bath induced non-adiabatic crossover. Geometric Berry phase~\cite{Berry} properties have been obtained in related experiments in quantum materials~\cite{Guillaume,Graphene,Kim}.

The SSE can also be used to investigate dissipative and driven spin chains with long-range forces \cite{recati_fedichev,orth_stanic_lehur,Carlos_scientific_reports,Loic1,Garry,Nalbach,Marquardt}, in analogy to a quantum Dicke model~\cite{Dicke,Marco}, and that find various experimental realizations~\cite{Maryland,Rey,Martinis,AntoineOptique,Rydberg}. Coupling to a common environment can induce long-time synchronization of two spins~\cite {Loic1,NRG2spins,JeanNoel}. Synchronization mechanisms have been observed in cold atoms~\cite{synchronization_salomon}. They can be theoretically adressed in the Kuramoto model \cite{Kuramoto}, which makes an interesting link between the physics of Josephson junctions, cold-atom Bose-Hubbard models and synchronization in neural networks~\cite{NeuroRecent}. 

We would like to stress that other results on light-matter systems obtained with the stochastic Schr\" odinger equation (SSE) have been summarized in a previous review by several of the authors~\cite{CRAS2016}.

\subsection{Bloch equations and spin-boson model : phenomenology at weak-coupling}

\label{sec:gener-argum-bloch}
We begin with the study of the dynamics of a spin-1/2 particle $\boldsymbol{\sigma} = (\sigma_x, \sigma_y, \sigma_z)$ subject to dissipation in an applied magnetic field ${\bf H}_{appl}$. Here, $\sigma_i$ are the Pauli matrices and the state of the system at time $t$ can be specified uniquely by giving their expectation values $\langle \sigma_i(t)\rangle = \text{Tr} [\sigma_i \rho(t)]$ (spin components). The (reduced) density matrix of the spin can be expressed as
\begin{equation}
\rho(t)
=
\left( \begin{array}{cc}
\frac{1}{2}(1+\langle \sigma_z\rangle) & \frac{1}{2}(\langle \sigma_x\rangle + i \langle \sigma_y\rangle)\\
\frac{1}{2}(\langle \sigma_x\rangle - i \langle \sigma_y\rangle) & \frac{1}{2}(1-\langle \sigma_z\rangle)
\end{array} \right).
\end{equation}
A pure state must satisfy $\rho^2=\rho$ implying that $\langle \sigma_x\rangle^2 + \langle \sigma_y\rangle^2 + \langle \sigma_z\rangle^2=1$. A pure state is thus represented by a point ${\bf n} = (n_x, n_y, n_z)$ with $|{\bf n}|= 1$ on the surface of the Bloch sphere $S^2$. In contrast, a mixed state is represented by a point in the interior of the Bloch sphere. The surface point ${\bf n}$ corresponds to the ground state of the Hamiltonian  $H = - \frac{\gamma}{2} {\bf H}_{appl} \cdot \boldsymbol{\sigma}$, where $\gamma$ is the gyromagnetic ratio and ${\bf H}_{appl} = |{\bf H}_{appl}| {\bf n}$. In the ground state the spin points along ${\bf n}$. 

If the spin-bath coupling is sufficiently weak, one expects the spin dynamics to be described by the celebrated Bloch equations, which describe memoryless and time-local relaxation and decoherence effects. For an applied field pointing along the $x$ direction ${\bf H}_{appl} = |{\bf H}_{appl}| {\bf x}$, they read
\begin{eqnarray}
\frac{d S_x}{dt} &=& - \frac{S_x - S_x^{(eq)}}{T_1} \\ \nonumber
\frac{d {\bf S}_{\perp}}{dt} &=& \omega_0 {\bf x}\times {\bf S}_{\perp} - {\bf S}_{\perp}/T_2 \,.
\label{eq:9}
\end{eqnarray}
Here, $S_i=\langle {\sigma_i} \rangle/2$, ${\bf S}_{\perp} = (S_y, S_z)$, $\omega_0 = \gamma |{\bf H}_{appl}|$ is the Larmor frequency associated with the applied field. Due to the coupling to a thermal bath the spin dynamics tends towards equilibrium 
\begin{equation}
\label{eq:2}
S_x^{eq} = \frac{1}{2}\tanh(\beta\omega_0/2)
\end{equation}
with $\beta=1/(k_B T)$. At zero temperature the spin is in the ground state and $S_x^{eq}(T=0)=1/2$, while thermal fluctuations lead to $S_x^{eq}<1/2$ when $T>0$. Note that a population inversion refers to $S_x^{eq}<0$ and corresponds effectively to a negative (absolute) temperature. This situation will occur below based on purely non-equilibrium dynamical protocols. 

The time scales $T_1$ and $T_2$ are the longitudinal and transverse spin relaxation times measured, for example, in nuclear magnetic resonance (NMR). They describe energy relaxation $(T_1)$ and dephasing $(T_2)$ processes due to the coupling of the spin to its environment. For a nuclear spin this occurs primarily via the hyperfine interaction to nearby electronic spins. A straightforward theory of $T_1$ and $T_2$ can be obtained within perturbation theory by considering a coupling to a Markovian (i.e. memoryless) fluctuating magnetic field $H_{spin-env} = \lambda {\bf S} \cdot {\bf H}_{env}$ and calculating golden rule transition rates \cite{leggett}
\begin{equation}
  \label{eq:1}
  \frac{1}{T_1}  = \frac{\lambda^2}{4} \int_{-\infty}^\infty d\tau e^{i \omega_0 \tau} \langle \{H^-_{env}(\tau), H^+_{env}(0)\} \rangle  = \frac12 J(\omega_0) \coth( \beta \omega_0/2) \,.
\end{equation}
Here, we have defined the spectral function of the environment $J(\omega)$, which within NMR is proportional to the imaginary part of the electronic spin susceptibility, and we used the fluctuation-dissipation theorem. Note that only the field components perpendicular to the static field ${\bf H}_{appl}$ lead to relaxation, because spin flips are necessary.

Fluctuations of the parallel field components $H^x_{env}$, however, still contribute to dephasing of the Larmor precessions as described by $T_2$. This can be understood intuitively by considering a field ${\bf H}_{env} = (H^x_{env},0,0)$ that is aligned with the $x$-axis and thus parallel to ${\bf H}_{appl}$. Let us start at time $t=0$ with the spin polarized along the $z$-direction. It is convenient to eliminate the effect of Larmor precession by going to a frame rotating with angular velocity $\omega_0$ around the $x$-axis. We define the complex quantity $S_y + i S_z = S_{+}e^{i \varphi}$, whose magnitude $|S_{\perp}|$ will remain constant (equal to 1/2 according to the definitions) for any realization of the noise. Its phase, however, will precess randomly as a result of the noisy environment (by analogy to the procedure described below Eq. (2)), with a new operator $B(t)$ that is proportional to ${\bf H}_{env}$:
\begin{equation}
\frac{d\varphi}{dt}  = B(t) \rightarrow \varphi(t) = \int_0^t dt' B(t')\,.
\end{equation}
Let us assume Gaussian statistics for $B$, which is satisfied for many of the models we consider below, for example, those that describe the bath by an ensemble of harmonic oscillators. It then holds that 
\begin{equation}
\Bigl\langle \exp \bigl[ i \int_0^t B(t') dt' \bigr] \Bigr\rangle = \exp\Bigl( -\frac{1}{2} \int_0^t dt' \int_0^t dt'' \langle B(t') B(t'')\rangle\Bigr)\,.
\end{equation}
For a Markovian environment, e.g., based on harmonic oscillators at high temperatures, where  $\langle B(t') B(t'')\rangle \propto k_B T \delta(t' - t'')$, the dominant behavior at long times $t$ is
\begin{equation}
S_+(t) \equiv \exp(-t/T_2),
\end{equation}
where (by analogy to Eq. (5))
\begin{equation}
\label{eq:6}
T_2^{-1} = J(\omega_0) \frac{k_B T}{\omega_0} \,.
\end{equation}
The environment produces dephasing of the off-diagonal elements in the spin reduced density matrix via both its parallel field components ${\bf H}_{env} \parallel {\bf H}_{appl}$ and its transverse ones ${\bf H}_{env} \perp {\bf H}_{appl}$. Relaxation of the diagonal elements (populations) to thermal equilibrium, as described by Eq.~\eqref{eq:2}, is caused by the transverse components only (see Eq.~\eqref{eq:1}). Dephasing here refers to the exponential relaxation in time of the off-diagonal matrix elements of the spin density matrix caused by the average of the fluctuating phase in Eq. (7). The $B(t)$ field fluctuates very rapidly in time suppressing the quantum phase information contained in the off-diagonal elements of the spin reduced density matrix. Furthermore, the particular form of $T_2$ with temperature and $J(\omega_0)$ can be seen
as an exemple of Johnson-Nyquist noise in an electrical circuit where a resistance induces voltage fluctuations (similar to Eq. (8)) proportional to the resistance and to $k_B T$. 

\subsection{Exact solution for pure dephasing}
\label{sec:exact-solution-pure}
Let us now go beyond a phenomenological discussion and calculate $T_2$ for a microscopic model, the paradigmatic (standard) spin-boson model shown in Eq.~\eqref{eq:11}. To obtain $T_2$, we consider the situation of pure dephasing and set $\Delta = 0$ \cite{Makhlin}. If we prepare a spin in the $x$-$y$-plane at $t=0$, it will perform undamped oscillations $\sigma_\pm(t) = e^{\pm i h_z t} \sigma_\pm(0)$ in the rotated frame, where $\sigma_{\pm} = \frac12 (\sigma_x \pm i \sigma_y)$. Returning to the original frame using $U \sigma_{\pm} U^\dag = \sigma_\pm e^{\mp i \phi}$ defined below Eq. (2), these oscillations are influenced by the environment according to $\sigma_\pm(t) = e^{\pm i \phi(t)} e^{\mp i \phi(0)} e^{\pm i h_z t} \sigma_\pm(0)$. Tracing over the Gaussian bath degrees of freedom and using the cumulant expansion, we find damped oscillations (at times $t > 1/T$)
\begin{equation}
  \label{eq:3}
   \sigma_{\pm}(t) = \langle e^{\pm i \phi(t)} e^{\mp i \phi(0)} \rangle_{bath} e^{\pm i h_z t} \sigma_\pm(0) = e^{-t/T_2} e^{\pm i \tilde{h}_zt} \sigma_{\pm}(0) \,.
\end{equation}
with a dephasing time $T_2$ that is in agreement with Eq.~\eqref{eq:6} if we recall that $J(\omega) = 2 \pi \alpha \omega$:
\begin{equation}
  \label{eq:4}
  T_2^{-1} = 2 \pi \alpha k_B T \,.
\end{equation}
Here, we have used the cumulant expansion to derive $\langle e^{\mp i \phi(t)} e^{\mp i \phi(0)} \rangle_{bath}= e^{- \frac{1}{\pi} Q_2(t) + \frac{i}{\pi} Q_1(t)}$ with the (Ohmic) bath correlation functions~\cite{leggett,weiss}
\begin{equation}
  \label{eq:5}
  Q_2(t) = \int_0^\infty d\omega \frac{J(\omega)}{\omega^2} (1 - \cos \omega t) \coth \frac{\omega}{2 k_B T} =  \pi \alpha \ln (1 + \omega_c^2 t^2) + 2 \pi \alpha \ln \Bigl( \frac{\sinh(\pi t k_B T)}{\pi t k_B T}  \Bigr) 
\end{equation}
and $Q_1(t) = \int_0^\infty d\omega J(\omega) \sin(\omega t)/\omega^2 = 2 \pi \alpha \tan^{-1} \omega_c t$. The function $Q_1$ leads to a renormalization of the oscillation frequency. 

\subsection{Spin-boson dynamics beyond weak-coupling}
\label{sec:stoch-schr-equat}

Now, let us address the quantum limit at zero temperature, where, if we naively apply the high-temperature result of Eq.~\eqref{eq:4}, we would predict $T_2\rightarrow \infty$ for the Ohmic bath. However, as shown below, the bath is subject to zero-point quantum fluctuations. In Eq. (2), the coupling to the environment is along the z direction, and we focus on situations with finite values of $\Delta$.

One straightforward approximate approach to the spin dynamics for non-zero transverse field $\Delta$ is to neglect the feedback of the spin on the environment. Under this assumption the bath operators evolve freely as $b_k(t) = b_k e^{- i \omega_k t}$. The exact Heisenberg equations of motion for the spin $i \dot{\boldsymbol{\sigma}} = [\boldsymbol{\sigma},H']$ read $\dot{\sigma}_z(t) = - i \Delta (\sigma_+(t) e^{i \phi(t)} + \text{h.c.} )$ and $\dot{\sigma}_+(t) = - i \frac{\Delta}{2} \sigma_z(t) e^{- i \phi(t)}$. Here, we have used the ``polaronic form'' of the spin-boson Hamiltonian introduced above $H' = U^\dag H U = \frac{\Delta}{2} ( \sigma_+ e^{i \phi} + \text{h.c.} ) + \sum_k \omega_k b^\dag_k b_k$ and set $h_z = 0$ for simplicity. Tracing over the bath degrees of freedom like in Sec.~\ref{sec:gener-argum-bloch}, we arrive at 
\begin{equation}
\label{eq:7}
\frac{d}{dt} \sigma_z = -\Delta^2 \cos(\pi \alpha) \int_0^t ds \exp\bigl(-Q_2(t-s)/\pi\bigr) \sigma_z(s)
\end{equation}
for $\omega_c t \gg 1$. This equation assumes weak-coupling or separable states because the function $Q_2$ is evaluated with the bath degrees of freedom only.
In the absence of noise from the environment ($\alpha = 0$), the $Q_2$ function vanishes and one recovers the usual Rabi formula $d^2 \langle \sigma_z\rangle/dt^2 = -\Delta^2 \langle \sigma_z(t)\rangle$ meaning undamped oscillations between state $|\uparrow\rangle_z$ to $|\downarrow\rangle_z$ with a frequency $\Delta$. The coupling to the environment causes both damping, as described by $Q_2(t)$, and a renormalization of the oscillation frequency $\Delta \rightarrow \Omega$, which is described by $Q_1(t)/\pi = 2 \alpha \tan^{-1} \omega_c t = \pi \alpha$ for $\omega_c t \gg 1$. At large temperatures $Q_2(t) = 2 \pi^2 \alpha k_B T t$ leading to exponential damping as discussed in the previous section. In contrast, at low temperatures there exist long-range non-Markovian memory effects in the spin dynamics that are mediated by the bath. The right-hand side of Eq.~\eqref{eq:7} then depends on the full history of the spin $\sigma_z(s)$ at times $s < t$. In particular at $T=0$ one finds $Q_2(t) = \pi \alpha \ln (1 + \omega_c^2 t^2)$ such that memory effects only decay algebraically. 

One can solve Eq.~\eqref{eq:7} through Laplace transformation and extract, for example, the exact quality factor of the damped oscillations $\langle \sigma_z(t)\rangle \sim \exp(-\gamma t)\cos(\Omega t)$ at $T=0$ as~\cite{leggett,weiss} (see Fig.~\ref{fig:2})
\begin{equation}
\label{eq:8}
\frac{\Omega}{\gamma} = \cot \frac{\pi \alpha}{2(1-\alpha)}.
\end{equation}
It is important to stress that the memory effects in the environment not only produces the decay of the Rabi oscillations, but also gives a visible Lamb shift for the spin frequency caused by the bath in the vacuum state. The effective Rabi frequency vanishes at the Toulouse limit. Both frequency $\Omega$ and damping rate $\gamma$ are proportional to the renormalized transverse field $\Delta_r = \Delta (\Delta/\omega_c)^{\alpha/1-\alpha}$~\cite{leggett,weiss}. This quality factor has also been obtained from the non-interacting-blip approximation (NIBA) within a real-time influence functional path integral description~\cite{leggett,weiss}. Interestingly, Rabi oscillations disappear at $\alpha = 1/2$, where the quality factor in Eq.~\eqref{eq:8} vanishes. The dynamics at this special ``Toulouse point'' $\alpha =1/2$~\cite{Toulouse} can be solved exactly $\langle \sigma_z(t) \rangle = e^{-\pi \Delta^2 t /(2 \omega_c)}$ via a mapping to a free fermion resonant level model~\cite{leggett, resonant}. The dynamics remains completely incoherent for larger dissipation strengths $1/2 \leq \alpha \leq 1$ \cite{KLH,Lesage}. At $\alpha_c = 1$ (and $T=0$) the system undergoes the localization quantum phase transition and for $\alpha \geq \alpha_c$ the spin thus remains trapped in its initial state $\ket{\uparrow}_z$ (or $\ket{\downarrow}_z$) even for finite $\Delta$ \cite{KLH}. 

Let us now go beyond the approximation of separable spin and bath states. It is important to realize that without this assumtion Eq.~\eqref{eq:7} turns into:
\begin{equation}
\langle \dot{\sigma}_z(t) \rangle = - \Delta^2 \cos(\pi \alpha) \int_0^t ds \bigl\langle \cos(\phi(t) - \phi(s)) \sigma_z(s) \bigr\rangle.
\end{equation}
Even if the full density matrix of the system initially at $t=0$ factorizes into a spin and a bath part, the presence of spin-bath coupling generates correlations between the spin trajectories and the environment over time. 
These spin-bath correlations are small at short times and for weak spin bath coupling $\alpha \ll 1$, which are exactly the conditions that justify the NIBA~\cite{leggett}. 
However, at long times or in the presence of time-dependent (bias) fields $h_z(t)$, and in particular at stronger spin-bath coupling, these correlations become important. In the path integral language these additional correlations are embodied in the longer range blip-blip (and blip-sojourn) interactions~\cite{leggett}. This shows a necessity to develop approaches to tackle these non-perturbative problems. 

A variety of analytical and numerical methods have been devised over the years to investigate the spin-boson model dynamics beyond the weak-coupling (Born) and Markovian approximation. Below, we describe in more detail the main idea behind the non-perturbative Stochastic Schroedinger Equation (SSE) method~\cite{PAK-1,PAK-2,Loic1,Loic2,Loic3}. Some results have been compared with those obtained with the Bloch-Redfield approach and Lindblad equations \cite{Maxim}. As mentioned in introduction, its development relied on important earlier works in the literature~\cite{S1,S2,S3,S4,S5}. Other numerical approaches are direct summations of the real-time path integral expression the quasi-adiabatic propagator path integral  approach QUAPI with recent extensions using tensor-networks techniques ~\cite{TEMPO}) or renormalization group (RG) techniques such as time-dependent numerical RG (TD-NRG), functional RG (FRG) (both in real-time and in frequency space), density matrix RG (DMRG) and Wegner's flow equations. Analytical approaches employ, for example, conformal field theory techniques or derive perturbative or even exact master equations beyond the Markov approximation. We also note current theoretical efforts to describe open random walks in terms of stochastic differential Lindblad equations~\cite{BDA}. Different methods are often complementary since they experience different strengths and limitations, i.e., parameter regimes where they work well and when they break down.

\subsection{Stochastic Schroedinger Equation approach}

The Stochastic Schroedinger Equation method begins with the classical representation of the spin variables on the Bloch sphere, called sojourn and blip \cite{leggett,weiss}, representing the spin reduced density matrix diagonal and off-diagonal elements. 
This enables us to encode the spin dynamics in time through classical variables and describe the time evolution of observables through the influence functional path integral representation. The non-perturbative Stochastic Schroedinger Equation approach~\cite{PAK-1,PAK-2,Loic1,Loic2,Loic3} relies on an exact Hubbard-Stratonovich (HS) transformation of the two-time non-local Keldysh path integral expression into the form of a local, time-ordered exponential. The resulting expression can be efficiently determined by solving a simple time-local Schroedinger-type differential equation for a given realization of the HS variables (and the equations for spin observables become then very similar to those of classical Bloch equations with random phases, as described below. This approach goes beyond Eq. (14) since we do not assume separable states between the spin and the bath). The spin expectation values $\langle \sigma_\alpha(t)\rangle$ are then obtained by averaging the resulting spin state trajectories over different realizations of the HS variables, which can be interpreted as Gaussian-distributed random noise of the environment. 

This interpretation becomes most transparent for an Ohmic bath at $\alpha < 1/2$ and large $\omega_c$, where one can make an analogy with the classical Bloch equations (see Eq. (4)) as the stochastic Schroedinger equation takes the exact form~\cite{PAK-1,PAK-2}
\begin{equation}
\label{eq:12}
\frac{d {\bf S}}{dt} = {\bf H}(t)\times {\bf S}(t)\,.
\end{equation}
The effective noisy magnetic field ${\bf H}(t)$ lies in the $x$-$y$ plane
\begin{equation}
\label{eq:13}
{\bf H} = h (\cos \varphi(t), \sin \varphi(t), 0) = (H_x, H_y, H_z),
\end{equation}
with amplitude $h = \Delta \sqrt{\cos (\pi \alpha)}$. The stochastic function $\varphi(t)$ is introduced as a HS field to decouple long-range correlations in time. It represents a classical random field with identical correlation functions in time given by the $Q_2$ function: $\langle \varphi(t) \varphi(s) \rangle_{S} \propto Q_2(t-s) + \text{const.}$, if it is averaged over the Gaussian HS random variables~\cite{PAK-1,PAK-2}. The spin expectation values are then obtained as $\langle \sigma_\alpha(t) \rangle = \langle S_\alpha(t) \rangle_{S}$. We observe that while the analogy with the classical Bloch equations only holds for the $\sigma_y$ and $\sigma_z$ components of the spin, where one can perform the summation over sojourn variables analytically, the SSE method is more general and has been used to obtain $\langle \sigma_x(t) \rangle$ as well as spin-spin correlation functions $\langle \sigma_z(t) \sigma_z(0)\rangle$~\cite{PAK-1,PAK-2}. It is important to mention that in the blip and sojourn approach~\cite{leggett,weiss}, the initial and final boundary conditions must be taken with care. We have done several efforts in this direction and applied the stochastic method to a variety of different situations which require a treatment that goes beyond a decoupled description of spin and bath.  The method was then developed further with two stochastic fields for the dissipative quantum Rabi model and two-spin models \cite{Loicthese}.

First, we have checked that the SSE method confirms the exact quality factor $\Omega/\gamma$ of Eq.~\eqref{eq:8} for $0 < \alpha < 1/2$ (see Fig.~\ref{fig:2})~\cite{PAK-1,PAK-2}, which is obtained both from the weak-coupling, non-interacting blip approximation (NIBA)~\cite{leggett,weiss} and non-perturbative conformal field theory \cite{Lesage}. Curiously, the SSE method cannot be applied to the exactly solvable Toulouse point $\alpha=1/2$. This can be understood from the fact that the point $\alpha=1/2$ is special, as the blips in the path integral become non-interacting~\cite{leggett}. Beyond $\alpha=1/2$, the SSE method can suffer from convergence problems (since the effective Hamiltonian involves exponential functions of the stochastic fields which may have both real and imaginary parts) and one must average over many more realizations of the HS noise to obtain reasonable error bars. 

One of the main advantages of the SSE method is that it allows one to consider arbitrary time-dependent bias fields $h_z(t)$. This was used to investigate dissipative versions of the classic Landau-Zener and Kibble-Zurek problems. In both situations, the bias field is changed linearly in time $h_z(t) = vt$~\cite{PAK-1,PAK-2,Loic1}. Similar time-dependent protocols can be used to map out the topology of a quantum spin on the Bloch sphere~\cite{PolkovnikovGritsev, Lehnert, Roushan, AndreasBerry}. The influence of dissipation on the spin topology is discussed in detail in the Section~\ref{sec:topol-diss-quant}. 

Using SSE the dissipative Landau-Zener problem was theoretically investigated directly in the universal scaling regime of a large bath bandwidth $\omega_c \gg \Delta_r$, which is complementary to previous studies in Ref. \cite{Kayanuma}. It was found that even long after the Landau-Zener level crossing has occured $\Delta_r \ll h_z(t) \ll \omega_c$, the spin experiences a \emph{universal} decay from the upper to the lower level. This occurs due to boson assisted spin transitions, which include the emission of a boson into the bath to carry away the energy, and is possible as long as $h_z(t) \lesssim \omega_c$.

Another application is to consider parameter sweeps across (second-order) phase transitions and the production of (topological) defects in the final phase as a function of sweep speed. A fast Landau-Zener transition can be described thanks to the Kibble-Zurek mechanism, which predicts the production of topological defects when sweeping dynamically through quantum phase transitions~\cite{Damski}. This description splits the dynamics into three consecutive stages: it is supposed to be adiabatic far away from the level crossing point $h_z(t) = 0$, then evolves in a non-adiabatic way near the crossing, and finally becomes adiabatic again. This interpretation permits to express the probability of a non-adiabatic transition, which is proportional to the density of topological defects, with respect to the sweep velocity and the energy gap at the transition point, recovering the Landau-Zener formula at large velocity. Using SSE we have shown that this interpretation still holds at the mean-field level in the case of $N$ spins coupled to a common ohmic bath, if one considers that the main effect of the environment is to induce a strong ferromagnetic Ising-like interaction between spins~\cite{Loic1}.

Let us finally note that that the derivation of stochastic Schr\" odinger type equations for models beyond the Ohmic spin-boson model, can require more than one stochastic field, as we have shown for example in the context of dissipative and driven light-matter systems~\cite{Loic2} and for two dissipative coupled spins~\cite{Loic1}. Spatial disorder problems in one dimension also require more than one stochastic fields to solve the problem~\cite{Zoran}. For two dissipative coupled spins, our results from SSE agree with previous results obtained with the numerical renormalization group~\cite{NRG2spins} and with quantum Monte Carlo \cite{QMC2spins}. In addition, novel results have been obtained regarding synchronization of spins, which we discuss below in Sec.~\ref{sec:synchr-reviv-from}, and the dynamics in the localized regime, as well as correlation functions between two spins~\cite{Loic1}. 

\subsection{Effective Boltzmann-Gibbs description}
 \label{sec:effect-boltzm-gibbs}
 
\begin{figure}[t]
\center
\includegraphics[width=.65\linewidth]{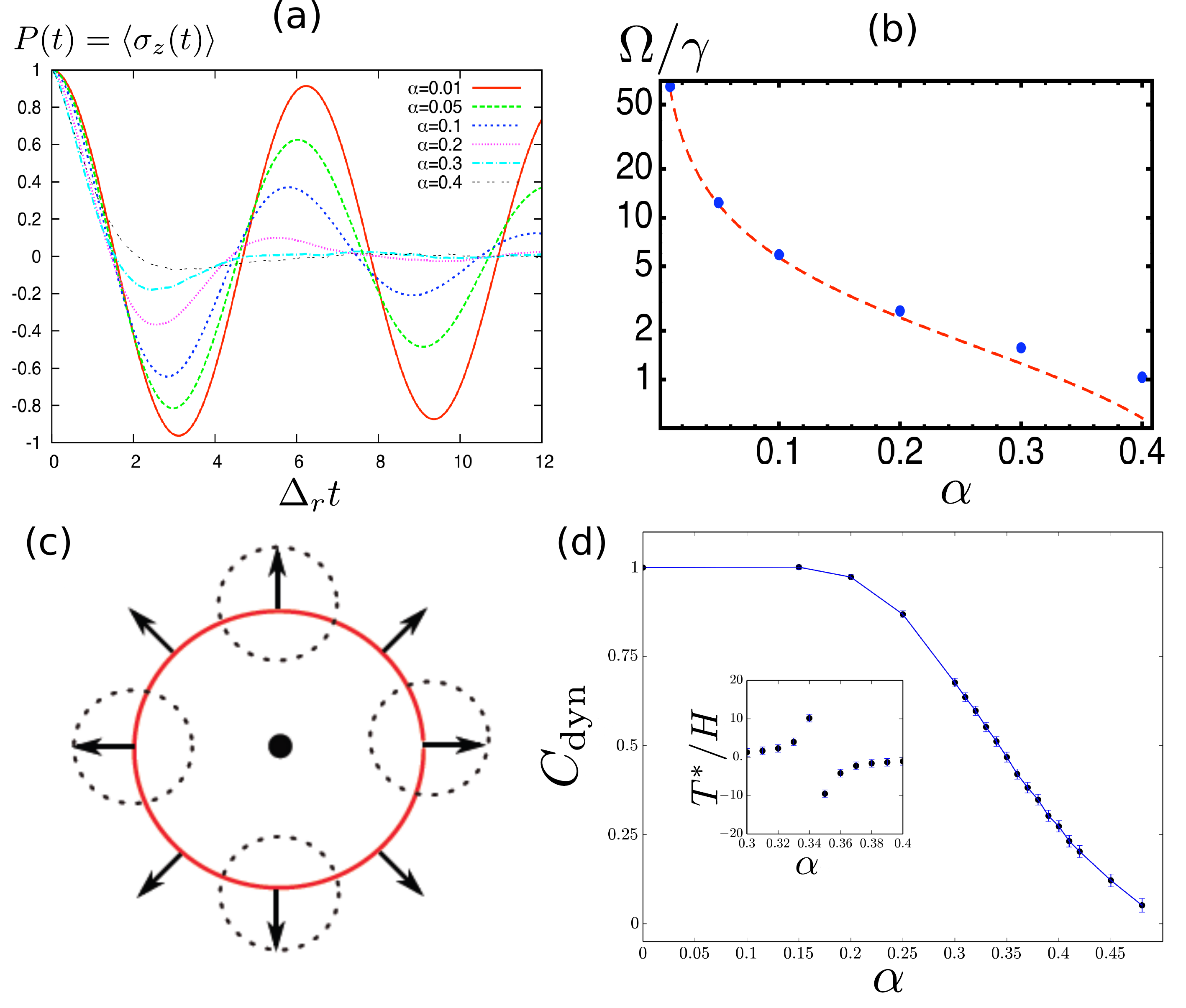} 
\caption{(a) Evolution of the Rabi dynamics $P(t)=\langle \sigma_z(t)\rangle$ as a function of the coupling with the environment with the first implementation of the stochastic Schr\" odinger equation approach with one stochastic field (see Eq.~\eqref{eq:12}). (b) Quality factor $\Omega/\gamma$ of the damped oscillations and comparison with Eq.~\eqref{eq:8} derived within the NIBA approximation. In Ref.~\cite{PAK-1,PAK-2}, there is a precise comparison between the effects of increasing the coupling with the environment at $T=0$ and of increasing the temperature. Recent developments with more than one stochastic fields allow to approach the point $\alpha=1/2$ more closely \cite{Loic3}. However, the method still suffers for $\alpha>1/2$ from convergence problems. For two spins, the quantum phase transition occurs for smaller coupling strengths allowing a very precise analysis of the dynamics in the localized phase~\cite{Loic2}. (c) Cartoon of spin winding on the Bloch sphere in relation with the d-vector ${\bf d} = (H\sin \theta \cos \phi , H \sin \theta \sin \phi, H \cos \theta)$ when $\theta$ vary. Note that it is sufficient to consider the variation of the ground state with $\theta$ to characterize the topology due to azimuthal symmetry of the spin-boson Hamiltonian. (d) Result obtained from the Stochastic Schr\" odinger Equation showing that the "non-adiabatic" Chern number $C_{dyn}$ can become non-quantized and be strongly reduced (compared to the equilibrium quantized Chern number $C=1$) when increasing the dissipation strength; here $v/H=0.08$ and $H/\omega_c=0.01$~\cite{Loic3}. The crossover occurs when $\Delta_r\sim v$ since $C=C_{dyn}+O(v/\Delta_r)$. (inset) We show the effective temperature $T_{eff}$ (defined as $T^*$ for this particular protocol). There is an inversion of population when $C_{dyn}=1/2$ and the environment produces an effective field compensating for the applied field. In Sec.~\ref{sec:dynam-meas-spin}, we describe a simple toy mean-field model to describe this quantum dynamo effect, which occurs for half a Floquet time period, and therefore on short time scales.}
\label{fig:2}
\end{figure}

 Here, we establish a relation between decoherence of Rabi oscillations, entanglement with the environment and the notion of an effective temperature $T_{eff}$, which will be useful to re-interpret certain of our results in Sec.~\ref{sec:topol-diss-quant} and~\ref{sec:quant-brown-moti}. Essentially, when we increase the coupling between spin and bath in the quantum limit, this produces more and more entanglement and therefore the small system is not at thermal equilibrium anymore \cite{Williams}. However, the problem can be re-interpreted as a statistical model in the grand canonical ensemble where the bath is at an effective temperature $T_{eff}$ and weakly-coupled to the small quantum system. This analogy is already useful in the study of Rabi oscillations (Fig.~\ref{fig:2}) since we observe an analogy in the decoherence effects between increasing the coupling with the environment at zero temperature and increasing the temperature in the weak-coupling limit \cite{PAK-1,PAK-2}. 
 
An analogy with a Boltzmann-Gibbs grand canonical ensemble can be then formulated as follows. Let us replace the quantum bath by a thermal bath described by an effective temperature $T_{eff}$, such that $\beta_{eff}=1/(k_B T_{eff})$,  and weakly coupled to the spin, allowing for a statistical analogy. The spin-1/2 then would be described by the effective Hamiltonian $\hat{H}=\frac{1}{2}(\Delta^* \sigma_x + h_z^* \sigma_z)$ with eigenvalues are $E_{\pm}= \pm \sqrt{(\Delta^*)^2 + (h_z^*)^2}$.  In this pseudo-equilibrium picture, the spin reduced density matrix is described by the two eigen-values $\lambda_{\pm} = \exp(-\beta_{eff} E_{\pm})$. Following Ref. \cite{Williams}, then one can identify the eigenvalues of the spin reduced matrix in the spin-boson model in the quantum limit with the effective thermal Gibbs weights $\lambda_{\pm}$ leading to $T_{eff} = \sqrt{(h_z^*)^2 + (\Delta^*)^2}/\ln (\lambda_+/\lambda_-)$ \cite{Williams}. From the spin-boson model at zero temperature, we also have $\lambda_{\pm} = 1/2(1 \mp p)$ where $p=\sqrt{\langle \sigma_x\rangle^2 + \langle \sigma_z\rangle^2}$ (here, $\langle \sigma_y\rangle=0$) and the eigenvalues are calculated with the Hamiltonian in the quantum limit.  In the weak-coupling limit of the spin-boson model, $\beta_{eff}=1/(k_B T_{eff})$ tends to infinity as $\lambda_+\rightarrow 0$. More precisely, the spin resides in its ground state with $p=1$. Increasing the coupling with the quantum bath produces some uncertainty and mixed state for the spin-1/2 such that $0\leq p <1$. We also note that a similar steady-state Gibbs description has been introduced for driven mesoscopic systems \cite{Hershfield,Natan,Doyon,Prasenjitcurrent} and in one-dimensional quantum systems  \cite{LauraLeticia,Laura,Cecile}. This way of reasoning has also resulted to the development of entanglement spectrum in the field of many-body quantum systems; when the sub-system is large enough one finds a universal effective temperature in the sub-system \cite{HaldaneLi,RANA}. It is important to note applications of these ideas in correlated quantum materials \cite{Gabi} and many-body localization \cite{Pollmann,Alet}.

Now, let us interpret the results of the Rabi dynamics at the Toulouse limit $\alpha=1/2$ in terms of an effective steady-state thermodynamics. When focussing on Rabi oscillations, we apply the external field $H_{appl}=\Delta$ along the $x$ direction (see Eq. (17)) and the spin is prepared in one state $|\uparrow\rangle_z$ or $|\downarrow\rangle_z$. At the Toulouse limit $\alpha=1/2$, we observe that similar to the high-temperature limit, at zero temperature $\langle \sigma_z(t) \rangle \sim \exp (-\gamma t)$, where formally $\gamma=1/T_2=\Gamma$ and $\Gamma$ can be interpreted as the width of an effective resonant level model $\Gamma \sim \Delta^2/\omega_c\sim \Delta_r$ \cite{leggett,weiss,KLH}. Important quantitative results have also been obtained using the NRG \cite{Andersschiller} and analytical renormalization group methods \cite{Schoeller}. $T_2$ traduces the typical time to form a resonance between the spin and the environment, and the spin will relax to the effective equilibrium with $\langle \sigma_z\rangle \rightarrow 0$ when $h_z\sim 0$. Computing the magnetization at equilibrium in the $x$ direction gives  $\langle \sigma_x\rangle \sim \Delta/\omega_c\rightarrow 0$ when $\omega_c\gg \Delta$ \cite{KLHannal}, which suggests by analogy to the thermal formula in Eq. (5), a temperature for $T_{eff}$ much larger than $\Delta_r$. In the pseudo-equilibrium picture, the two states become equally populated as $\lambda_+ \sim \lambda_-$. In the quantum problem, this traduces the entrance to a strongly entangled and incoherent regime (we note the absence of Rabi oscillations even at short-time scales). This regime is also related to the Kondo regime of the spin-boson model, which concerns $\alpha$ closer to unity \cite{Blume}. It is perhaps relevant to mention the close analogy between increasing temperature and increasing the coupling with the environment at zero temperature, in the context of Rabi dynamics, has been carefully analyzed with the stochastic Schr\" odinger equation approach \cite{PAK-1,PAK-2}. It should be noted that a similar (and universal) incoherent dynamics has been obtained in the synchronized regime of two spins using the stochastic Schr\" odinger approach \cite{Loic1}.

\subsection{Topology of a dissipative quantum spin on the Bloch sphere}
\label{sec:topol-diss-quant}

It is well-known that a spin in a cyclically varying magnetic field ${\bf B}(t+t_0) = {\bf B}(t)$ acquires a geometric Berry phase $\varphi_{Berry}$~\cite{Berry} after period $t_0$. For an adiabatic process, it corresponds to the phase difference acquired by the (ground) state of the spin $|g\rangle$ during one cycle. A phase $\varphi_{Berry}$ is accumulated during adiabatic evolution along any closed loop in parameter space. Being a geometric quantity, the Berry phase is simply given by half of the area subtended by the closed path followed by the spin on the Bloch sphere. 

Another way of thinking about the topology of a spin is to consider the geometric properties of a Hamiltonian mapping between the parameter space ${\bf d}$ of a Hamiltonian $H[{\bf d}]$ and the Bloch vector ${\bf n} = (\sin \theta \cos \phi , \sin \theta \sin \phi,  \cos \theta)$ characterizing the ground state $|g\rangle$ of the Hamiltonian (of course, one could also consider the excited state manifold). For a spin-$1/2$ in a magnetic field ${\bf d}$, the Hamiltonian reads
\begin{equation}
\hat{H} = -\frac{1}{2} {\bf d}\cdot \boldsymbol{\sigma},
\end{equation}
where the direction of the ${\bf d}$ vector in Fig.~\ref{fig:2}(c) (defined in caption of Fig.~\ref{fig:2}(c) and proportional to ${\bf n}$) depends on two parameters $\theta_d \in [0, \pi)$ and $\phi_d \in [0, 2 \pi)$ in a periodic manner. One can show that the Berry phase accumulated along a closed path can be related to the infinitesimal changes of the ground state wavefunction when varying $\theta_d$ and $\phi_d$. Defining the components of the Berry connection $A_{\phi_d} = \langle g| i \partial_{\phi_d} |g\rangle$ and $A_{\theta_d} =  \langle g| i \partial_{\theta_d} |g\rangle$, one finds more precisely that the Berry phase corresponds to the circulation of the Berry connection along the loop $\varphi_{\text{Berry}} = \int_{C} d \boldsymbol{l} \cdot \boldsymbol{A}$. The (gauge invariant) Berry curvature, defined by
\begin{equation}
F_{\phi_d \theta_d} = \partial_{\phi_d} A_{\theta_d} - \partial_{\theta_d} A_{\phi_d},
\end{equation}
allows for a characterization of the evolution of $|g\rangle$ upon variation of both variables, and the integral of the Berry curvature over the parameter variables $0 \leq \theta_d \leq \pi, 0\leq \phi_d < 2 \pi$ amounts to an integer, the Chern number. In particular, when ${\bf d} = H (\sin \theta_d \cos \phi_d , \sin \theta_d \sin \phi_d, \cos \theta_d)$, we have the mapping $\theta =\theta_d $ and $\phi =\phi_d $ corresponding to a Chern number
\begin{equation}
C = \frac{1}{2\pi} \int_0^{2\pi} d\phi_d \int_0^{\pi} d\theta_d F_{\phi_d\theta_d} = 1.
\end{equation}
A non-zero Chern number $C = 1$ reflects the fact that the total Berry (or magnetic) flux through the Bloch sphere is non-zero. This Berry flux can be thought as being caused by a magnetic monopole that is trapped inside the Bloch sphere at the origin $\boldsymbol{d} =0$, where the system become gapless and the energy is twofold degenerate. In other words, having the Chern number $C = 1$ is due to the fact that the ground state manifold $\{|g(\theta_d, \phi_d)\rangle\}$, which clearly depends on ${\bf d}$, wraps the Bloch sphere once when $\theta_d$ and $\phi_d$ are varied in their full domains. Due to azimuthal symmetry ($F_{\phi_d \theta_d}$ does not depends on $\phi_d$), this corresponds to a winding around the Bloch sphere as $\theta_d$ is varied from $0$ to $\pi$ (see Fig.~\ref{fig:2}(c)). Other choices of ${\bf d}$ can lead to other values of $C$, e.g., a field ${\bf d} = (0,0,H_0) + H(\sin \theta_d \cos \phi_d ,  \sin \theta_d \sin \phi_d,  \cos \theta_d)$ will lead to $C=0$ for $H_0 > H$ corresponding to no winding. More generally one can show~\cite{Loic3} that in the presence of azimuthal symmetry, we have the expression of the Chern number 
\begin{align}
  \label{eq:14}
  C=-\frac12 \Bigl( \langle \sigma^z\rangle(\theta_d=\pi)-\langle \sigma^z\rangle (\theta_d=0) \Bigr) \,.
\end{align}
For an isolated spin-1/2 particle, this formula reproduces $C=1$ for a radial magnetic field.
The Chern number and Berry phase of a quantum spin-1/2 have been measured in recent experiments with superconducting circuits~\cite{Lehnert,Roushan,AndreasBerry}. Experiments in ultra-cold atoms have measured the Chern number related to Bloch bands, where the pseudo-spin can refer to the lattice properties with two inequivalent sites in the unit cell \cite{atomChern2,Esslinger}.

Let us now investigate the robustness of the Chern number under the influence of dissipation in the Ohmic spin-boson model for ${\bf d} = H (\sin \theta_d \cos \phi_d , \sin \theta_d \sin \phi_d, \cos \theta_d)$. Remarkably, we find that the Chern number remains unchanged by the coupling to the environment until one reaches the localization quantum phase transition of the spin-boson model at strong-coupling $\alpha_c \sim 1$ (much beyond the Toulouse limit $\alpha\sim 1/2$ which usually means maximal entanglement between spin and bath \cite{KLHannal,Kopp}). More precisely, in the presence of the bath, the ground state can be generally written as~\cite{Loic3}:
\begin{equation}
|g \rangle = \frac{1}{\sqrt{p^2 + q^2}}\left( p e^{- i \phi} |\uparrow\rangle_z \otimes | \chi_{\uparrow}\rangle + q |\downarrow\rangle_z \otimes | \chi_{\downarrow}\rangle \right).
\label{ground_state_bath}
\end{equation}
Here, $|\chi_{\uparrow} \rangle$ and $|\chi_{\downarrow} \rangle$ correspond to the two bath states associated with the two spin states. The first observation is that $\langle g| \sigma_z | g \rangle = (p^2 - q^2)/(p^2 + q^2)$. 
Now, we can use exact calculations \cite{KLHannal} to check that $q=0$ close to $\theta_d=0$ and $p=0$ close to $\theta_d = \pi$, due to the fact that the transverse field $\Delta=H\sin\theta_d=0$ vanishes for both $\theta_d = 0, \pi$. The bath then has no effect (meaning $\langle \sigma_z \rangle =\pm 1$), and from the definition of the Chern number in Eq.~\eqref{eq:14}, it follows that as long as one is in the delocalized phase of the spin-boson model $(0\leq\alpha<1)$, the Chern number $C$ remains invariant and is equal to its value at 
$\alpha = 0$. In a recent paper~\cite{Loic3}, we have checked this explicitly using a variational approach that expands the bath states $\chi_{\uparrow, \downarrow}$ in terms of coherent states~\cite{variational_1,variational_2}. One can also show from Eq.~(\ref{ground_state_bath}) that $F_{\phi_d\theta_d}=-\partial_{\theta_d} \langle g| \sigma_z |g\rangle$ and we observe that the quantum phase transition in the ohmic spin-boson model at $\alpha=1$ is characterized by a divergence of the Berry curvature at the equator, where the field $h_z$ is small. More precisely, the Fermi liquid ground state of the ohmic spin-boson model is characterized by $\langle g| \sigma_z | g\rangle \sim h_z/\Delta_r = H\cos\theta_d/\Delta_r$ at small fields $h_z$ \cite{KLH}, with $\Delta_r = \Delta(\Delta/\omega_c)^{\alpha/1-\alpha}$ and $\Delta=H\cos\theta_d$. Close to the equator $\theta_d=\pi/2$, one finds~\cite{Loic3}:
\begin{equation}
- \partial_{\theta_d} \langle g| \sigma_z | g \rangle \sim \left(\frac{\omega_c}{H}\right)^{\frac{\alpha}{1-\alpha}}. 
\end{equation}
The divergence of the spin susceptibility at the quantum phase transition~\cite{Meirong1} leads to a divergence of the Berry curvature at the equator, at the quantum phase transition $\alpha\sim 1$. This reflects the gap closing that occurs at the phase transition due to $\Delta_r \rightarrow 0$. We note that such a divergence of the spin susceptibility is expected to subsist for sub-ohmic environments~\cite{Bulla,subPhilippe}. At finite temperature, the Kosterlitz-Thouless transition at $\alpha\sim 1$ is replaced by a smooth crossover. The effect of the temperature on the results remain to be investigated. Some aspects have been addressed, for example, in Refs.~\cite{BudichDiehl,Uhlmann} with recent experimental progress~\cite{Shi}. Some other theoretical aspects have been addressed on studying properties of the Berry phase with an environment in Ref. \cite{Gefen}. 

Next, we focus on how to measure these dissipative effects in the realistic time-dependent protocol of Refs.~\cite{Lehnert,Roushan,AndreasBerry, Hamburg}.

\subsection{Dynamical measurements of spin topology and influence of the environment: Quantum Dynamo Effect}

\label{sec:dynam-meas-spin}

Let us address a realistic (Landau-Zener) protocol where one changes the polar angle linearly with time $\theta_d = v t$ from the initial time $t_{\text{init}}=0$. The final time of the protocol is $t_{final} = \pi/v$. By increasing the velocity of the protocol $v$, one expects an adiabatic to non-adiabatic crossover when the final time $t_{final}$ is equal to $1/\Delta_r$ (see Fig.~\ref{fig:2}(c)). More precisely, at $t=0$ the spin starts in the $|\uparrow\rangle_z$ state. The typical time to flip the spin from $|\uparrow\rangle_z$ to $|\downarrow \rangle_z$ based on Sec.~\ref{sec:stoch-schr-equat} is of the order of $1/\Delta_r$; therefore, one expects a crossover in the measurement  $-1/2[\cos \theta_d]_0^{\pi} = -1/2 \left(\langle \sigma_z \rangle (\theta_d=\pi) - \langle \sigma_z \rangle (\theta_d=0)\right)$ from $1$ to $0$ when $t_{final}\sim 1/\Delta_r$ or equivalently $v\sim \Delta_r$. This point has been explicitly recovered using the stochastic Schr\" odinger equation method \cite{Loic3}. It is important to note that formally there is no disagreement with the fact that the equilibrium Chern number remains quantized to one. More precisely, defining the dynamical observable 
\begin{align}
  \label{eq:15}
  C_{\text{dyn}}=-\frac12 \Bigl( \langle \sigma_z \rangle (t_{\text{final}} = \pi/v) - \langle \sigma_z \rangle (t_{\text{init}} = 0) \Bigr)
\end{align}
as a dynamical or non-equilibrium ``Chern number'' $C_{\text{dyn}}$. Importantly, one can show that $C=C_{\text{dyn}} + O(v/\Delta_r)$ \cite{PolkovnikovGritsev,Loic3}. 

The effect of the coupling to the bath $\alpha>0$ is to renormalize $\Delta \rightarrow \Delta_r$, which poses stricter requirements on $v$ to remain adiabatic. In particular, for a given fixed value of $v$, we observe a breakdown of the dynamic measurement protocol of $C$ when $\alpha$ is increased. One observes a crossover between $C_{\text{dyn}}=C=1$ to $C_{\text{dyn}}=0 \neq C$ when $v \sim \Delta_r$, corresponding to a crossover to a non-adiabatic regime. Clearly, as soon as $v \sim \Delta_r$ contributions to $C_{\text{dyn}}$ from terms of the order of $O(v/\Delta_r)$ become important. The point corresponding to $C_{\text{dyn}}=1/2$ implies that the environment screens the applied field such that $\langle\sigma_z \rangle (\theta=\pi)=0$ and the effective temperature $T_{eff}$ at the end of the protocol (referred to as $T^*$ in Fig.~\ref{fig:2} for this particular protocol) becomes infinity. Increasing $\alpha$, then $C_{dyn}$ becomes zero and there is an inversion of population, meaning that one observes negative (absolute) temperature. The entanglement entropy of the spin with the environment then shows a maximum when tracing out the environment and focusing on $S=-\hbox{Tr}(\hat{\rho}\log \hat{\rho})$ \cite{Loic3}. The scaling of $C_{dyn}$ close to the Toulouse point  $\alpha=1/2$ was investigated using the Schwinger-Keldysh formalism \cite{Loic3}, in relation with a dynamical type Fermi liquid behavior \cite{Loic3}.

Note that reproducing quantitatively these results using a master equation in the weak-coupling sense is not so straightforward, since the $Q_2(t)$ function cannot be evaluated using the ground state of the environment. There is production of 'photons' in the environment, compensating for the applied magnetic field. We refer to this effect as a quantum dynamo effect (rotating the spin on the Bloch sphere produces excitations, or artificial photons, in the environment). Let us now describe physically this effect in a one-mode representation.

We can interpret the effect of the boson bath as an effective magnetic field 
$\langle \sum_k \lambda_k(b^{\dagger}_k + b_k)  \rangle$ on the spin.  Since the spin is driven at the frequency $v$, then we can expect that the relevant bosonic modes will also be produced at this frequency. Therefore, 
we can build an effective one-mode model to describe more quantitatively this photon emission in the environment, following Ref. \cite{Loic3}:
\begin{equation}
\hat{H}_{eff} = \frac{H}{2}\cos(vt)\sigma_z +  \frac{H}{2}\sin(vt)\sigma_x + \frac{\lambda}{2} \sigma_z (b+b^{\dagger}) + v b^{\dagger} b.
\end{equation}
We obtain coupled equations of motion for the spin and the bosonic mode operator. The spin obeys classical Bloch equations similar to Eq. (17), and the bosonic environment satisfies
\begin{equation}
\frac{1}{v^2} \partial_t^2 h_{ind} + h_{ind} = - \frac{\lambda^2}{v} \sigma_z(t),
\end{equation}
and $h_{ind}=\lambda \langle b + b^{\dagger} \rangle$. 

The driving of the spin triggers the excitation of bath modes at the resonant frequency $v$. Due to the resonance condition, a large number of excitations are created and this induces an effective magnetic field $h_{ind}$ for the spin along the $\hat{z}-$direction. The
strength of this bath-induced field increases with the spin-bath coupling. Therefore, by increasing the spin-bath coupling, one expects to have $|h_{ind}|>H$, i.e., a bath-induced field that compensate the external magnetic field, thus preventing the spin to flip during the dynamical protocol. Fixing the spectral function of the environment or $\lambda=\sqrt{2\alpha v H}$, we find numerically that $|h_{ind}|$ becomes larger than the applied field $H$ for a critical value of $\alpha$ of the order of 1/2. 

\subsection{Mean-Field Stochastic Approach: quantum spin relaxation}
\label{sec:solvable-mean-field}
Now, we would like to address the question of the spin relaxation to equilibrium  in dissipative quantum spin chains. We note that various works have studied theoretically the phase diagram and dynamical properties of dissipative quantum Ising models \cite{Loic3,Lesanosky,Werner,Werner2,Subir,Pankov,Garry,Nalbach}. Let us consider a solvable mean-field quantum Ising spin model. We focus on a system of $M$ interacting spins, which are coherently coupled to one common bath of harmonic oscillators:
\begin{align}
\hat{H} =&\frac{\Delta}{2}\sum_{p=1}^M  \sigma_p^x+\sum_{p=1}^M \sum_{k} \lambda_{k} e^{ik x_p} \left(b^{\dagger}_{-k}+b_k \right) \frac{\sigma_p^z}{2} -\frac{K}{M}\sum_{p \neq r}\sigma^z_p \sigma^z_{r}+\sum_{k} \omega_{k} b^{\dagger}_{k} b_{k}.
\label{ising_1}
\end{align}
This model can be simulated in ultra-cold atoms \cite{recati_fedichev,orth_stanic_lehur,Carlos_scientific_reports}, ion traps \cite{Maryland}, with polar molecules \cite{Rey} and in quantum electrodynamics circuits \cite{Martinis,Marquardt}. In Fig.~\ref{fig:1}(e), we show the setup that we have suggested in Ref.~\cite{orth_stanic_lehur}. We have also shown how the bath degrees of freedom can affect critical exponents at the quantum phase transition between paramagnetic and ferromagnetic phase of the one-dimensional quantum Ising model (or two-dimensional classical Ising model) in this geometry, following Refs. \cite{Werner,Werner2,Subir}. Here, we discuss real-time dynamics close to the quantum phase transition using a solvable mean-field dynamics and with the stochastic Schr\" odinger equation approach which has already provided conclusive results for Kibble-Zurek physics \cite{Loic3}. Without dissipation, this model exhibits a mean-field like second-order transition at zero temperature when $K=\Delta/2$. The quantum phase transition separates a paramagnetic phase to a ferromagnetic phase with Ising long-range correlations in the $z$ direction. Again, we assume an ohmic form for the spectral function and take $\omega_k=v |k|$, where $v$ represents the velocity of the sound modes. Using a unitary transformation \cite{orth_stanic_lehur}, we find that the environment has two effects on the Hamiltonian, a renormalization of the transverse field into $\Delta_r$ similar to the one spin situation, and a renormalization of the spin-spin interaction.  The renormalization of the spin exchange $K_r$ due to the bosonic environment is analogous to the Ruderman-Kasuya-Kittel-Yosida interaction in Kondo lattices \cite{KLH,Hewson} and can occur in light-matter systems \cite{Igor}. The strength of the induced interaction can be controlled through properties of the bath \cite{orth_stanic_lehur}. We assume induced long-range forces, which allows one to formulate a giant spin description.

\begin{figure}[t]
\center
\includegraphics[width=\linewidth]{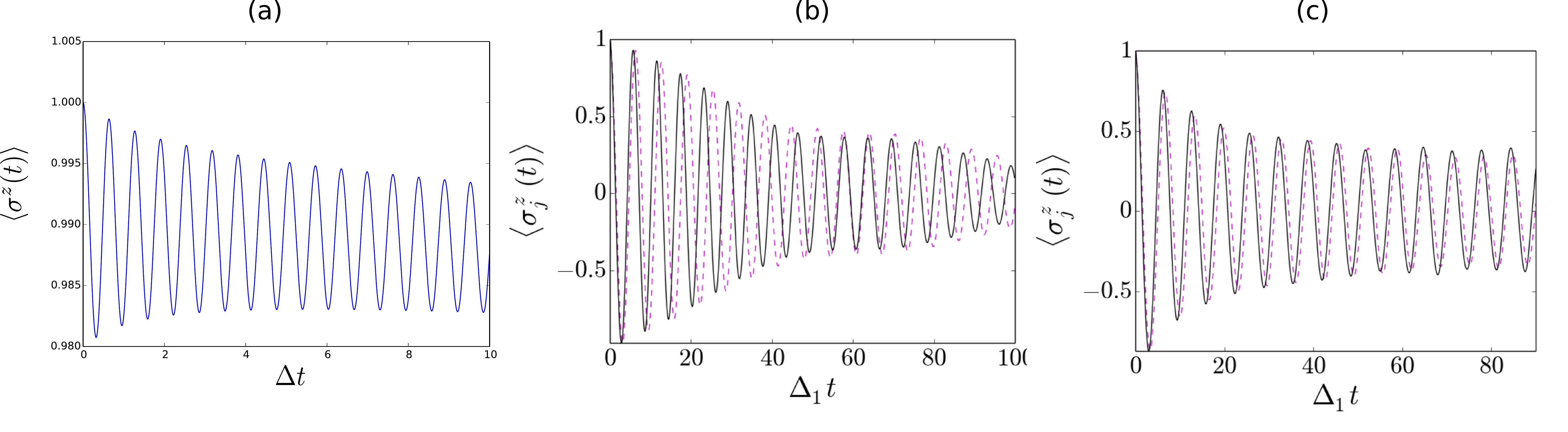}
\caption{(a) Magnetization $\langle \sigma^z\rangle = \langle \sigma_p^z\rangle$ in the mean-field Ising spin chain far in the ferromagnetic phase using the NIBA approximation of Eq. (30). Parameters are fixed to $\Delta=1$, $w_c=3$ and $K_r=10$. The transverse field produces small oscillations around the equilibrium value. (b-c): Synchronization dynamics of $\langle \sigma^z_1 \rangle$ and $\langle \sigma^z_2 \rangle$ for (b) $\alpha=0.01$, (c) $\alpha=0.05$ with $\Delta_2/\Delta_1=1.1$, $\omega_c=20 \Delta_1$, $K=0$.}
\label{fig:3}
\end{figure}

To make a link with the Dicke model in quantum optics \cite{Dicke}, assuming long-range coupling between spins, it is convenient to envision the model as a giant collective spin: $\tau_z  = \sum_{p} \frac{\sigma_p^z}{2}$
and $\tau_x = \sum_p \frac{\sigma_p^x}{2}$. The quantum phase transition can be understood defining an effective Hamiltonian $\hat{H}_{eff} = \Delta_r \tau_x - \frac{K_r}{M} (\tau_z)^2$. We can then introduce the Holstein-Primakoff transformation: 
$\tau_x = b^{\dagger} b - M/2$ and $\tau_z = \sqrt{M}(b + b^{\dagger})$, such that the Hamiltonian takes the simple form (up to a constant)
\begin{equation}
\hat{H}_{eff} = \Delta_r b^{\dagger} b - K_r (b + b^{\dagger})^2 = (\Delta_r - 2K_r) b^{\dagger} b - K_r (b^2 + (b^{\dagger})^2).
\end{equation}
For $2K_r>\Delta_r$, $\langle b^{\dagger} b \rangle \neq 0$ and $\langle \tau_x\rangle$ will progressively diminish in the Ising ordered phase (the transverse magnetization $\langle \tau_x\rangle$ shows a crossover at the transition). Note that a coupling with a bosonic Gaussian bath can be included in such Dicke-type models allowing for a rigorous treatment of non-Markovian effects at low temperatures \cite{Marco}. Dicke type models can also result in novel glassy phases \cite{Strack} and exotic relations between fluctuations and entanglement at a quantum phase transition \cite{Pierre}. The dissipative quantum Ising model can be studied using an effective $\phi^4$ theory and Monte Carlo numerical approaches \cite{Werner,Werner2,Subir,Pankov}. In the case of short-range spin correlations, the dissipation is expected to deeply affect the dynamical exponent $z$ at the transition \cite{Werner,Werner2,Subir}; in contrast, here the transition is mean-field like due to the long-range interactions \cite{Loic3}.  Coupling the spin array to the environment following Ref. \cite{orth_stanic_lehur} and integrating out the Gaussian environment, then we can derive the closed equation starting from the paramagnetic phase (where the choice of mapping is appropriate):
\begin{equation}
\frac{d^2 \langle \tau_z(t)\rangle}{dt^2} \approx -\Delta_{eff}^2 \langle \tau_z(t)\rangle - \int_0^t dt' \alpha(t-t') \frac{d}{dt'} \langle \tau_z(t') \rangle,
\end{equation}
and $\alpha(t-t')$ is related to an effective kernel defining dissipation in relation with the spectral function \cite{orth_stanic_lehur}. In the paramagnetic phase, then we predict a renormalization of the Rabi frequency:
\begin{equation}
\Delta_{eff} \sim \Delta_r\left(1-2\frac{K_r}{\Delta_r}\right).
\end{equation}
Treating the last term in a Markovian manner as $\sim -\eta d \langle \tau_z\rangle/dt$ would suggest an exponential relaxation  of the Rabi oscillations in the paramagnetic phase. Approaching the transition, this analysis would suggest a purely exponential
relaxation towards equilibrium with a relaxation time that would diverge for $\alpha=0$ (in accordance with mean-field theories close to the critical temperature $T=T_c$). A similar analysis has been pushed forward in relation with ultra-fast dynamics in high-Tc superconductors \cite{Luca}. By Fourier transform, dissipation effects in Eq. (30) scale linearly with frequency (in agreement with the Korringa-Shiba relation \cite{leggett,SassettiWeiss,Shiba}, which will be studied in detail in Sec. 3). 

To study the dynamics of $\langle \sigma^z(t)\rangle= \langle \sigma_p^z\rangle$ in the ferromagnetic phase and to make a connection with preceding sub-sections,  we can also derive a closed equation of motion for the mean-field spin dynamics in the framework of the well-known NIBA approximation \cite{leggett,weiss}. One gets  
\begin{align}
\partial_t \langle \sigma^z (t) \rangle=-\Delta^2 \int_{t_0}^t ds \langle \sigma^z (s) \rangle  \cos\left[\frac{Q_1(t-s)}{\pi} \right] \exp\left[-\frac{Q_2(t-s)}{\pi} \right] \cos\left(K_r \int_{s}^t ds' \langle \sigma^z (s') \rangle \right).
\label{Mean_field_EOM_SB}
\end{align}
Numerical results suggest that small oscillations around the equilibrium magnetization value can still take place in the ferromagnetic phase (see Fig.~\ref{fig:3}(a)). Approaching the quantum phase transition from the ferromagnetic side, the spin relaxation time becomes larger than typical time scales accessible in the simulations. 

\subsection{Synchronization: revival from an environment}
\label{sec:synchr-reviv-from}

Now, we show an interesting effect of the environment in the context of spin dynamics and synchronization of two spins, obtained with the stochastic approach \cite{Loic3}. Synchronization mechanisms have been studied in the context of two spins \cite{NRG2spins,JeanNoel} and an ensemble of Josephson-coupled harmonic oscillators by analogy with circuit QED arrays \cite{Kuramoto}. Many-Body versions of synchronization have also been explored in superconducting systems \cite{Barankov}.
 It is important to mention that synchronization mechanisms have attracted a lot of attention recently, at the frontier between physics and neuroscience, and more precisely in the context of neural synchronization and the Kuramoto model, with some relation with Josephson junction arrays and the Brownian motion, as well as spin networks \cite{Kuramoto,NeuroRecent}. Artificial neural networks could also help understanding many-body phenomena, through the development of useful methods \cite{Carleo}.
 
A synchronization regime was recently observed in Ref.~\cite{synchronization_salomon} in the oscillatory dynamics of a mixture of bosonic and fermionic species. The authors suggested that the appearance of this synchronization regime was due to the coupling of the relative motion of the two clouds to a dissipative environment. We build a toy model related to this problem by restricting the dynamics of each species to only two motional states (essentially, the left top and right top of an harmonic trap). The resulting Hamiltonian is analogous to two coupled spins-1/2 impurities and reads:
\begin{align}
\hat{H}=\frac{\Delta_1}{2}\sigma^x_1+\frac{\Delta_2}{2}\sigma^x_2-K\sigma^z_1\sigma^z_2+ \sum_k\left[ \left(\sigma^z_1-\sigma^z_2\right)\frac{\lambda_k}{2} (b_k +b_k^{\dagger})+\omega_k b_k^{\dagger} b_k\right],
\label{hamiltonien}
\end{align}
where the operator $\sigma^z_p$ describes the position of the cloud $p\in\{1,2\}$ (either left or right), while the tunneling term $\sigma^x_p$ switches the position from left to right or from right to left. $\Delta_1 \neq \Delta_2$ are the two bare frequencies of the two species. $K$ denotes the interaction strength between the two spins. We assume that the relative motion of the two species is coupled to an ohmic bath.\\

Fig.~\ref{fig:3}(b-c) shows the dynamics of the two spins, with initial state $|\uparrow_z,\uparrow_z\rangle$ for increasing strengths of the system-environment coupling (from left to right and top to bottom). At very weak coupling 
($\alpha<0.02$), we observe an asynchronous decay of the two spin oscillations towards an equilibrium state with $\langle \sigma^z_1 \rangle=\langle \sigma^z_2 \rangle=0$. Interestingly, we remark that the damping of the oscillations is temporarily smaller when the spins (oscillators) are in phase, signaling the onset of synchronization. In this regime of very weak coupling, the life-time of the oscillations diminishes with increasing $\alpha$. Then, above a certain (small) coupling strength of the order of 0.05, we observe the appearance of long-lived synchronized oscillations of the two spins (this phenomenon occurs until $\alpha=0.15$ without attenuation). In this regime, the two spins oscillate at the same frequency and these oscillations acquire a seemingly infinite life-time. Other initial conditions do not lead to the same synchronized regime, signaling the presence of an attractor in the phase space. An extended study and other results have been presented in Refs. \cite{Loic3,Loicthese} and the limit of large $K_r$ has been discussed in Ref. \cite{NRG2spins}.
  
\section{Realization of spin-boson and quantum impurity models}
\label{sec:real-spin-boson}

Here, we discuss more specifically several geometries that were only briefly introduced in the introductory Sec.~\ref{sec:quant-simul-exper}.

\subsection{Spin-boson model in BECs and Mott-superfluid transition}
\label{sec:spin-boson-model-1}

In relation with Fig.~\ref{fig:1}(e)~\cite{recati_fedichev}, here we discuss engineering of the spin-boson model in ladder systems which may be realized in ultra-cold atoms, Josephson junction arrays and quantum circuits. We build a relation between the theoretical predictions of Fig.~\ref{fig:1}(e) and the Mott-superfluid transition in one dimension. The first realization of the spin-boson model in ultra-cold atoms has been proposed by Recati {\it et al.}~\cite{recati_fedichev} and generalized by us in Ref.~\cite{orth_stanic_lehur}. A generalization to mobile impurities has been addressed in Ref.~\cite{lamacraft}. Here, inspired by Ref.~\cite{mobile1}, we propose a slightly different version on a lattice where the spin-1/2 (a double well impurity) is used to probe the Mott-superfluid transition~\cite{Fisher,Jaksch,Greiner,Maurice} in a ladder system. This geometry also makes a connection with the Josephson-Kondo circuit of light, which will be discussed in Sec. 3.2. We note current efforts to address Kondo physics in ultra-cold atoms with fermionic environments~\cite{Demler}. 

\begin{figure}[t]
\begin{center}
\includegraphics[width=.7\linewidth]{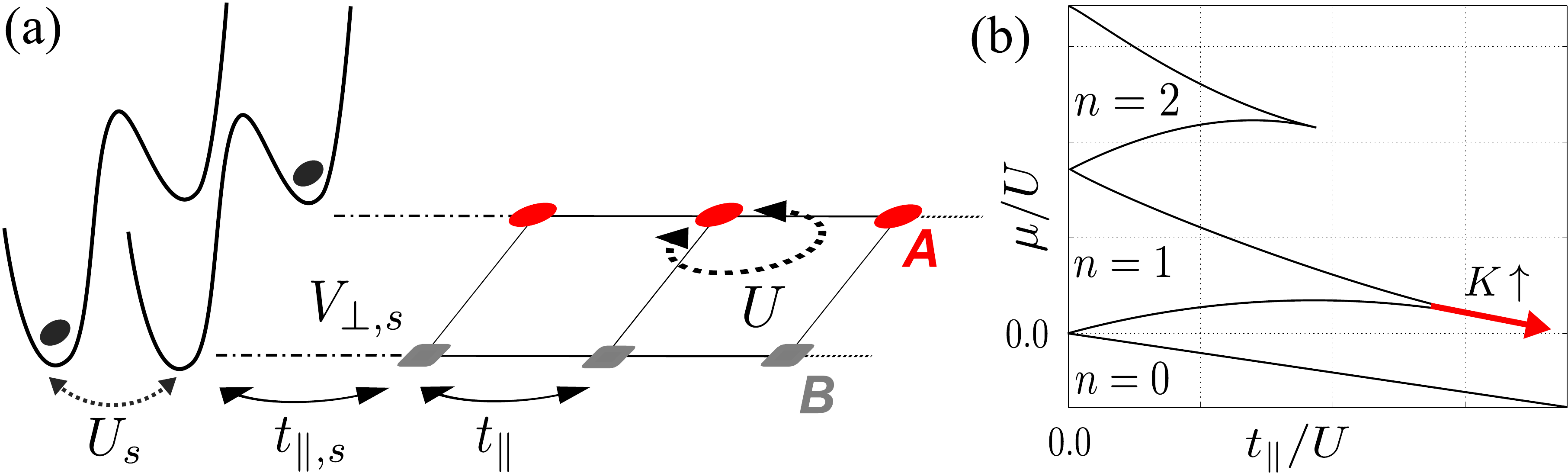}  
		\caption{\textbf{(a)} Model of the spin-impurity (quantum microscope) coupled to the low-energy excitations of the two-leg bosonic ladder. \textbf{(b)} Schematic representation of the Mott-Superfluid quantum phase transition of Berezinskii-Kosterlitz-Thouless type (for density $n=1$, the transition occurs at $t_{\parallel}/U\sim 0.3$ and $K=2$). Here, $\mu$ denotes the chemical potential in the ladder ($\mu_b$ in the text).}
	\label{fig:4}
	\end{center}
\end{figure}

We consider the system defined on the ladder geometry in Fig.~(4a) (a similar spin-boson Hamiltonian can be derived in a single chain geometry). At one extremity of the ladder we create a quantum impurity (microscope, to probe locally the dynamics at one boundary of the chain), a double-well system with occupancy one, as shown in Fig.~\ref{fig:4}. We suppose that interactions are sufficiently strong such that one realizes a state with one atom on the two wells (the two locations of the double well then correspond to the two polarization states of a spin-1/2 particle). The rest of the ladder forms the environment (bath). More precisely, we suppose that particles can hop to and from the impurity with the amplitude $t_{\parallel,s}$. We eliminate states with no occupancy on the microscope by fixing its chemical potential $\mu_s$ to be much bigger than the chemical potential of the bath $\mu_b$, such that $\mu_s - \mu_b = \mu$, and we fix interaction terms $U_s$ and $V_{\perp, s}$ large compared to the other energy scales of the system. Thus, single particle jumps to or from the impurity sites require high energy. We can then build second-order processes coupling the microscope with the ladder, and we obtain the Hamiltonian:
\begin{align}
& \hat{H} = 
\sum\limits_q v |q| b^\dag_{a,q} b_{a,q} +
\frac{\sigma_z}{2} \sum\limits_{q \neq 0}
\lambda_{q}
\left( b^\dag_{a,q} + b_{a,-q} \right) +
\left( n V_{x} + \Delta_0 \right) \sigma_x
\;, \quad \text{where} \notag \\ &
\lambda_{q} = v \sqrt{\frac{\pi|q|}{L}} \left(
\frac{aV_z}{2\pi v}\sqrt{K} -
\frac{1}{\sqrt{K}}
\right)
\;.
\end{align}

(Similarly to Sec. 3.2 and Eq. (39), only the anti-symmetric bosonic mode $b_{a,q}$ couples to the impurity.) In addition, we suppose a direct tunnel coupling $\Delta_0$ between the two wells. 
Here $a$ is the spacing between lattice sites, $L$ -- the total length of the ladder, $n$  -- the filling of the bath, $U$ -- the strength of on-site Bose-Hubbard interactions, $v$ -- the speed of sound associated to the antisymmetric mode of the bath and 
$K$ -- the Luttinger parameter \cite{Haldane,Giamarchi}
\begin{align}
v = a\sqrt{2 n t_\parallel U}
\;, \quad
K = 2 \pi
	\sqrt{ \frac{nt_\parallel}{2U} }
\;.
\end{align}
$t_{\parallel}$ and $t_{\parallel,s}$ are amplitudes of hopping between different sites in the bath and of the particle exchange between the spin and the bath respectively. Coupling amplitudes $V_z$ and $V_x$ are defined as follows:
\begin{align}
V_z &= - t_{\parallel, s}^2 \left(
\frac{1}{\mu + U n} - 
\frac{1}{V_{\perp, s} - \mu} +
\frac{1}{U_{s} - \mu}
\right)
\;, \notag \\
V_x &= - t_{\parallel, s}^2 \left(
\frac{1}{\mu + U n} - 
\frac{1}{V_{\perp, s} - \mu}
\right)
\;.
\end{align}
We identify an ohmic spin-boson model. 
Since low-energy excitations in the ladder system are phonon-like only for wavelengths larger then the healing length $\xi = aK/n$, the high energy cut-off can be expressed as
\begin{equation}
\omega_c = v / \xi = n v / \left( a K\right)
\;.
\end{equation} 
Finally, one can deduce the dimensionless dissipative parameter $\alpha$
\begin{equation}
\label{eq:alpha}
\alpha = 
\frac{1}{2K} \left( \frac{V_z}{2U} - 1 \right)^2 =
\frac{1}{2K} \left[ 
\frac{V_z}{t_\parallel n} \left( \frac{K}{2 \pi} \right)^2 - 1 
\right]^2
\;.
\end{equation}
Dynamics of the spin is strongly related to the value of the dissipation parameter $\alpha$. It is important to emphasize that such a formula stems from sound mode effects in a BEC. 

As an application, we discuss the evolution of the quantum microscope in time in the vicinity of the Mott-Superfluid transition, when fixing the mean density $n=1$ (this can be achieved by fixing the chemical potential in the ladder in Fig.~(4b). In one dimension, such a transition is also of Berezinskii-Kosterlitz-Thouless type, then resulting in a cusp like profile in the phase diagram \cite{Giamarchi}. In the Mott regime, one must be careful when applying Eq. (38) due to strong renormalization effects and formally the dissipative parameter $\alpha = 0$ (phonons become suppressed due to charge quantization): the Luttinger parameter of the bath is renormalized to zero and the spin is completely decoupled from the bath (see Fig.~\ref{fig:4}(b)). In this case we recover perfect Rabi oscillations of the microscope. If we now modify dissipation and reach the tip of a Mott lobe, one can apply Eq. (38) in the superfluid regime, and the dissipation parameter increases (the Luttinger parameter $K = 2$ at the Mott-superfluid transition \cite{GiamarchiMillis}). The Rabi oscillations remain underdampled as long as $\alpha < 1/2$. When $\alpha = 1/2$ the system would leave the coherent regime, in agreement with the arguments in Sec. 2 (for rigorous figures obtained with the stochastic approach, see for example Fig.~\ref{fig:1}). The different regimes are shown in Fig.~\ref{fig:4}(b). It is important to mention the actual efforts both experimentally and numerically to localize the tip of Mott lobes in one dimension \cite{Laurent}, which requires the development of concepts such as the bi-partite fluctuation measurements (see Ref. \cite{Stephan} and a comparison with various existing methods, and Refs. \cite{Francis,AlexReview,Nicolas-PhysRep} for a review). We note recent experimental progress to measure such  spin dynamics in ultra-cold atoms \cite{Optiqueatom,Munichmicro,Harvardmicro,Stony}. This could also be realized in cQED systems and Josephson junction arrays. The quantum microscope dynamics is also expected to measure novel glassy quantum dynamics induced by disorder \cite{Juliette,ManyBodyLoca}.

We also note that driven Floquet protocols on these spin-boson systems have been addressed recently theoretically \cite{Angelakis}. In fact, similar spin impurity effects emerge in bosonic ladders in the Mott regime, in relation with the Meissner and Josephson physics \cite{EdmondThierry,Alex} and more generally in relation with artificial gauge fields \cite{Fallani,Atala,Zoller,Artificial1,L13,L14,Goldman,Cayssol}. This will be discussed in Sec. 4, in relation with Sec. 2.6 and the engineering of dissipative quantum spin arrays through the Hamiltonian (28). The novelty will be that the transverse field can be controlled by artificial gauge fields (magnetic flux) similar to the circuit of Fig.~\ref{fig:1}(b) and the spin dynamics will be related to Meissner currents. 

\subsection{Spin-boson model in a Josephson circuit: Kondo resonance of light and phase shift}
\label{sec:spin-boson-model-2}

The Kondo effect \cite{Kondo,AndersonRG,Wilson,Nozieres} has been first observed in metals and heavy fermions through an upturn of resistivity at low temperatures \cite{Hewson}, as a signature of the presence of magnetic impurities. In mesoscopic systems, the Kondo effect rather results in a unitary conductance \cite{Leo,David}. The spin-1/2 impurity is then simulated with an odd number of electrons on an artificial atom (mesoscopic quantum dot or island). At low temperatures, this effective spin-1/2 is screened by the spin of the conduction electrons in the reservoirs. The unitary conductance traduces the formation of a many-body resonance peaked at the Fermi level. The unitary DC conductance can also be re-interpreted as a Friedel's $\pi/2$ phase shift acquired by a reflected electron wavepacket from the mesoscopic island (dot) in the low-temperature Kondo regime, as a result of the Pauli principle and the screening of the impurity spin \cite{Leonid}. Here, we show that the Josephson circuit of Fig.~\ref{fig:1} realizes an analogue of the Kondo effect for photons and we propose an AC analogue of the Friedel's phase $\delta=\pi/2$, in the context of microwave light (This proof was not presented in the previous Ref. \cite{Karyn}). The formation of the Kondo resonance for light here incorporates both the effect of renormalization of the qubit frequency due to zero-point fluctuations or bosonic excitations as well as information of the ohmic dispersion of the bath through the width of the resonance. The Kondo resonance for microwave light increases the width of the Rayleigh transmission peak, which then leads to a larger bandwidth with perfect transmission one in the frequency space. In this sense, this geometry not only builds connections with Kondo physics in general, but also finds applications in light transport. 

Now, we describe in more detail the Josephson circuit of Fig.~\ref{fig:1}(d)~\cite{Karyn,Goldstein,Camalet} (again, this is directly related to experimental progress \cite{Gross} and the strong-coupling limit has been recently achieved \cite{Lupascu}). The two-level system corresponds to the two-charge states on a mesoscopic superconducting system (an additional pair on the left or on the right box, see Fig.~\ref{fig:1} for an illustration; an equivalent implementation can be realized with a single box at resonance, as realized in Refs. \cite{Gross,Lupascu}).  Quantum excitations in the two long transmission lines are described by collections of harmonic oscillators (zero-point fluctuations); $b_{lk}$ and $b_{rk}$ destroy an 'excitation' in mode $k$ in the left and right transmission lines, respectively. This produces zero-point fluctuations reminiscent of vacuum in free space; the main difference is that in one-dimensional waveguides, the quantization of energy requires only `one-polarization' of bosons.  Photons can experience strong scattering (coupling) effects already with one atom (or artificial two-level system) in weak-coupling \cite{Astafiev,Chalmers}. We introduce the symmetric $b_{sk}$ and antisymmetric $b_{ak}$ combinations of $b_{lk}$ and $b_{rk}$ \cite{Karyn}. Similar to a quantum dot coupled to electron leads in the Kondo regime \cite{Leonid}, one can engineer that only one combination (here, the antisymmetric one), couples to the effective spin describing the two charge states on the mesoscopic island.  After a unitary transformation, the Hamiltonian takes the form (similarly to Eq. (1)) \cite{Karyn}:
\begin{equation}
\hat{H} = \sum_{k>0}  v_a |k| \left( b^{\dagger}_{ak} b_{ak} +\frac{1}{2}\right) - \frac{h_z}{2}\sigma_z - \frac{E_J}{2}\sigma_x - \sum_{k>0} \lambda_k(b_{ak} + b_{ak}^{\dagger})\frac{\sigma_z}{2}.
\end{equation}
In Ref. \cite{Karyn}, the effective spin-1/2 particle is built from two equivalent charge states of a double-dot Cooper pair box at resonance (and $h_z\rightarrow 0$). The effective tunneling of Cooper pairs from left to right results in an effective Josephson coupling $E_J$ (in the spin-boson language, this term mimics the transverse field $\Delta$ in Eq. (2)). We focus on the propagation of a field from left to right and focus on the right-moving modes with $k>0$.  The spectral function of the environment is ohmic $J(\omega)=\pi \sum_{k>0} \lambda_k^2 \delta(\omega-\omega_k)= 2\pi\alpha\omega e^{-\omega/\omega_c}$ where $\omega_c\gg E_J$ represents the high-frequency plasma frequency cutoff of the transmission lines and the dissipative parameter $\alpha$ is given by
\begin{equation}
\alpha=\frac{2R}{R_Q}(\gamma_l^2 + \gamma_r^2).
\end{equation}
Here, $R_Q=h/(2e)^2=2\pi/(2e)^2$ denotes the quantum of resistance where $2e$ is a charge of a Cooper pair, $R$ is the resistance of each transmission line and $\gamma_l$ and $\gamma_r$ represent effective dimensionless
couplings of the qubit with the left and right transmission lines. In fact, in this geometry, one finds that $\gamma_l^2$ and $\gamma_r^2$ are of the order unity after a unitary transformation \cite{Karyn}. Such a strong-coupling limit with a transmission line has been achieved experimentally in Ref. \cite{Lupascu}. 

In the coherent regime, essentially for a range of $\alpha$ parameters $(0\ll \alpha\ll 0.5)$ similar to the underdamped Rabi oscillations of Fig.~\ref{fig:2}, the spin is described by the following dynamical spin susceptibility which describes the response to an input AC field
\begin{equation}
\chi(\omega) = \frac{\omega_K}{\omega_K^2 - \omega^2 - i \gamma(\omega)},
\end{equation}
where $\omega_K = T_K$ characterizes the many-body shift of the qubit frequency induced by the zero-point fluctuations.  Here, $T_K$  can be seen as the Kondo energy scale or equivalently the effective Josephson coupling. This equation has been derived in the supplementary material of Ref. \cite{Karyn}. This form also agrees with numerical renormalization group arguments  \cite{Costi}.  Using a combination of input-output theory \cite{Clerk} and Bethe-Ansatz approach \cite{KLHannal} to describe the transport of input and output photon fields, we predict that the broadening takes the form \cite{Karyn} $\gamma(\omega)=\omega_K J(\omega)$ until frequencies of the order of $\omega_K$, in agreement with the Korringa-Shiba relation in the low-frequency domain \cite{SassettiWeiss,Shiba}
\begin{equation}
\Im \chi(\omega)|_{\omega\rightarrow 0} = 2\pi\alpha\omega (\Re \chi)^2(\omega=0).
\end{equation}
In particular, one can reach analytically the equation:
\begin{equation}
\langle \sigma_z(\omega)\rangle (-\omega^2 + \omega_K^2 - i\gamma(\omega)) =  \omega_R \gamma_l \langle V_l^{in}\rangle.
\end{equation}
The input field $V_l^{in}$ in the left transmission line is defined as a coherent superposition $(b_{lk} + b_{lk}^{\dagger})$, and  $\langle V_l^{in}(\omega)\rangle = V_{ac}\cos(\omega t)$ \cite{Karyn}. 
The spin susceptibility can be measured through the transport of one photon in the circuit, leading to a many-body resonance in the elastic
transmission \cite{Karyn}
\begin{equation}
t(\omega,P_{in}) = - \frac{2i \gamma_r \gamma_l}{\gamma_l^2 + \gamma_r^2} J(\omega) \chi(\omega,P_{in}).
\end{equation}
Here, $P_{in}$ is the averaged input power related to $\langle V_l^{in}\rangle$. For the spin-boson model with an emergent Fermi liquid behavior at low energies, we note $J(\omega_K)\chi(\omega_K)=i$ for $P_{in}\rightarrow 0$.

Close to resonance, when $\omega=\omega_K$, an analogy with the unitary DC conductance in the Kondo regime of mesoscopic systems can be formulated as follows.  Let us first remind the form of the DC conductance through a quantum dot in the Kondo regime \cite{Leonid}
\begin{equation}
G_{DC} = 2\frac{e^2}{h} \sin^2 (2\theta) \sin^2 \delta,
\end{equation}
where $\delta=\pi/2$ is the Friedel's phase acquired for a plane wave (here, at the Fermi energy or $\omega=0$) backscattered by the magnetic impurity, and $\theta$ measures the anisotropy between left and right probes (tunnel couplings). The factor $\sin^2 (2\theta) \sin^2 \delta$ can be re-interpreted as a transmission coefficient $|t|^2$ according to the Landau-B\" uttiker pinciple. It is perhaps important to remind that from the T-matrix approach, one relates $-\pi T (\omega=0) = (1/2i)(e^{2i\delta} - 1)$ \cite{Leonid}. The maximum of DC conductance then corresponds to a phase $2\delta(\omega=0)=\pi$ between an incident and a scattered wave packet. The phase $2\delta=\pi$ is reminiscent of the Friedel's phase such that due to the screening of the impurity and the Pauli principle in the case of electron reservoirs, then the wave function should have a node at the impurity site $x=0$, implying the rule between left and right going waves \cite{Ian} $\psi(0) = \psi_L(0) + \Psi_R(0) = 0$. More precisely, when the spin is screened by a first electron, then the Pauli principle prevents another electron to go close to the impurity. Then, we have $\Psi_L(0) = e^{2i\delta} \Psi_R(0) = - \Psi_R(0)$. The spin channel is taken into account in the factor $2$ in the DC conductance. 
In the Josephson geometry, from Eq. (44), we also check the form $\sin(2\theta)=2\gamma_l \gamma_r/(\gamma_l^2 + \gamma_r^2)$. For the microwave light, the transmission $t$ is a transmission amplitude, referring to the ratio between the output and input signals (AC fields), and for frequencies close to $\omega_K$ \cite{Karyn}:
\begin{equation}
t(\omega\sim \omega_K,P_{in}\rightarrow 0) = |t|  =  \sin(2\theta).
\end{equation}  
The phase of the transmitted light signal is zero. For symmetric conditions, the transmission reaches unitarity.  Now, if instead we focus on the reflection coefficient (ratio of the output and input fields in the left transmission line)
\cite{Karyn}:
\begin{equation}
r(\omega,P_{in}) =  |r| e^{i2\delta} =  1 + 2i \frac{\gamma_l^2}{\gamma_l^2 + \gamma_r^2} J(\omega)\chi(\omega,P_{in}),
\end{equation}
and at resonance, we find:
\begin{equation}
|r| e^{i2\delta(\omega=\omega_K)} =  -\cos(2\theta)  = \frac{\gamma_r^2 - \gamma_l^2}{\gamma_l^2 +\gamma_r^2}.
\end{equation}
For $\gamma_l=0$, we find the phase of an open line $\delta=0$ (and formally, $2\theta=\pi$) whereas for $\gamma_r < \gamma_l$, we find a phase of $2\times \pi/2=2\delta(\omega=\omega_K)$ and $\theta=0$. 
Writing the input and output signals in the left transmission line in terms of $b_{lk}$ and $b_{lk}^{\dagger}$, then this phase can also be seen as the phase taken by a bosonic excitation in the left transmission line with wave-vector defined as $vk=\omega_K$. We conclude that the phase $2\delta=\pi$ occurs at resonance for the photon reflection (when the left line which is coupled to the measurement reflection device), emphasizing the duality between zero-point fluctuations and electron-hole pairs in one dimension through bosonization \cite{Haldane,Giamarchi}. It might appear surprising at first sight that for bosons such a phase occurs since for fermions this phase is attributed to the Pauli principle (as discussed above). The occurrence of such a $\delta=\pi/2$ phase can be understood due to blockade effects: the light frequency is fixed at resonance with the spin (frequency) and therefore one cannot absorb two photons at the same time. 

To summarize this part, the observation of a Kondo-type resonance in these microwave circuits can be established as follows. First, observe a many-body shift of the spin frequency:
\begin{equation}
\omega_K = E_J\left({E_J}/{\omega_c}\right)^{\frac{\alpha}{1-\alpha}} \ll {E_J}.
\end{equation}
This form of many-body frequency shift requires formally that the spectral function of the environment has an ultra-violet cutoff $\omega_c\gg E_J$ and an infra-red cutoff smaller than $\omega_R$ (see Eq. (21) of supplementary material of Ref. \cite{Karyn}). In the absence of Kondo physics (meaning $\alpha=0$), the light resonance condition is fixed by the gap $\sim E_J$ separating the two states $\sigma_x=+1$ and $\sigma_x=-1$ of the spin (qubit). Recent experimental results report on the observation of such many-body effects in Josephson systems \cite{Lupascu}. In addition, the Friedel' s phase of $2\delta=\pi$ in the reflected light signal at the resonance frequency $\omega=\omega_K$ would also be a strong indication of Kondo physics. Some efforts have been performed to observe such a phase shift in hybrid systems, as described in Sec. 3.4. 

The broadening of the elastic Rayleigh peak is much larger than in the weak dissipation limit \cite{Astafiev,Gross}. By increasing the input power power $P_{in}$, the evolution of the Mollow triplet \cite{Astafiev} observed for weak dissipation 
remains an open question for larger dissipation regimes. The scattering matrix around $\omega=\omega_K$ becomes non-unitary since $J(\omega_K)\Im \chi(\omega_K,P_{in})<1$,
which hides the presence of additional inelastic corrections \cite{Karyn}. These inelastic Fermi-liquid corrections have been studied by Goldstein {\it et al.} in a larger regime of $\alpha$ parameters (ultra-strong coupling limit close to the Kosterlitz-Thouless transition) \cite{Goldstein} and in a Rabi-Kondo laser-driven system \cite{Sbierski}.  Theoretical efforts to describe quantitatively the scattering of bosonic waves in a many-body sense in solvable models have been performed \cite{Leclair,Shen,Busch,Yudson}.  Furthermore, the exact non-Markovian dynamics of transmon qubits in open multi-mode resonators can be obtained from reduced Heisenberg-Langevin equations of motion, in which the effects of the (possibly lossy) electromagnetic environment are present via a classical Green's function \cite{HakanAlex}. Extensions of this model could include realizations of two-channel Kondo physics \cite{NozieresBlandin} with light, as recently observed for electrons \cite{Stanford,LPNMarcoussis}, or a single effective spin-1/2 in a cQED array \cite{Houck,CRAS2016}. 

As described below in Sec. 3.4, hybrid systems comprising reservoirs of electrons as well as a cavity also offer a novel platform where light can probe the Kondo effect entangling a spin and conduction electrons. We summarize a few theory steps
\cite{MarcoKaryn,Olesia,Audrey} in relation with current experiments \cite{ENSTakis,DengChina}.  

\subsection{Generalized Impurity RC systems}
\label{sec:driven-quantum-rc}
After the discussion above of two implementations of the spin-boson model in Josephson circuits and ultra-cold atoms in relation with actual technology and with results presented in Sec. 2, we generalize the discussion to other quantum impurity models in particular to fermion systems starting from quantum RC circuits. We investigate the AC regime more carefully in the limit of low-frequency (long-time limit). The spin dynamics discussed previously here is embodied by the low-frequency charge dynamics.

\begin{figure}[t]
\center
\includegraphics[width=.8\linewidth]{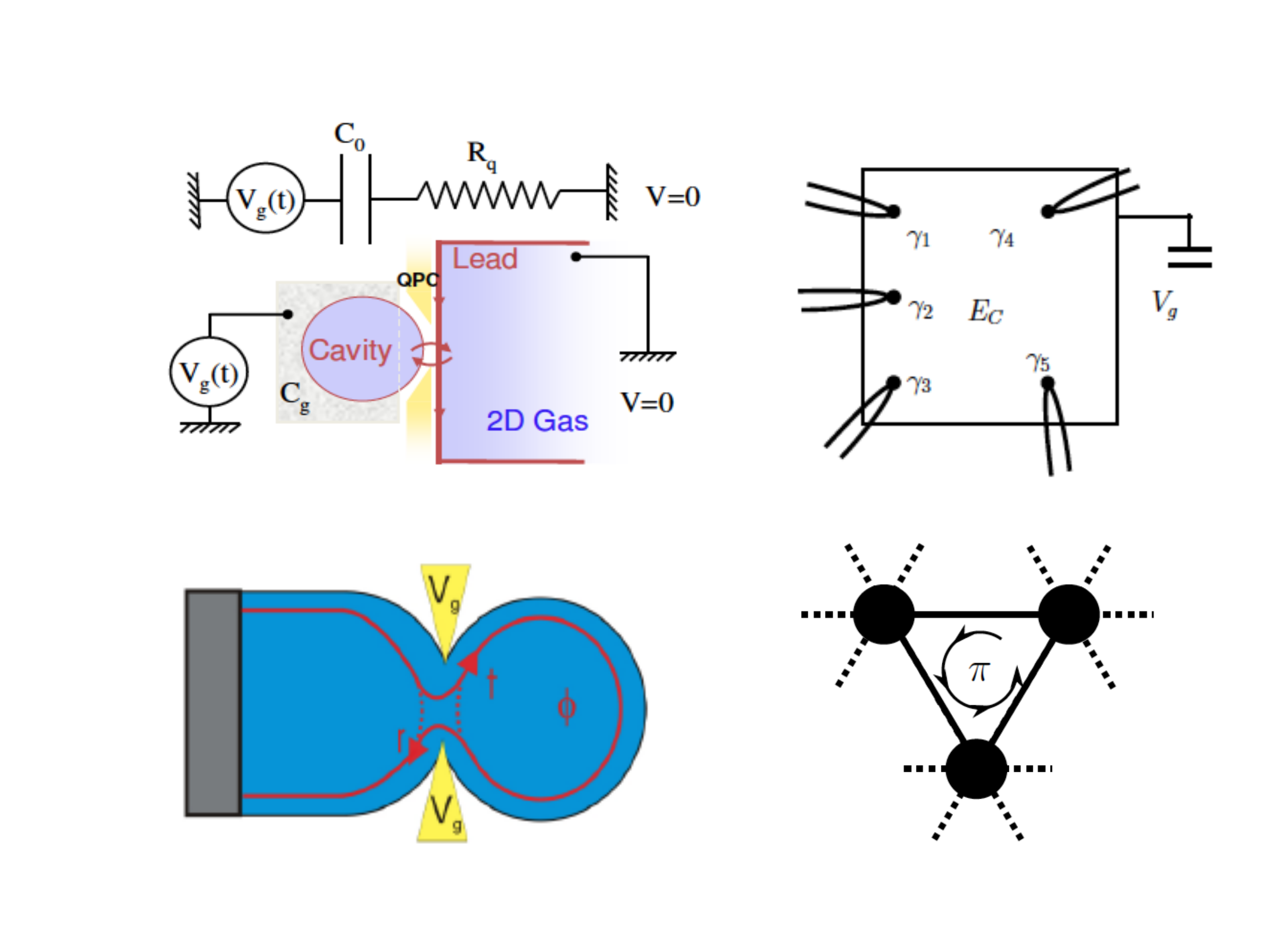}  
\caption{Examples of Mesoscopic Circuits. (Top left) Quantum RC circuit \cite{MarkusRC} (from Ref. \cite{ChristopheKaryn}) coupling a quantum dot (box or island, cavity) with an electron reservoir driven by an AC gate voltage as realized in $GaAs$ \cite{Gabelli}. For a metallic quantum dot, charge Kondo physics was recently realized with quantum Hall edge states \cite{LPN}, in an analogue circuit as Bottom Left, following theory \cite{Matveev,Matveev2}. Universal quantized resistances can emerge for all transparencies as Fabry-Perot effects, maintaining the phase coherence $\phi$. These large boxes also lead to a novel relation between charge relaxation resistance and Korringa-Shiba relation at low frequency \cite{ChristopheKaryn}. (Top Right) Topological Mesoscopic Box with Majorana Fermions allowing to realize exotic multi-channel Kondo physics \cite{NozieresBlandin}, at the charge degeneracy points \cite{LoicHerviou,Michaeli} and away from charge degeneracy \cite{BeriCooper,EggerAltland,Erik}. This multi-channel Kondo physics also allows one to make a connection with the quantum Brownian motion \cite{YiKane} and dual lattices with artificial gauge fields, which may reveal $\pi$ phases due to pseudo-spin effects \cite{LoicHerviou} (Bottom Right). The effect of artificial gauge fields leads to complex phenomena such as topological phenomena and Meissner physics \cite{Atala,Artificial1}, as discussed in Sec.~\ref{sec:diss-array-spin}.}
\label{fig:5}
\end{figure}

A typical experimental system is a quantum RC circuit, shown in Fig.~\ref{fig:5}, consisting of a mesoscopic metallic island (cavity) tunnel-coupled to a metal (essentially, a two-dimensional electron gas or a one-dimensional Luttinger liquid, edge channel of the quantum Hall effect). The study of AC coherent transport was pioneered in a scattering approach by B\" uttiker, Pr\^etre and Thomas in 1993, where a charge relaxation resistance of $R_q=h/2e^2$ was predicted for a single-mode resistor \cite{MarkusRC}. In this experiment, the averaged charge $\langle Q\rangle= e\langle N\rangle$ on the island  is measured in response to a small AC gate potential:
\begin{equation}
\frac{\langle Q(\omega)\rangle}{V_g(\omega)} = C_0(1+i\omega C_0 R_q)+O(\omega^2).
\end{equation}
Formally, we focus on time scales much longer than the typical $RC$ time and we assume that the voltage is of the form $V_g(\omega)=V_{dc}+V_{ac}\cos(\omega t)$ with $(V_{ac}, \omega) \rightarrow 0$.
The quantum RC circuit is drawn in Fig.~\ref{fig:5} and has been successfully implemented in a two-dimensional electron gas and a quantized resistance of $R_q=h/2e^2$ was measured in $GaAs$ \cite{Gabelli}. This quantized  resistance must be thought as a contact resistance between the mesoscopic island and the reservoir lead. Note that the factor $2$ does not come from spin effects. This resistance must be thought
as a quantum coherent effect analogous to a Fabry-Perot resonator, and occurs in fact for any transmission amplitude $t$ (or reflection $|r|$) at the junction. 

Coulomb blockade effects were first ignored and later they have been partially taken into account in an Hartree-Fock theory \cite{Nigg}. Using bosonization approach, we can include interactions in a non-perturbative manner. We also show a crossover at finite frequency $\omega$, where the charge relaxation resistance changes from $h/2e^2$ to $h/e^2$ regardless of the mode transmission. All the technical details can be found in Ref. \cite{ChristopheKaryn}. Here, we give a simple understanding to this crossover and we then make an analogy with the Kondo effect.  Let us model the system by a one-dimensional line (see Fig.~\ref{fig:5}), the lead is between $-\infty$ and $-L$ and the quantum dot between $-L$ and $0$. The level spacing on the quantum dot (cavity) is $\Delta=\pi v_F/L$, where $v_F$ is the Fermi velocity.  The Coulomb blockade phenomena are treated exactly in an action and bosonization formalism \cite{Giamarchi} after integrating out all irrelevant modes. At perfect
transmission, the system is then described by the action \cite{ChristopheKaryn}:
\begin{equation}
S_0 = \frac{1}{\pi} \sum_n \phi_0(\omega_n) \phi_0(-\omega_n) \left[\frac{|\omega_n|}{1 - e^{-2 |\omega_n| L/v_F}} + \frac{E_c}{\pi}\right],
\end{equation}
where $\omega_n=2\pi T n$ denotes bosonic Matsubara frequencies ; the Boltzmann constant $k_B$ is set to unity and $E_c=e^2/2C_g$ corresponds to the (bare) charging energy (Fig.~\ref{fig:5}). Here, the bosonic field $\phi_0$ means $\phi(-L)-\phi(0)$, which is related to the charge on the dot through $C_0 V_g/e + \phi_0/\pi$. From this bosonic action, one can read the Green's function of the charge field $\phi_0$, and immediately infer the response
\begin{equation}
\frac{Q(\omega)}{V_g(\omega)} = \frac{C_g}{1-\frac{i\omega \pi/E_c}{1-e^{2i\omega \pi/\Delta}}}.
\end{equation}
The response vanishes each time the frequency $\omega$ hits a multiple of $\Delta$, {\it i.e.} an eigenstate of the isolated island. At low frequency, we extract $C_0=C_{\mu}$, where $C_{\mu}$ is the electrochemical
capacitance corresponding to the classical capacitance $C_g$ in series with the quantum correction $e^2/\Delta$, such that the Coulomb blockade vanishes for perfect transparencies at the junction. In the low-frequency limit, for $\omega\ll \Delta$, we also recover the universal resistance $R_q=h/2e^2$ (in units of $\hbar=1$ then $h=2\pi$). By increasing the size of the quantum dot, the response shows an oscillatory behavior for $\omega> \Delta$. We thus average over a finite bandwidth $\delta \omega$, such that $\omega\gg \delta\omega \gg \Delta$, and 
\begin{equation}
\frac{Q(\omega)}{V_g(\omega)} = \frac{C_g}{1-i\omega \pi/E_c}.
\end{equation}
This leads to $R_q=h/e^2$,  equal to the quantum of resistance $R_K$ (this result can also be obtained directly from the action in imaginary time when $L\rightarrow \infty$). 

This result can also be recovered using an analogy with the Kondo effect, following Matveev \cite{Matveev,Matveev2}, where two successive charge states on the metallic island can be identified to a macroscopic pseudo-spin. The two effective spin polarizations of conduction electrons must be associated to the 'metallic island' and 'lead' locations respectively. Using standard linear response theory, then one identifies $Q(\omega) = e^2 K(\omega) V_g(\omega)$ where \cite{ChristopheKaryn}
\begin{equation}
K(t-t') = i \theta(t-t') \langle [ N(t), N(t') ] \rangle.
\end{equation} 
The spin analogy allows one to relate $K$ with the spin susceptibility $\chi$ of the underlying Kondo and Fermi liquid theory \cite{Nozieres,Shiba}.  The isotropic Kondo model is equivalent to the spin-boson model in the limit $\alpha\rightarrow 1$. Taking 
$\alpha=1$ in the Korringa-Shiba relation of Eq. (42), one immediately recovers $R_q=h/e^2$ in the low-frequency domain. Such a charge relaxation resistance in large metallic dots with $\Delta\rightarrow 0$ has not yet been observed. We note that recent experimental progress has allowed to observe such a charge Kondo effect in DC transport \cite{LPN}.  This experiment involves (integer) quantum Hall edge states. If we generalize the calculation for an Abelian quantum Hall edge state with a filling factor $\nu$ (see Fig.~\ref{fig:5}), this gives the two (generalized) resistances $R_q=h\nu /(2 e^2)$ and $R_q=h \nu / e^2$ \cite{ChristopheKaryn}. The effect of a small zeeman field has been studied in Ref. \cite{Georg}.

Other important results have been established by Hamamoto {\it et al.} \cite{Hamamoto}, related to Luttinger liquids and fractional quantum Hall edges, using perturbative renormalization group arguments, bosonization and quantum Monte Carlo.  The quantum RC circuit also allows one to formulate close analogies with dissipative mesoscopic rings \cite{Ledoussal}. A generalization to the Anderson model has also been done  in Refs. \cite{Michele1,Michele2} as well as for several conducting channels \cite{Prasenjit}. Moreover, considering a quite general cavity geometry with a single-channel lead, Ref. \cite{MicheleChristophe} has shown that the universal resistance $R_q= h/e^2$ is in fact more general than the specific form of the action (51))  or the Anderson or Kondo underlying models, but applies generally as soon as the system is described at low energy by a Fermi liquid fixed point. In this case, the Korringa-Shiba formula can be derived on general grounds leading to the resistances $R_q = h/e^2$ and $R_q = h/2 e^2$ depending only on whether the cavity has a dense spectrum or not. For a two-channel Kondo model subject to non-Fermi liquid corrections, using a Majorana description for the spin \cite{Coleman}, one finds an increase of the charge relaxation resistance induced by the non-Fermi liquid behavior at larger frequencies \cite{Christophe2}. The charge relaxation resistance has also been computed in mesoscopic topological superconducting circuits \cite{Grosfeld} and at the edges of two-dimensional topological insulators \cite{Muller}. We also note recent efforts to consider real-time dynamics and charge relaxation in these interacting systems \cite{Michele3}.

One-dimensional leads also allow, through a dissipative action proportional to $|\omega|$, a connection to the quantum Brownian motion addressed by Yi and Kane \cite{YiKane} (The analogy with the quantum Brownian motion occurs via bosonization through the dynamics of the phase variable by analogy to Eq. (51)). Let us consider a specific recent example of Fig.~\ref{fig:5} with $M$ one-dimensional leads described by a Luttinger theory (with Luttinger parameter $K$ \cite{Haldane,Giamarchi})) tunnel coupled through a mesoscopic box,  which comprises several topological superconducting wires possessing Majorana fermions at their extremities \cite{Oreg,Lutchyn}. The Majorana fermions act as impurities that already allow for electron tunneling (instead of Cooper pair tunneling). The Hamiltonian takes the form
\cite{LoicHerviou,Michaeli}: 
\begin{equation}
\hat{H} = \hat{H}_{leads} + \hat{H}_{box} + \sum_{j=1}^M t_j e^{-i \chi} \psi^{\dagger}_j(0) \gamma_j + h.c.
\end{equation}
Here, $t_j$ is the tunneling amplitude between each lead and the box, $\psi_j(0)$ describes the electron operator in a lead $j$ in the entrance of the mesoscopic box, $\gamma_j$ represents a Majorana fermion in the mesoscopic box and in this expression $\chi$ is the superconducting phase conjugate to the number of Cooper pairs. The charging energy is hidden in $H_{box}$ following Refs. \cite{LoicHerviou,Michaeli}.  Away from charge resonance in the mesoscopic system, single-particle transport is blockaded and one can apply a standard Schrieffer-Wolff transformation to re-write the tunneling terms in the bosonized language as \cite{LoicHerviou}
\begin{equation}
\hat{H}_{SW} = - \sum_{j \neq k} \lambda_{j,k} \cos(\theta_j - \theta_k).
\end{equation}
The phases $\theta_j$ describe the superfluid-type phases in the entrance of the box in each metallic lead (wire) \cite{Giamarchi}. Majorana fermions and tunneling electrons combine to form a purely bosonic 
particle through the Klein factor of bosonization \cite{Haldane,Beri}. The statistical properties of the charge carriers in the island are effectively changed.

This problem is also referred to as the topological Kondo model first introduced in Refs. \cite{BeriCooper,EggerAltland}: one can identify an effective pseudo-spin as a product of two Majorana fermions \cite{Coleman} and therefore in the fermion picture $H_{SW}$ makes link with a multi-channel Kondo model of the form $\sum_{j,k}  \tilde{\lambda}_{j,k} \psi^{\dagger}_k(0) \psi_j(0) \gamma_j \gamma_k$. The word topological here can be understood from the fact that the spin is built from Majorana excitations and from the fact that a channel anisotropy is irrelevant. By considering attractive interactions in the leads (implying a Luttinger parameter $K>1$), the system flows to a strong-coupling Kondo fixed point. An analogy with the quantum Brownian motion can be understood at the strong coupling fixed point of the Kondo model. The global mode $\theta=(1/\sqrt{M}) \sum_{j=1}^M \theta_j$ decouples from the Hamiltonian \cite{LoicHerviou}, and has a free evolution reflecting the charge quantization. From there, by analogy with the quantum Brownian motion, one can build an effective model for the remaining phase variables $\theta_j$. It consists in a massless particle subject to dissipation (here, stemming from the electron-hole pairs in each lead) in a $(M-1)$ dimensional potential. The minima of the potential form a (hyper)triangular lattice, and must satisfy that $\theta_j-\theta_k=2n\pi$ where $n\in \mathbb{N}$. The (fractional) DC conductance at this strong-coupling fixed point is also in agreement with an incoming wave equally flowing in all the leads.

At the charge degeneracy point, one can formulate a similar analogy to the Kondo model and obtain the conductance as a function of the phae shift $\delta$ similarly to Eq. (45). More explicitly, according to the recent Ref. \cite{LoicHerviou}, we introduce a pseudo-spin acting on the charge space in the box as $\tau^- |N+1\rangle = | N\rangle$. The model is mapped onto a $M$ channel Kondo model \cite{NozieresBlandin}. In the strong-coupling limit (that can be reached for free electrons), as a reminiscence of the topological Kondo effect \cite{BeriCooper,EggerAltland,Erik}, one can build a connection to the same quantum Brownian motion. However, there is a difference: each minimum of the dual lattice is characterized by a (pseudo-)spin wave function: 
$(e^{- i \pi/M} |\uparrow \rangle_z + e^{i \pi/M} |\downarrow \rangle_z)$.  Performing a loop around a unit cell in the dual lattice produces then an overall Berry phase $e^{i \pi} = -1$ (a site has $M$ neighbors) \cite{LoicHerviou}. 
This phase is drawn in Fig.~\ref{fig:5}. it is equivalent to the presence of artificial gauge fields in the dual lattice. It is relevant to note that recent experiments in quantum cQED transmon circuits \cite{Artificial1} and ultra-cold atoms \cite{Artificial2} have engineered similar gauge fields. These gauge fields also emerge as artificial phase factors (such as Jordan-Wigner strings) in quantum field theories \cite{Feng,NewPRBpaper,Sedrakyan}.  At this strong-coupling fixed point, the DC conductance (which is proportional
to $\sin^2\delta$ as in Eq. (45)) is in agreement with an in-coming electron wave equally flowing in all the leads. We also identify a possible intermediate Kondo fixed point, for a Luttinger parameter $K=1/2$, described by an intermediate phase shift $\delta=\pi/(M+2)$ obtained from the $M$ channel Kondo model \cite{LoicHerviou,Michaeli}. 

This research is directly motivated by current efforts to realize and manipulate Majorana fermions in superconducting systems \cite{MajoranaReview}. We note the very recent related theoretical developments \cite{Sela,Zazunov}. 
It is also important to remind that $H_{SW}$ is connected to tunnel junction dissipative Kane-Fisher models and dynamical Coulomb blockade physics \cite{IngoldNazarov,Kane-Fisher}. As an application studied in Ref. \cite{Tal}, we would like to mention that a Ramsey protocol in time allows one to probe the charging energy at the junction, already in the small resistance limit for a dissipative ohmic environment. One can then draw a similar analogy with a dissipative quantum Brownian motion.

\subsection{Hybrid Systems}

\label{sec:quant-brown-moti}

Let us now generalize the discussion to hybrid mesoscopic systems, comprising bosons (cavity and transport channels) and electrons (mesoscopic quantum dot system). We study the interplay between quantum electron transport and a circuit-QED environment  \cite{TakisReview}. In particular, we derive a stochastic classical Langevin equation for the cavity field and analyze feedback effects from the mesoscopic circuit (see Eq. (61)). We find a dynamical crossover for the quantum Brownian motion as a function of the bias voltage applied across the mesoscopic system where the diffusion coefficient (follows the current and) saturates and the friction coefficient is progressively suppressed. We also discuss recent experimental observations of Kondo physics in these hybrid systems. The relation with Sec. 3.2 will occur through dynamical correlation functions and the Korringa-Shiba relation. In relation with Sec. 2.3, we study the notion of effective temperature (in the steady-state limit) and its evolution in the non-equilibrium limit when increasing the bias voltage. These systems have attracted some attention in the context of nano machines and thermoelectricity \cite{AndrewReview,LoicNano}.  It is relevant to mention progress realized in quantum optoelectronics, by applying ideas and concepts of quantum optics to quantum electronics \cite{LPA}. Below, we illustrate such a relation between these two communities by relating quantum electronic transport and light measurements in the microwave regime.

A specific model is based on the Anderson-Holstein Hamiltonian \cite{MarcoKaryn}
\begin{equation}
\hat{H} = \sum_{kl} \omega_k b^{\dagger}_{kl} b_{kl} + (a+a^{\dagger}) \sum_{k l} g_{kl} (b^{\dagger}_{kl} + b_{kl}) + \omega_0 a^{\dagger} a + g x(N-1) + H_{Anderson}.
\end{equation} 
A cavity described by the creation operator $a^{\dagger}$ is coupled to two transmission lines (here, the two ohmic bosonic environments) carrying the microwave signal. Here, the sum $l$ acts on the left and right transmission lines. 
The Hamiltonian $H_{Anderson}$ describes a  quantum dot coupled to two electronic reservoir leads. The coupling with the cavity $g$ is assumed to be a capacitive coupling where $N$ represents the dot occupancy and $x\propto (a+a^{\dagger})$ the electric displacement field in the cavity.  Following Refs. \cite{Karyn,Clerk}, we can generalize the input-output theory for this situation and we obtain the transmission coefficient \cite{MarcoKaryn}:
\begin{equation}
t(\omega) = i J(\omega) \chi_{xx}(\omega),
\end{equation} 
where $J(\omega)$ is of ohmic type and describe photon dissipation in the transmission lines and the (retarded) photon propagator $\chi_{xx}(t) = -i\theta(t) \langle [x(t),x(0)] \rangle$ can be evaluated using the
Schwinger-Keldysh approach in certain limiting cases. At a general level, one can write $\chi_{xx}$ as:
\begin{equation}
\chi_{xx}(\omega) = \frac{\omega_0}{\omega^2 - \omega_0^2 - \omega_0 \Pi^R(\omega) + i\omega_0 \kappa};
\end{equation}
the retarded function $\Pi^R$ includes both the effects of frequency renormalization and damping due to the electronic environment, by analogy to the NMR situation, and $\kappa\sim 2\pi\alpha\omega_0$ describes dissipation in the photon field due to
the small coupling with the transmission lines (photon wave-guides) \cite{MarcoKaryn}. Treating the coupling between light and matter (the quantum dot, here) to second order in perturbation, then we can establish again a connection with the Korringa-Shiba relation:
\begin{equation}
\Pi^R(t,t') = \Lambda_R(t,t') = -g^2 K(t-t'),
\end{equation}
and $K(t-t')$ describes the charge fluctuations on the quantum dot similarly to Eq. (54). Let us first consider a solvable non-interacting resonant level model with a width $\Gamma$; here, the resonant level width is given by a sum of the coupling parameters $\Gamma_L$ and $\Gamma_R$. Then, the Keldysh calculation can be done rigorously \cite{MarcoKaryn}. We choose the light frequency  $\omega_0 \sim \Gamma$ (the analogue of the Kondo energy scale). The position of the level here is at the Fermi energy of the leads when $\mu_L=\mu_R=0$. Other parameters regimes, in particular the low-frequency regime, have been studied in Ref. \cite{Olesia}. 

We study the effect of a large bias voltage $V$ across the mesoscopic system, producing a non-linear $I-V$ characteristics across the quantum dot \cite{Prasenjitcurrent,MarcoMC,Thomas,WernerOM}.  Similar nonlinear effects have also been addressed in tunnel junctions coupled to a cavity \cite{UdsonChristophe}. 

For $1/eV\ll 1/\omega_0$, the electron motion becomes fast compared to the photon dynamics, and therefore we observe that the photon damping vanishes rapidly $\Re \Pi^R(\omega_0,V)\gg \Im \Pi^R(\omega_0,V)$ \cite{MarcoKaryn}. This implies that the phase associated with the transmitted  signal converges to $\pi/2$. The origin of this phase $\pi/2$ here simply comes from the small dissipation limit (electrons are fast and the cavity is good). Note that such a phase $\pi/2$ implicitly assumes very small dissipation from the transmission lines carrying the AC signal. 

\begin{figure}[t]
\center
\includegraphics[width=.7\linewidth]{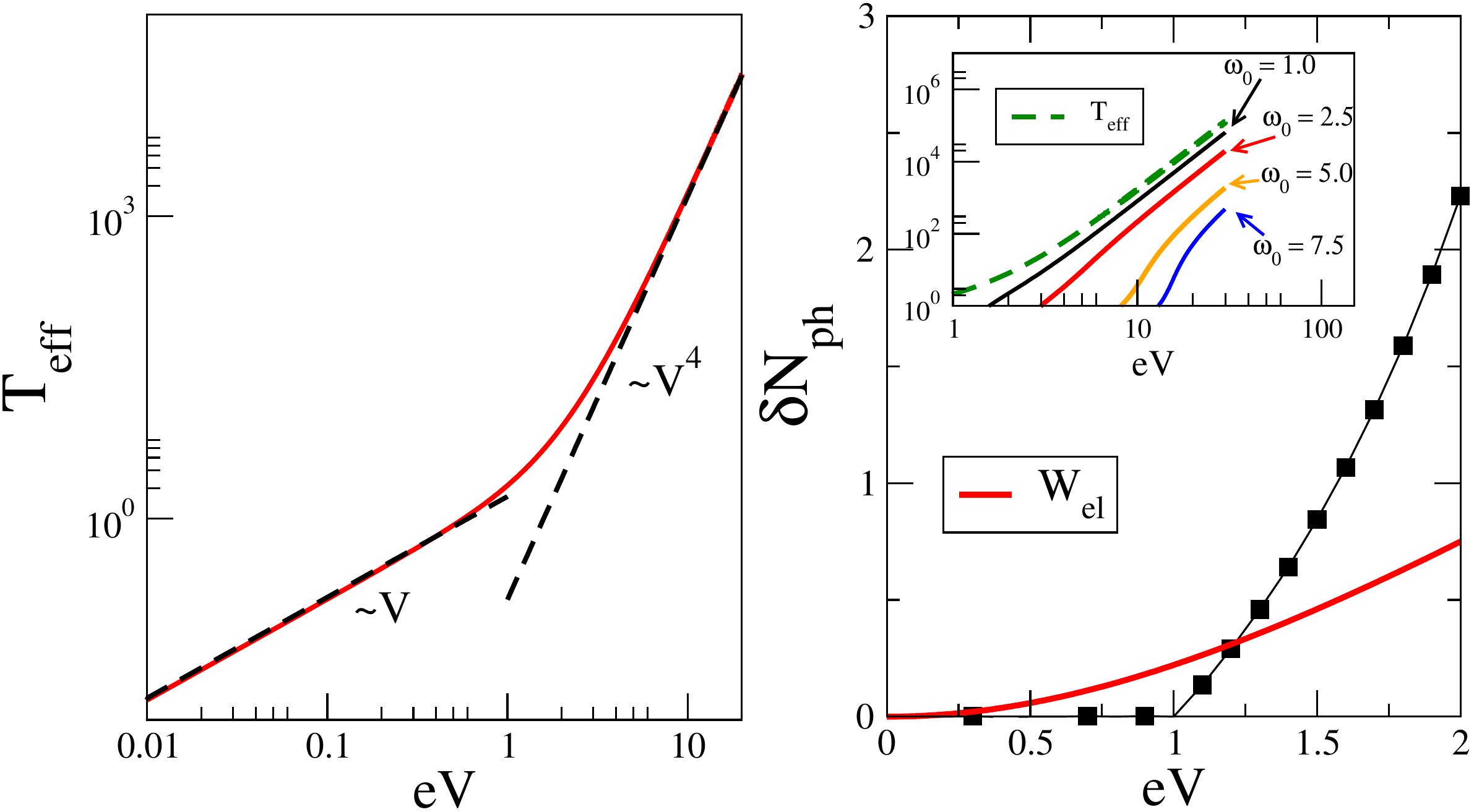}  
\caption{Effective temperature $T_{eff}$ for the cavity coupled to a resonant level model driven by a biased voltage $V$. Units are chosen such that $\omega_0\sim \Gamma=1$ and $g=0.5\Gamma$. Mean Photon Number $\delta N_{ph}$ or Effective Thermal Occupancy obtained with the Keldysh approach, and comparison with the dissipated power $W_{el}=I.V$ where $I$ is the current flowing through the quantum dot system \cite{MarcoKaryn}.}
\label{fig:6}
\end{figure}

In addition, one can build a precise analogy with a stochastic classical Langevin equation. In this large-bias regime, after integration of the matter (electrons in the mesoscopic system), one obtains an equation for the classical component of the cavity field in the Keldysh sense \cite{MarcoKaryn}:
\begin{equation}
\ddot{x}_c = -\omega_0 x_c - F(x_c) - \gamma(x_c) \dot{x}_c + \xi(t),
\end{equation}
and $\langle \xi(t) \xi(t') \rangle = D(x_c) \delta(t-t')$ where the diffusion coefficient depends on the bias voltage. The force $F(x_c)$ can be obtained rigorously for the resonant level situation on the dot, leading to an anharmonic potential. This analogy suggests an effective temperature $T_{eff}\sim D(V)/(4\gamma(V))$ in the cavity. By increasing the bias voltage across the mesoscopic system, we report a novel crossover in the light dynamics: at small bias voltages, the diffusion coefficient $D(V)\sim V$ and the friction coefficient $\gamma$ is constant, leading to $T_{eff}\sim V$, whereas at large bias voltages compared to $\Gamma$, we report $T_{eff}\sim 1/\gamma \sim V^4$. In this limit, the motion of the electrons
is fast enough such that the dissipation (friction) becomes negligible in agreement with the $\pi/2$ phase arguments above and the diffusion coefficient follows the current which saturates at large bias voltages \cite{MarcoKaryn}.
It is also relevant to note that at this crossover, there is the production of a mean number of (thermal) photons in the cavity $\delta N_{ph}\sim  T_{eff}/\omega_0\sim V^4$ \cite{MarcoKaryn}. In Fig.~\ref{fig:6}, we draw the effective temperature as well as the
induced (mean) photon number as a function of the bias voltage. In relation with the Korringa-Shiba relation, we note the following interesting property : at low frequency, we observe $\Lambda^R(\omega) = i \gamma(x_c) \omega$ even at large bias voltages \cite{MarcoKaryn}. 

Let us now make a link to recent discoveries in hybrid carbon systems, reporting some observation of Kondo physics with light measurements. The Kondo physics here refers to the formation of an entangled and
Fermi liquid ground state between the effective spin in the mesoscopic system and electron reservoirs \cite{Leonid}. In Ref. \cite{ENSTakis}, results in the spin Kondo regime of a carbon nanotube are consistent with a phase of the light signal following the DC conductance across the mesoscopic system in the low-frequency regime, in agreement with theory results (see Ref. \cite{ENSTakis} and Eq. (60) of Ref. \cite{Olesia}) based on admittance theory. Next, we describe a recent circuit in graphene double-dot quantum dot where the charge resonance condition described in Sec. 3.2 has also been achieved \cite{DengChina}.  The circuit has been introduced and characterized in Refs. \cite{Guo1,Guo2}. The circuit-QED environment here couples to the left lead only, and by applying a gauge transformation we can reach a similar capacitive coupling as in Eq. (54) involving one of the two dots only \cite{DengChina}.  The theory described above then can be adapted. 

The reflection coefficient in the circuit-QED environment has been measured, by analogy to Eq. (47), and the resonance condition for light $(\omega_0^*(V))^2=(\omega_0)^2+ \omega_0 \Re \Pi^{R}\left[\omega_0^*(V) \right]$ has been realized (these measurements are done at the light resonance condition).  Datas shown in Ref. \cite{DengChina} seem consistent with an emergent Kondo fixed point, predicted by the theory in these double-dot charge qubits \cite{Halperin}. Again here, the effective spin corresponds to two macroscopic charge states on the mesoscopic system 
similar to Sec. 3.2 and the fixed point is assumed to yield an emergent $SU(4)$ symmetry as a result of entangled pseudo-spin (orbital-like) and spin degrees of freedom of conduction electron. The $SU(4)$ Kondo fixed point has also been measured in $GaAs$ nano-circuits \cite{Keller}. An analogy between a resonant level model and a Kondo low-energy fixed point can be formulated in the light of the emergent Fermi liquid ground state \cite{ChristopheSUN}. First, decoherence effects induced by the bias voltage produce a resonance width which becomes bias dependent \cite{Achim}:
\begin{eqnarray}
\Gamma= T_K  \textrm{ for } eV \ll  T_K,  \nonumber \\
\Gamma \sim e V/\ln^2 (eV/ T_K) \textrm{ for } e V \gg T_K.
\end{eqnarray}
Here, $T_K$ is the Kondo temperature scale (which is estimated around $550mK$ \cite{DengChina}). We have applied renormalization group arguments for the SU(4) Kondo effect, and obtained similar decoherence rates \cite{DengChina}.
In addition, the $SU(4)$ symmetry moves the position of the effective resonance at an energy $\sim T_K$ \cite{KarynLoss}. By analogy with the photonic Kondo effect discussed in Sec. 3.2, the reflected microwave signal at the resonance condition for light 
$\omega=\omega_0^*$ again seems to reveal a phase shift of $2\delta(\omega_0^*)=\pi$, which is consistent with theoretical calculations \cite{DengChina}. Note that the DC phase shift is equal to $\delta(\omega=0)=\pi/4$ for this $SU(4)$ Kondo fixed point \cite{Ian,ChristopheSUN}, due to the position of the Kondo resonance above the Fermi energy. These recent important observations \cite{ENSTakis,DengChina} are promising steps towards observing many-body features with light. 

\section{Dissipative Arrays}
\label{sec:diss-array-spin}
In relation with Sec. 3.1, we generalize the discussion to systems subject to a magnetic flux (see Fig.~7a) starting in the Mott regime. Even though the discussion below is presented for ladder systems, it works equally for chain arrays. Our motivation is two-fold: first, propose a controllable architecture of spin-boson arrays where the transverse field can be tunable and function of a magnetic flux (similar to Fig.~\ref{fig:1}(b)), and second relate the local spin dynamics of XXZ spin chains \cite{Alex} with the one of a dissipative spin-boson array studied in Sec. 2.6. Such a local dynamics has been studied and measured in ultra-cold atomic ladders in relation with Meissner currents \cite{EdmondThierry,Atala}. Similar systems can be built with Josephson arrays. 
 
We introduce couplings to bosonic baths which prompt relaxation of the boson population and illustrate based on the Bloch-Redfield master equation for the reduced density matrix \cite{breuer_petruccione_2002} that relaxation to exotic ground states with chiral current and low charge fluctuations can be achieved starting with featureless initial states, such as a Mott insulator at unit site filling.

\subsection{Relaxation to a Mott Insulator}
\label{sec:relax-dynam-rung}
We consider the two--leg ladder lattice with an odd number of bosons per rung in Fig. (7a). The two-leg ladder consists of one-dimensional chains with inter- and intrachain kinetic and interaction terms and it is described by the Hamiltonian
\begin{eqnarray}
\label{Eq:Hbt}
H &=& -t \sum_{\alpha=1}^2 \sum_{i=1}^{L-1}  e^{i a A^\alpha_{i,i+1}}b_{\alpha i}^\dagger b_{\alpha , i+1}  - g \sum_{i=1}^{L} e^{-i a' A_{\perp i}} b^\dagger_{2 i} b_{1 i} + \text{H.c.},  \nonumber \\
&+& \frac{U}{2} \sum_{\alpha=1}^2 \sum_{i=1}^{L} n_{\alpha i} (n_{\alpha i} - 1) + V_\perp \sum_{i=1}^{L} n_{1 i}n_{2 i}.
\end{eqnarray}
In Eq.~(\ref{Eq:Hbt}), $b_{\alpha i}^\dagger$ creates a boson at site $i=1,..., L$ in chain $\alpha = 1,2$. The phase $a A^\alpha_{i,i+1}$ is acquired by a boson on chain $\alpha=1,2$ traversing a horizontal bond ($a' A_{\perp i}$ is acquired by a particle crossing from one chain to the other along a vertical bond). Here, $U$ and $V_\perp$ are repulsive on-site and rung interactions. 

\begin{figure}[ht!]
\begin{center}
\includegraphics[width=0.5\linewidth]{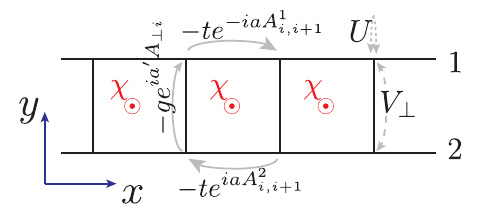} 
\includegraphics[width=0.6\linewidth]{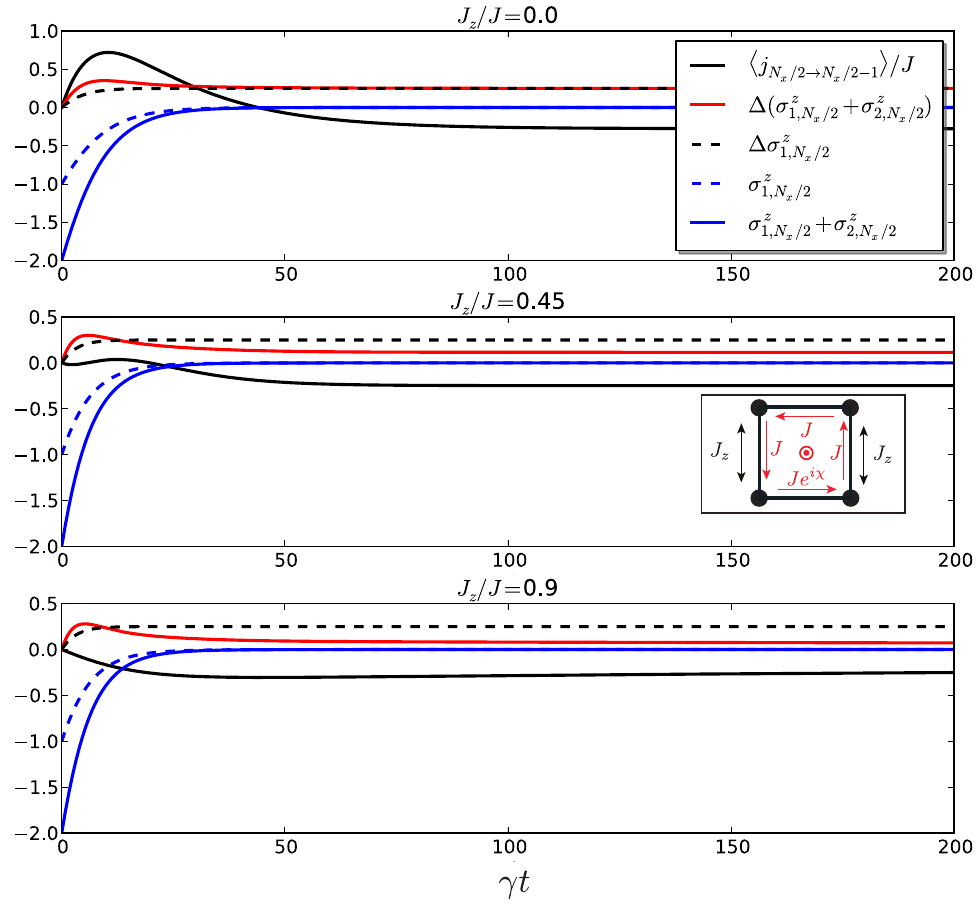} 
\caption{(a) Energy scales of bosonic two-leg ladder. Flux $\chi$ threads each square plaquette. Hopping integrals (\textit{solid arrows}) $t, g$ and repulsive interaction strengths (\textit{dashed arrows}) $U, V_\perp$ correspond to Eq.~(\ref{Eq:Hbt}).  (b) At flux $\chi = \pi / 8$, plots of the expectation value of horizontal current on the top bond (black), and fluctuations of total magnetization on rungs (red) and on-site (black-dashed), as well as site magnetization (blue-dashed) and rung magnetization (blue) for $J_z / J = 0.0, 0.45, 0.9$, versus time in units of the inverse decay constant $1/\gamma$. At $L=2$, the lattice is merely a square plaquette pierced by flux (inset of middle panel). As the vertical bond Ising exchange is increased, fluctuations of total magnetization on rungs subside. }
\end{center}
\label{fig:7} 
\end{figure}

At vanishing flux at $V_\perp = 0$ and density of one boson per site, the model transitions from a Mott insulator to a superfluid as $g$ increases \cite{DonohueGiamarchi}. At arbitrary density there is a low-dimensional Meissner phase at low flux and a vortex phase \cite{EdmondThierry} at high flux. The low-field model with $V_\perp=0$ at unit filling exhibits a superfluid with Meissner currents and a Mott insulator with Meissner currents for weak enough $U$ \cite{TokunoGeorges}. At $V_\perp = 0$ and $U \to \infty$ (hard core bosons) the ground state is the rung Mott insulator at half-filling (one boson per rung) \cite{Crepin}. Recent numerical investigations cover the phase diagram versus filling and flux, containing Meissner and vortex phases, as well as Meissner and vortex Mott insulators at half-filling \cite{piraud_et_al_2014,greschner_et_al_2015,greschner_et_al_2016,petrescu_et_al_2017}, as well as a low-dimensional precursor of the Laughlin \cite{Laughlin} state \cite{Grusdt2014,Alex}.

We consider the model~(\ref{Eq:Hbt}) in the hard core limit $U \to \infty$, which maps bosonic operators to spin-1/2 operators via the Matsubara-Matsuda mapping $\sigma^{+} = b^\dagger$, $\sigma^- = b$ and $\sigma^z = 2 b^\dagger b - 1$ \cite{matsubara_matsuda_1956,batyev_braginskii_1984}. Moreover, we allow uniform flux $\chi$ per plaquette, and set $t=g=J$ so that the model becomes
\begin{eqnarray}
  \label{Eq:Hbt2}
  H &=& -J \sum_{\alpha=1}^2 \sum_{i=1}^{L-1}  e^{i (\alpha-1) \chi}\sigma_{\alpha,i+1}^+ \sigma_{\alpha, i}^-  - J \sum_{i=1}^{L} \sigma^+_{2 i} \sigma^-_{1 i} + \text{H.c.} + J_z \sum_{i=1}^{L} \sigma^z_{1,i} \sigma^z_{2,i}.
\end{eqnarray}  
The Ising coupling $J_z = V_\perp / 4$, acting only on the rungs, favors spin density waves with zero magnetization per rung. To connect this back in the language of hardcore bosons, note that since $\sigma^z_{i\alpha} = 2 b^\dagger_{i\alpha} b_{i\alpha} - 1$, to go from Eq.~(\ref{Eq:Hbt}) to Eq.~(\ref{Eq:Hbt2}) we added the boson chemical potential term $H_{\mu} = - 2J_z \sum_{i\alpha} b_{i\alpha}^\dagger b_{i\alpha}$. This enforces a bosonic ground state in a rung Mott state with one boson per rung. As $J_z/J$ is increased, the ground state $|GS\rangle$ is expected to turn into a rung Mott insulator (at large $L$), i.e. a zero magnetization state $\langle \sum_{\alpha,i}\sigma^z_{\alpha,i}\rangle = 0$ with suppressed fluctuations of the rung magnetization $\sigma_{1,i}^z+\sigma_{2,i}^z$ for $i=1,...,L$ but not of the relative magnetization $\sigma^z_{1,i}-\sigma^z_{2,i}$ (in the figure captions, we denote fluctuations as $\Delta O = \langle O^2 \rangle - \langle O \rangle^2$). 

Additionally, each spin in Eq.~(\ref{Eq:Hbt2}) can relax via a coupling to a bosonic bath
\begin{eqnarray}
  \label{Eq:BosonsBathC}
  H_{\textit{system--bath}} = \sum_{i,\alpha} \sigma^x_{i,\alpha} \sum_l g_{i,\alpha,l} (a_{i,\alpha,l} + a_{i,\alpha,l}^\dagger), H_{\textit{bath},i,\alpha} = \sum_l \nu_{i,\alpha,l} a_{i,\alpha,l}^\dagger a_{i,\alpha,l}.
\end{eqnarray}
The dissipation here again could be realized by coupling the system to one-dimensional BECs. For this particular case, we choose a bath coupling with the spin along the $x$ axis, such that it can also change the number of particles (it
would therefore refer to Raman couplings with the BECs, as drawn in Fig.~\ref{fig:1}(e) as red arrows. 
We take for our analysis a flat bath spectral function at zero temperature $J(\omega) = \gamma \theta(\omega)$ where $\theta$ is the Heaviside function and $\gamma$ is the characteristic decay constant that derives from a Markov approximation treatment of~(\ref{Eq:BosonsBathC}). We consider the evolution of the $L = 2$ system, which is a single plaquette of the ladder. In a Bloch--Redfield treatment \cite{breuer_petruccione_2002} the bath is expected to relax the system to its single non-degenerate ground state, $\rho(t \to \infty) \propto | GS \rangle \langle GS |$. 

This is illustrated in Fig.~(7b) where the $L=2$ system is initialized in the unit filled Mott insulating state $\psi(0) = |\uparrow\uparrow\uparrow\uparrow\rangle$, where all magnetization fluctuations and currents are suppressed. For all values of $J_z/J$ shown, the dynamics evolves to the ground state of $H$ of Eq.~(\ref{Eq:Hbt2}). More generally, we expect similar relaxation to the ground state via the spin decay mechanism~(\ref{Eq:BosonsBathC}) to occur in larger systems. As $J_z/J$ is increased, fluctuations of rung magnetization subside, which is a feature of the rung-Mott phase.

\subsection{Rung-Mott insulating phase and the effect of dephasing}
\label{sec:effective-spin-model}

To study the effect of dephasing on the Mott-Meissner phase, we commit to the Hilbert subspace with one boson per rung and derive an effective spin model in that subspace. 

At weak longitudinal hopping rates, $t \ll U,V_\perp$, the ground state is a rung Mott insulator, protected by a gap to the creation or annihilation of one bosonic particle \cite{petrescu_et_al_2017,Alex}. In this ground state fluctuations in the boson density at every rung $i$, $b_{1,i}^\dagger b_{1,i} + b_{2,i}^\dagger b_{2,i}$, are suppressed. There remains a pseudospin degree of freedom, in which the original bosons of Eq.~(\ref{Eq:Hbt}) play the role of Schwinger bosons: $\sigma^z_i = b_{1,i}^\dagger b_{1,i} - b_{2,i}^\dagger b_{2,i},\; \sigma^x_i = b_{1,i}^\dagger b_{2,i} + \text{H.c.},\; \sigma^y_i = - i b_{1,i}^\dagger b_{2,i} + \text{H.c.}$ The pseudospin sector is described by the low--energy Hamiltonian \cite{Alex,svistunov_kuklov_2003,altman_et_al_2003,duan_et_al_2003,isacsson_et_al_2005}
\begin{eqnarray}
\label{Eq:XXZ}
H_\sigma = -\sum_{i=1}^{L-1} \left[ 2 J_{xx} (\sigma_i^+ \sigma_{i+1}^- e^{i a A_{i,i+1}^\sigma} + \text{H.c.}) - J_z \sigma_i^z \sigma^z_{i+1} \right] 
-g \sum_{i}^{L} \left[ \sigma_i^x \cos(a' A_{\perp i}) - \sigma_i^y \sin(a' A_{\perp i}) \right],
\end{eqnarray}
with couplings $J_{xx} = \frac{t^2}{V_\perp}$, $J_z = t^2\left(-\frac{2}{U} + \frac{1}{V_\perp}\right)$. We have let $A_{i,i+1}^\sigma \equiv A_{i,i+1}^1 - A_{i,i+1}^2$. If $V_\perp = U/2$ this is the $XY$ model, whereas the Heisenberg antiferromagnet is reached for $U\rightarrow +\infty$. At vanishing flux, the $XY$ term is ferromagnetic.

The dynamics of the boson population imbalance at rung $i$ is related to pseudospin current via the Heisenberg equation. We pick a gauge such that all Peierls phases are acquired along horizontal bonds $a A_{i,i+1}^\sigma = \chi$ and $a' A_{\perp,i}=0$ and then
\begin{eqnarray}
  \frac{d\sigma^z_i}{dt} &=&  j_{\sigma, (i-1) \to i} + j_{\sigma,(i+1) \to i} + j_{\perp,i}, \nonumber \\
  j_{\sigma,(i-1) \to i} &\equiv& -8 i J_{xx} e^{i \chi} \sigma^+_{i-1}\sigma_{i}^-  + \text{H.c.} \nonumber \\
  j_{\sigma,(i-1) \to i} &\equiv& -8 i J_{xx} e^{-i \chi} \sigma^+_{i+1}\sigma_{i}^-  + \text{H.c.} \nonumber \\
  j_{\perp,i}          &\equiv& - 2 g \sigma_i^y.
\end{eqnarray}
The parallel current $j_{\sigma,(i-1)\to i}$, for a range of energy scales, is expected from variational arguments to take on expectation values $j_{\sigma,(i-1)\to i} \propto - 2 J_{xx} \sin(\chi)$ for fluxes $\chi$ less than some critical field $\chi_c$ \cite{Alex}, beyond which the vortex phase ensues and there is a suppression of the antisymmetric current.

\begin{figure}[t!]
\begin{center}
  \includegraphics[width=0.6\linewidth]{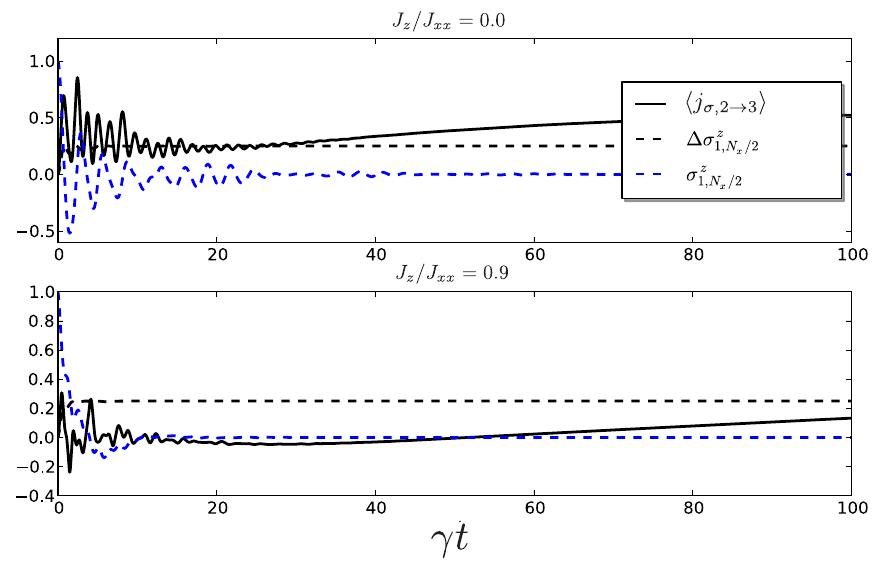}
  \caption{Antisymmetric current (black), fluctuations (black-dashed) and expectation value (blue dashed) of antisymetric boson density $\sigma^z_{i} = b^\dagger_{1,i}b_{1,i} - b^\dagger_{2,i}b_{2,i}$.}
\label{fig:8}
\end{center}
\end{figure}

To phenomenologically model the effect of dephasing, we consider a coupling of $H_\sigma$ to a bosonic bath via
\begin{eqnarray}
  \label{Eq:SpinBathC}
  H_{\sigma,\textit{bath}} = \sum_i \sigma_{i}^z \sum_l g_{i,l} (a_{i,l} + a_{i,l}^\dagger), H_\textit{bath} = \sum_{l,i} \nu_{i,l} a_{i,l}^\dagger a_{i,l}.
\end{eqnarray}
Here, $g_{i,l}$, $\nu_{i,l}$ denote the coupling strength and the frequency of the $l^\textit{th}$ bosonic bath mode to the $i^{\textit{th}}$ rung pseudospin. The form of dissipation is similar to the one studied in Sec. 3.1. 
In Fig.~\ref{fig:8} we show the dependence of the antisymmetric current on time using a Bloch--Redfield solution for $L = 4$ rungs. The spin system is initialized in fully polarized state $|\uparrow \uparrow \uparrow \uparrow \rangle$, which corresponds to $\langle n_{\alpha i} \rangle = 1$ for $\alpha=1$ and $0$ for $\alpha=0$. As in the previous section, times are rescaled with respect to the decay constant $\gamma$ that derives from Eq.~(\ref{Eq:SpinBathC}), and as before a flat spectral function is chosen. The effect of dephasing on the Hamiltonian~(\ref{Eq:XXZ}) is to relax to the ground state, which in the cases considered in Fig.~\ref{fig:8} is always in the $\langle \sigma_z \rangle = 0$ sector in the absence of Zeeman fields to break the symmetry between chains 1 and 2 in Fig.~7(a). 


 We emphasize the numerous current efforts to realize novel correlated phases in these ladders systems with artificial gauge fields, related to flux phases \cite{Paramekanti,TokunoGeorges}, vortex phases \cite{,EdmondThierry} and Laughlin phases \cite{Alex,petrescu_et_al_2017,Leonardo}. We note the numerous theoretical proposals to realize exotic quantum Hall phases \cite{L1,L2,L3,L4,L5,L6,L7,L8,L9,L10,L11,L12,L15}, in relation with wire-constructions of topological phases \cite{Kane,TeoKane,Neupert}. It is also important to mention novel efforts to predict and build Thouless pump type experiments \cite{petrescu_et_al_2017,Leonardo,Pump1,Pump2,Angelakistopo}.

\section{Conclusion}
\label{sec:conclusion}
Here, we summarize our main findings related to the dynamics of spin-boson systems and more precisely to the dynamics of driven and dissipative quantum impurity systems. 

When studying the real time dynamics of spin observables, we have made connections with classical Bloch equations and a statistical approach described by an effective temperature. At the same time, the stochastic dynamics of the quantum trajectories can produce novel phenomena in terms of many-body physics, such as localization and dissipation-induced quantum phase transitions, synchronization revivals, and a quantum dynamo effect when probing the topology of a driven spin-1/2 particle on the Bloch sphere. It is important to note that one is able to measure Berry phase properties of a spin-1/2 particle with the stochastic Schr\" odinger equation approach even though paths are encoded in classical blip and sojourn variables. Here, the Berry phase or Chern number is directly related to spin observables.  Then, we have shown how the stochastic aspect can be
generalized to dissipative spin chains, making some links towards Dicke-type models and mean-field approaches.
 
An environment can also serve for bath (or dissipation) engineering. We have illustrated this point by studying Kondo physics of light in Josephson circuits in the microwave limit, emergent Kondo physics in quantum RC circuits, multi-channel Kondo fixed points in topological Josephson junctions with Majorana fermions, and hybrid systems comprising a cQED environment coupled to a biased mesoscopic system. A dissipative environment engenders fluctuations and disorder in time; we have analyzed this aspect by building connections between Kondo and Quantum Brownian motion physics. These quantum impurity systems can also be useful as sensors of quantum many-body phenomena. We have illustrated this concept in ladder systems where the Rabi dynamics of a spin-1/2  can probe the Mott-Superfluid transition. We have also studied the relaxation of particle densities and currents, in the language of spin-boson systems.  Studying the fast dynamics in quantum materials has also attracted some attention recently, and  our work could stimulate new frontiers.
\\

{\it This work has been initiated by lectures given at Yale University, at Jouvence in Quebec and at CIFAR meetings, and at small classes at Ecole Polytechnique. We would like to dedicate this review to our dear friend and collaborator Adilet Imambekov whose original ideas initiated this work on the stochastic approach to the spin-boson model.  The work on spin-boson and Kondo models have also been inspired from discussions with Markus B\" uttiker and Bernard Coqblin in Geneva, Sherbrooke and Paris. K.L.H. acknowledges support from the DOE DE-FG02-08ER46541, the National Science Foundation DMR-0803200, the Yale Center for Quantum Information Physics (DMR-0653377), the German DFG Forschergruppe 2414. L. Henriet, L. Herviou, K. Plekhanov acknowledge support from Ecole Polytechnique and EDPIF Universit\' e Paris-Saclay. K.L.H. and T.G. acknowledge funding from the Labex Palm Paris-Saclay. P.P.O. acknowledges support from Iowa State University Startup Funds.}
\\
\\

\bibliographystyle{apsrev4-1}

\begin{thebibliography}{99}

\bibitem{NielsenChuang-QC}
M. A. Nielsen, I. L. Chuang, Quantum Computation and Quantum Information (2nd ed.), Cambridge University Press, (2010). 

\bibitem{Feynman-QuantumSimulators}
R. P. Feynman, Int. J. Theor. Phys. \textbf{21}, 467 (1982). 

\bibitem{Cappallaro-RMP}
C. L. Degen, F. Reinhard, and P. Cappellaro, Rev. Mod. Phys. \textbf{89}, 035002 (2017).

\bibitem{Kondo}
J. Kondo, Prog. Theor. Phys. Oxford Journals. \textbf{32} (1) 37-49 (1964).

 \bibitem{Hewson}
A. C. Hewson, The Kondo Problem to Heavy Fermions, Cambridge University Press 1997; B. Coqblin, Proceedings of the ARW Nato Workshop Hvar, Croatia (2002). 

\bibitem{Schofield-NFL} 
A. J. Schofield, Non-Fermi liquids, Contemp. Phys., \textbf{40:2}, 95-115, (1999).

\bibitem{cavity1}
C. Cohen-Tanoudji, J. Dupont-Roc and G. Grynberg, Photons and atoms, introduction to quantum electrodynamics, Wiley (1997)

\bibitem{cavity2}
S. Haroche and J.-M. Raimond, Exploring the Quantum : Atoms, Cavities, and Photons, Oxford University Press (2006).

\bibitem{Yale}
R. J. Schoelkopf and S. M. Girvin, Wiring up quantum systems, Nature \textbf{451}, 664 (2008).

\bibitem{DIW}
I. Bloch, J. Dalibard and W. Zwerger, Rev. Mod. Phys. \textbf{80}, 885 (2008).

\bibitem{FeynmanVernon}
R. P. Feynman and F. L. Vernon, Ann. Phys. (N. Y.) \textbf{24}, 118 (1963).

\bibitem{Hanggi}
P. H\" anggi and G.L. Ingold, Fundamental Aspects of quantum Brownian motion, Chaos, vol. \textbf{15}, ARTN 026105 (2005).

\bibitem{CaldeiraLeggett}
A. O. Caldeira and A.J. Leggett, Phys. Rev. Lett. \textbf{46} 211-214 (1981).

\bibitem{AndersonRG}
P. W. Anderson, J. Phys. C: Solid St. Phys. \textbf{3} 2436-2441 (1970).

\bibitem{Wilson}
K. Wilson, Rev. Mod. Phys.\textbf{47} 773-840 (1975).

\bibitem{Nozieres}
Ph. Nozi\`eres, Jour. Low Temp. Phys. \textbf{17}, 31 (1974).

\bibitem{IanCFT}
I. Affleck, Acta Phys. Polon. B\textbf{26} 1869-1932 (1995).

 \bibitem{NozieresBlandin}
 P. Nozi\`eres  and A. Blandin, J. Physique \textbf{41} 193 (1980).
 
\bibitem{Blume}
M. Blume, V. J. Emery and A. Luther, Phys. Rev. Lett. \textbf{25} 450-453 (1970).

\bibitem{leggett}
A. J. Leggett, S. Chakravarty, A. T. Dorsey, M. P. A. Fisher, A. Garg and W. Zwerger, Rev. Mod. Phys, \textbf{59} 1 (1987).

\bibitem{weiss}
U. Weiss, Quantum dissipative systems, World Scientific, Singapore (2002).

\bibitem{BrayMoore} 
A. J. Bray and M. A. Moore, Phys. Rev. Lett. \textbf{49}, 1545 (1982). 

\bibitem{anderson}
P. W. Anderson, G. Yuval, and D. R. Hamann, Phys. Rev. B \textbf{1} 4464-4473 (1970).

\bibitem{anderson2}
P. W. Anderson and G. Yuval, J. Phys. C \textbf{4}, 607-620 (1971).

\bibitem{Schoeller-FRG} M. Keil and H. Schoeller, Phys. Rev. B \textbf{63}, 180302 (2001).

\bibitem{Schoeller-FRG-Review}
H. Schoeller, Eur. Phys. J.: Spec. Top. \textbf{168}, 179 (2009).

\bibitem{Metzner-FRG}
W. Metzner, M. Salmhofer, C. Honerkamp, V. Meden, and K. Schonhammer, Rev. Mod. Phys. \textbf{84}, 299 (2012).

\bibitem{Breuer-MasterEquations}
H.-P. Breuer and F. Petruccione, \emph{The Theory of Open Quantum Systems} (Oxford University Press, Oxford, UK, 2002).

\bibitem{variational_1}
R. Silbey, and R. A. Harris, J. Chem. Phys. \textbf{80}, 2615 (1984).

\bibitem{variational_2}
D. P. S. McCutcheon, A. Nazir, S. Bose and A. J. Fisher, Phys. Rev. B \textbf{81}, 235321 (2010).

\bibitem{variational_3}
S. Bera, S. Florens, H. U. Baranger, N. Roch, A. Nazir, and A. W. Chin, Phys. Rev. B \textbf{89}, 121108(R) (2014). 

\bibitem{AlexeiPaul}
A. Tsvelick and P. Wiegmann, Advances in Physics, \textbf{32}, Issue 4, p.453-713 (1983).

\bibitem{NatanJohannesson}
N. Andrei and H. Johannesson, \textbf{100} 108-112 (1984).

\bibitem{NRG-Review}
R. Bulla, T. A. Costi, and T. Pruschke, Rev. Mod. Phys. \textbf{80}, 395 (2008). 

\bibitem{Hofstetter-NRGDynamics}
W. Hofstetter, Phys. Rev. Lett. \textbf{85}, 1508 (2000).

\bibitem{Meirong1}
M-R. Li, K. Le Hur, and W. Hofstetter, Phys. Rev. Lett. {\bf 95}, 086406 (2005).

\bibitem{Bulla}
M. Vojta, N.-H. Tong,  and R. Bulla, Phys. Rev. Lett. {\bf 94}, 070604 (2005).

\bibitem{Andersschiller}
F. B. Anders and A. Schiller, Phys. Rev. B {\bf 74}, 245113 (2006).

\bibitem{subPhilippe}
K. Le Hur, Ph. Doucet-Beaupr\' e, and W. Hofstetter, Phys. Rev. Lett. {\bf 99}, 126801 (2007).

\bibitem{NRG2spins}
P. P. Orth, D. Roosen, W. Hofstetter and K. Le Hur, Phys. Rev. B {\bf 82}, 144423 (2010).

\bibitem{CostiNRG}
H. T. M. Nghiem and T. A. Costi, Phys. Rev. B \textbf{90}, 035129 (2014).

\bibitem{Schollwoeck}
U. Schollw\"ock, Rev. Mod. Phys. \textbf{77}, 259 (2005).

\bibitem{White-TDDMRG}
S. R. White and A. E. Feiguin, Phys. Rev. Lett. \textbf{93}, 076401 (2004). 

\bibitem{Schmitteckert-TDDMRG}
P. Schmitteckert, Phys. Rev. B \textbf{70}, 121302 (2004).

\bibitem{QMC2spins}
A. Winter and H. Rieger, Phys. Rev. B {\bf 90}, 224401 (2014).

\bibitem{MarcoMC}
M. Schir\`o and M. Fabrizio, Phys. Rev. B {\bf 79} 153302 (2009). 

\bibitem{Thomas}
T. L. Schmidt, P. Werner, L. Muehlbacher and A. Komnik, Phys. Rev. B \textbf{78}, 235110 (2008).

\bibitem{WernerOM}
P. Werner, T. Oka and A. J. Millis, Phys. Rev. B \textbf{79}, 035320 (2009).

\bibitem{Maxim}
C. Xu, A. Poudel and M. G. Vavilov, Phys. Rev. A \textbf{89}, 052102 (2014).

\bibitem{BDA}
M. Bauer, D. Bernard and A. Tilloy,  J. Stat. Mech. P09001 (2014).

\bibitem{Weichselbaum}
A. Weichselbaum, F. Verstraete, U. Schollwock, J. I. Cirac, and Jan von Delft
Phys. Rev. B \textbf{80}, 165117 (2009).

\bibitem{TEMPO}
  A. Strathearn, P. Kirton, D. Kilda, J. Keeling and B.~W. Lovett, arXiv:1711.09641, (2017).

\bibitem{Cirac}
C. Schoen, K. Hammerer, M. M. Wolf, J. I. Cirac and E. Solano, Phys. Rev. A \textbf{75}, 032311 (2007).

\bibitem{Lesanosky}
M. Marcuzzi, E. Levi, S. Diehl, J. P. Garrahan and I. Lesanovsky, Phys. Rev. Lett. {\bf 113}, 210401 (2014).

\bibitem{AMRey}
M. L. Wall, A. Safavi-Naini and A. M. Rey, Phys. Rev. A \textbf{94}, 053637 (2016).

\bibitem{Rossini}
J. Jin, A. Biella, O. Viyuela, L. Mazza, J. Keeling, R. Fazio and D. Rossini, Phys. Rev. X \textbf{6}, 031011 (2016).

\bibitem{Blunden-Codd}
Z. Blunden-Codd, S. Bera, B. Bruognolo, N.-O. Linden, A. W. Chin, J. von Delft, A. Nazir and S. Florens,  Phys. Rev. B \textbf{95}, 085104 (2017).

\bibitem{stoch1}
J. Dalibard, Y. Castin, and K. Molmer, Phys. Rev. Lett. \textbf{68}, 580 (1992).

\bibitem{stoch2}
C. W. Gardiner, A. S. Parkins, and P. Zoller, Phys. Rev. A \textbf{46}, 4363 (1992).

\bibitem{stoch3}
N. Gisin and I. C. Percival, J. Phys. A \textbf{25}, 5677 (1992).

\bibitem{Breuer}
H.-P. Breuer, B. Kappler and F. Petruccione, Phys. Rev. A \textbf{59}, 1633 (1999).

\bibitem{S1}
G.  B.  Lesovik,  A.  O.  Lebedev,   and  A.  O.  Imambekov, JETP Lett. {\bf 75}, 474 (2002).

\bibitem{S2}
J.  Cao,  L.  W.  Ungar,   and  G.  A.  Voth,  J.  Chem.  Phys. {\bf 104}, 4189 (1996).

\bibitem{S3}
W. T. Strunz, L. Diosi,  and N. Gisin, Phys. Rev. Lett., \textbf{82}, 1801 (1999).

\bibitem{S4}
J.  T.  Stockburger  and  C.  H.  Mac,  J.  Chem.  Phys. {\bf 110}, 4983 (1999).

\bibitem{S5}
J.  T.  Stockburger  and  H.  Grabert,  Phys.  Rev.  Lett. {\bf 88}, 170407 (2002).

\bibitem{S6}
Y. Zhou and J. Shao, J. Chem. Phys. {\bf 128}, 034106 (2008).

\bibitem{S7}
A. Imambekov, V. Gritsev,  and E. Demler, in Proceedings of the 2006 Enrico Fermi Summer School on Ultracold Fermi gases, edited by M. Inguscio,  W. Ketterle,  and C. Salomon (Varenna, Italy, 2006).

\bibitem{PAK-1}
P. P. Orth, A. Imambekov and K. Le Hur Phys. Rev. A \textbf{82}, 032118 (2010). 
\bibitem{PAK-2}
P. P. Orth, A. Imambekov, and K. Le Hur,  Phys. Rev. B {\bf 87}, 014305 (2013).

\bibitem{Loic1}
L. Henriet and K. Le Hur, Phys. Rev. B \textbf{93}, 064411 (2016).

\bibitem{Loic2}
L. Henriet, Z. Ristivojevic, P. P. Orth, and K. Le Hur, Phys. Rev. A {\bf 90}, 023820 (2014).

\bibitem{Loic3}
L. Henriet, A. Sclocchi, P. P. Orth and K. Le Hur, Phys. Rev. B \textbf{95}, 054307 (2017).

\bibitem{Loicthese}
Loic Henriet, PhD thesis, Non-equilibrium dynamics of many body quantum systems, https://hal.archives-ouvertes.fr/tel-01525432v1.

\bibitem{DMFT}
A. Georges, G. Kotliar, W. Krauth and M. Rozenberg Rev. of Mod. Physics. \textbf{68} 13-125 (1996). 

\bibitem{Werner}
P. Werner, K. Volker, M. Troyer, and S. Chakravarty, Phys. Rev. Lett. \textbf{94} , 047201 (2005).

\bibitem{Werner2}
P. Werner, M. Troyer, and S. Sachdev, J. Phys. Soc. Jpn. Suppl. \textbf{74}, 67 (2005).

\bibitem{Subir}
S. Sachdev, P. Werner, M. Troyer, Phys. Rev. Lett. {\bf 92}, 237003 (2004).

\bibitem{Pankov}
S. Pankov, S. Florens, A. Georges, G. Kotliar and S. Sachdev, Phys. Rev. B {\bf 69}, 054426 (2004).

\bibitem{Olivier}
A. Georges, O. Parcollet and S. Sachdev, Phys. Rev. Lett., \textbf{85}, 840-843 (2000).

\bibitem{Vojta}
H. Weber and M. Vojta, Eur. Phys. J. B \textbf{53}, 185 (2006).

\bibitem{Michel}
W. Wu, M. Ferrero, A. Georges and E. Kozik,  Phys. Rev. B \textbf{96}, 041105 (2017).

\bibitem{Walter}
U. Bissbort, R. Thomale and W. Hofstetter, Phys. Rev. A \textbf{81}, 063643 (2010).

\bibitem{Jonathan}
G. Kulaitis, F. Kr\" uger, F. Nissen and J. Keeling, Phys. Rev. A \textbf{87}, 013840 (2013).

\bibitem{Cristiano}
A. Le Boit\' e, G. Orso and Cristiano Ciuti,  Phys. Rev. Lett. \textbf{110} 233601 (2013). 

\bibitem{Berry}
M. V. Berry, Proceedings of the Royal Society A. \textbf{392} (1802): 45Ð57 (1984).

\bibitem{PolkovnikovGritsev}
A. Polkovnikov and V. Gritsev, PNAS {\bf 109}, 6457 (2012).

\bibitem{Review}
M. Kolodrubetz, P. Mehta and A. Polkovnikov,  arXiv:1602.01062.

\bibitem{JeanNoel}
Y. Liu, F. Piechon and J. N. Fuchs, Europhysics Letters \textbf{103}, 17007 (2013).

\bibitem{synchronization_salomon}
M. Delehaye, S. Laurent, I. Ferrier-Barbut, S. Jin, F. Chevy, C. Salomon, Phys. Rev. Lett. \textbf{115}, 265303 (2015).


 \bibitem{Markus}
P. Cedraschi and M. B\" uttiker, Annals of Physics (NY) \textbf{289}, 1 - 23 (2001).

\bibitem{recati_fedichev}
A. Recati, P. O. Fedichev, W. Zwerger, J. von Delft and P. Zoller, Phys. Rev. Lett. \textbf{94}, 040404 (2005).

\bibitem{orth_stanic_lehur}
P. P. Orth and I. Stanic and K. Le Hur, Phys. Rev. A {\bf 77}, 051601(R) (2008).

\bibitem{Karyn2}
K. Le Hur, Phys. Rev. Lett \textbf{92}, 196804 (2004).

 \bibitem{Houck}
  A. A. Houck, H. T\" ureci and J. Koch, Nature Physics \textbf{8}, 292-299 (2012).

  \bibitem{CRAS2016}
  K. Le Hur, L. Henriet, A. Petrescu, K. Plekhanov, G. Roux and M. Schiro, C. R. Physique \textbf{17} 808-835 (2016).

\bibitem{Goldstein}
M. Goldstein, M. H. Devoret, M. Houzet and L. I. Glazman,  Phys. Rev. Lett. \textbf{110}, 017002 (2013).

  \bibitem{Clerk}
A. Clerk, M.-H. Devoret, S. M. Girvin, F. Marquardt and R. Schoelkopf, Rev. Mod. Phys. \textbf{82}, 1155 (2010).


  \bibitem{Frank}
  F. Verstraete, M. M. Wolf and I. Cirac, Nature Physics \textbf{5}, 633 - 636 (2009).

\bibitem{ZollerPeter}
S. Diehl, A. Micheli, A. Kantian, B. Kraus, H. P. B\" uchler and P. Zoller, Nature Physics \textbf{4}, 878 -883 (2008).

\bibitem{Torre}
E. G. Dalla Torre, E. Demler, T. Giamarchi and E. Altman, Nat. Phys. \textbf{6}, 806-810 (2010).

\bibitem{Bardyn}
C.-E. Bardyn, M. A. Baranov, C. V. Kraus, E. Rico, A. Imamoglu, P. Zoller and S. Diehl,  New Journal of Physics, \textbf{15}, Issue 8, 085001 (2013).

\bibitem{MobilTopo}
F. Grusdt, N. Y. Yao, D. A. Abanin, M. Fleischhauer and E. A. Demler, Nature Communications \textbf{7}, 11994 (2016).

\bibitem{cQEDWalk}
E. Flurin, V. V. Ramasesh, S. Hacohen-Gourgy, L. S. Martin, N. Y. Yao and I. Siddiqi,  Phys. Rev. X \textbf{7}, 031023 (2017).


\bibitem{Berezinskii}
V. L. Berezinskii, Sov. Phys. JETP, \textbf{32} (3): 493-500, (1971); V. L. Berezinskii, Sov. Phys. JETP, \textbf{34} 610 (1972).

\bibitem{KT}
J. M. Kosterlitz and D. Thouless, Journal of Physics C: Solid State Physics, \textbf{6} (7): 1181-1203 (1973).

\bibitem{KLH}
K. Le Hur, Understanding Quantum Phase Transitions, edited by Lincoln D. Carr (Taylor and Francis, Boca Raton, 2010) and arXiv:0909.4822.

\bibitem{KLHannal}
K. Le Hur, Ann. Phys. N.Y. \textbf{323} 2208-2240 (2008) and arXiv:0711.2301.

\bibitem{Reppy}
D. J. Bishop and J. D. Reppy. Phys. Rev. Lett. \textbf{40}, 1727 (1978).

\bibitem{Dalibard}
Z. Hadzibabic, P. Kr\" uger, M. Cheneau, B. Battelier and J. B. Dalibard, Nature \textbf{441} 1118 (2006).

\bibitem{Thouless}
D. J. Thouless, Phys. Rev. \textbf{187} 732-733 (1969).

\bibitem{Kopp}
A. Kopp and K. Le Hur, Phys. Rev. Lett. \textbf{98}, 220401 (2007).

\bibitem{Gross}
M. Haeberlein {\it et al.}  arXiv:1506.09114.

\bibitem{Lupascu}
P. Forn-D\' iaz, J. J. Garcia-Ripoll, B. Peropadre, M. A. Yurtalan, J.-L. Orgiazzi, R. Belyansky, C. M. Wilson, and A. Lupascu, Nature Physics \textbf{13}, 39 (2017). 

\bibitem{Martinez}
Javier Puertas Martinez, Sebastien Leger, Nicolas Gheereart, Remy Dassonneville, Luca Planat, Farshad Foroughi, Yuriy Krupko, Olivier Buisson, C\' ecile Naud, Wiebke Guichard, Serge Florens, Izak Snyman and Nicolas Roch,  arXiv:1802.00633

\bibitem{StanfordNadya}
A. Kapitulnik, N. Mason, S. A. Kivelson and S. Chakravarty, Phys. Rev. B \textbf{63}, 125322 (2001).

\bibitem{Fitzpatrick}
M. Fitzpatrick, N. M. Sundaresan, A. C. Y. Li, J. Koch, and A. A. Houck, Phys. Rev. X \textbf{7}, 011016 (2017).
 
 \bibitem{LPN}
S. Jezouin, M. Albert, F. D. Parmentier, A. Anthore, U. Gennser, A. Cavanna, I. Safi and F. Pierre, Nat. Commun. \textbf{4} 1802 (2013).

  \bibitem{Finkelstein}
H. T. Mebrahtu, I. V. Borzenets, D. E. Liu, H. Zheng, Y. V. Bomze, A. I. Smirnov, H. U. Baranger and G. Finkelstein, Nature \textbf{488}, p. 61 (2012).

\bibitem{huelga_plenio} 
S. F. Huelga, M. B. Plenio, Contemp. Phys. \textbf{54}, 181 - 207 (2013)


\bibitem{SchmidtKoch}
S. Schmidt and J. Koch, Annalen der Physik, \textbf{525} 395-412 (2013).

\bibitem{Hartmann}
M. J. Hartmann,  J. Opt. \textbf{18}, 104005 (2016).

  
  \bibitem{Angelakisreview}
  C. Noh and D. G. Angelakis, Report of Progress in Physics, \textbf{80} 016401 (2016).
  
  \bibitem{CRIA}
  I. Carusotto and C. Ciuti, Rev. Mod. Phys. \textbf{85}, 299 (2013).

\bibitem{Toulouse}
G. Toulouse, C. R. Acad. Sci. {\bf 268}, 1200 (1969).

\bibitem{resonant}
F. Guinea, V. Hakim,   and A. Muramatsu, Phys. Rev. B {\bf 32}, 4410 (1985).

\bibitem{Schoeller}
O. Kashuba, D. M. Kennes, M. Pletyukhov, V. Meden and H. Schoeller, Phys. Rev. B {\bf 88}, 165133 (2013).

\bibitem{Zoran}
Z. Ristivojevic, A. Petkovic, P. Le Doussal and T. Giamarchi, Phys. Rev. Lett. {\bf 109}, 026402 (2012).

\bibitem{Berkeley}
S. J. Weber, A. Chantasri, J. Dressel, A. N. Jordan, K. W. Murch and I. Siddiqi, Nature {\bf 511}, 570-573 (2014).

\bibitem{NicolascQED}
N. Roch, M. E. Schwartz, F. Motzoi, C. Macklin, R. Vijay, A. W. Eddins, A. N. Korotkov, K. B. Whaley, M. Sarovar and I. Siddiqi, Phys. Rev. Letters, \textbf{112} (17), 170501 (2014).

\bibitem{atomChern2}
N. Fl\" aschner, B. S. Rem, M. Tarnowski, D. Vogel, D.-S. L\" uhmann, K. Sengstock, and C. Weitenberg, Science {\bf 352}, 1091 (2016).

\bibitem{Ian}
S. Sugawa, F. Salces-Carcoba, A. R. Perry, Y. Yue and Ian B. Spielman, arXiv:1610.06228.


\bibitem{Nicolas}
N. Roch, M. E. Schwartz, F. Motzoi, C. Macklin, R. Vijay, A. W. Eddins, A. N. Korotkov, K. B. Whaley, M. Sarovar and I. Siddiqi, Phys. Rev. Letters, \textbf{112} (17), 170501 (2014).


\bibitem{Portier}
 O. Parlavecchio, C. Altimiras, J.-R. Souquet, P. Simon, I. Safi, P. Joyez, D. Vion, P. Roche, D. Esteve, and F. Portier, Phys. Rev. Lett.,  \textbf{114} (12) (2015).

\bibitem{IngoldNazarov}
 G.-L. Ingold and Yu. V. Nazarov,  arXiv:cond-mat/0508728, in: Single Charge Tunneling, edited by H. Grabert and M. H. Devoret, NATO ASI Series B, Vol. 294, pp. 21-107 (Plenum Press, New York, 1992).
 
 \bibitem{Kane-Fisher}
 C. L. Kane and M. P. A. Fisher, Phys. Rev. Lett. \textbf{68}, 1220 (1992).
 
\bibitem{SafiSaleur}
I. Safi and H. Saleur, Phys. Rev. Lett.  \textbf{93} 126602 (2004). 

 \bibitem{GiamarchiSchulz}
 T. Giamarchi and H. J. Schulz, Phys. Rev. B \textbf{37}, 325 (1988). 
 

\bibitem{Haldane}
F. D. M. Haldane, J. Phys. C \textbf{14}, 2585 (1981).

\bibitem{Giamarchi}
T. Giamarchi, Quantum Physics in One Dimension, Clarendon Press Oxford, 2004.

\bibitem{Glattli}
J. Dubois, T. Jullien, F. Portier, P. Roche, A. Cavanna, Y. Jin, W. Wegscheider, P.  Roulleau and D.C. Glattli, \textbf{502} (7473) 659-63 (2013).

\bibitem{LPA}
E. Bocquillon, V. Freulon, F. D. Parmentier, J.-M Berroir, B. Pla\c{c}ais, C. Wahl, J. Rech, T. Jonckheere, T. Martin, C. Grenier, D. Ferraro, P. Degiovanni and G. F\`eve, Annalen der Physik, \textbf{526}, 1 (2014).

\bibitem{Julien}
J. Gabelli and B. Reulet, Phys. Rev. B \textbf{87}, 075403 (2013).

\bibitem{Tal}
T. Goren, K. Le Hur, and E. Akkermans arXiv:  arXiv:1611.06738.

\bibitem{Eric}
 E. Akkermans and G. V. Dunne, Phys. Rev. Lett. \textbf{108}, 030401 (2012).

\bibitem{Immanuel}
 M.  Knap,  A.  Kantian,  T.  Giamarchi,  I.  Bloch,  M.  D. Lukin,   and  E.  Demler,  Phys.  Rev.  Lett. \textbf{111},  147205 (2013).

\bibitem{Munich}
 M.  Atala,  M.  Aidelsburger,  J.  T.  Barreiro,  D.  Abanin, T. Kitagawa, E. Demler,   and I. Bloch, Nature Physics \textbf{9}, 795 (2013).
 
 \bibitem{MarcoAditi1}
M. Schir\' o and A. Mitra, Phys. Rev. Lett. \textbf{112}, 246401 (2014).

\bibitem{MarcoAditi2}
M. Schir\' o and A. Mitra, Phys. Rev. B \textbf{91}, 235126 (2015).


\bibitem{MatveevFurusaki}
A. Furusaki and K. A. Matveev, Phys. Rev. Lett. \textbf{88}, 226404 (2002).

\bibitem{LiHur}
K. Le Hur and M.-R. Li, Phys. Rev. B \textbf{72}, 073305 (2005).

\bibitem{Neel}
T. Wei\ss l, B. K\" onig, E. Dumur, A. K. Feofanov, I. Matei, C. Naud, O. Buisson, F. W. J. Hekking and W. Guichard, Phys. Rev. B \textbf{92}, 104508 (2015).

\bibitem{Pop}
I. M. Pop, I. Protopopov,  F. Lecocq,  Z. Peng,  B. Pannetier, O. Buisson and W. Guichard, Nature Phys. \textbf{6}, 589 (2010).

\bibitem{Fluxonium}
V. E. Manucharyan, J. Koch, L. Glazman and M. Devoret, Science {\bf 326}, 113-116 (2009). 


\bibitem{Carlos_scientific_reports}
C. Sabin, A. White, L. Hackermuller and I. Fuentes, Nature Scientific Reports, Vol. \textbf{4}, id. 6436 (2014).

\bibitem{lamacraft}
A. Lamacraft, Phys. Rev. B \textbf{79}, 241105(R) (2009). 

\bibitem{Fischer}
  P. O. Fedichev and U. R. Fischer, Phys.Rev.Lett. \textbf{91} 240407 (2003).
  
\bibitem{Anna}
A. Posazhennikova and W. Belzig, EuroPhysics Letters \textbf{87}, 56004 (2009).

\bibitem{Stevelecturenotes}
S. M. Girvin, Quantum Machines: Measurement and Control of Engineered Quantum Systems: Lecture Notes of the Les Houches Summer School: Volume 96, July 2011
Michel Devoret, Benjamin Huard, Robert Schoelkopf, and Leticia F. Cugliandolo, Oxford 2014. 

\bibitem{Makhlin}
Y. Makhlin, G. Sch\" on and A. Shnirman, New Directions in Mesoscopic Physics (Towards Nanoscience), pp. 197-224. Eds. R. Fazio, V. F. Gantmakher, and Y. Imry, Kluwer, Dordrecht, 2003;  arXiv:cond-mat/0309049.

\bibitem{Pashkin}
Y. A. Pashkin, T. Yamamoto, O. Astafiev, Y. Nakamura, D. V. Averin, and J. S. Tsai, Nature \textbf{421}, 823 (2003).

\bibitem{Wiel}
W. G. van der Wiel, S. De Franceschi, J. M. Elzerman, T. Fujisawa, S. Tarucha and L.P. Kouwenhoven, Rev. Mod. Phys. Vol. \textbf{75} No. 1, 1-22 (2003). 

\bibitem{Optiqueatom}
Y. R. P. Sortais, H. Marion, C. Tuchendler, A.M. Lance, M. Lamare, P. Fournet, C. Armellin, R. Mercier, G. Messin, A. Browaeys and P. Grangier, Phys. Rev. A \textbf{75}, 013406 (2007).

\bibitem{Pascal}
L. Borda, G. Zarand and P. Simon, Phys. Rev. B \textbf{72}, 155311 (2005).

\bibitem{Transmon}
J. Koch {\it  et al.} Phys. Rev. A \textbf{76}, 042319 (2007); A. A. Houck {\it et al.} Quant. Inf. Proc. \textbf{8}, 105 (2009).

\bibitem{Xmon}
R. Barends {\it et al.} Phys. Rev. Lett.\textbf{111}, 080502 (2013).

\bibitem{Finkelstein2}
H. T. Mebrahtu, I. V. Borzenets, H. Zheng, Y. V. Bomze, A. I. Smirnov, S. Florens, H. U. Baranger and G. Finkelstein, Nature Physics \textbf{9}, 732 (2013).

\bibitem{Chung-Hou}
C.-H. Chung, K. Le Hur, G. Finkelstein, M. Vojta and P. Woelfle, Phys. Rev. B \textbf{87}, 245310 (2013).

\bibitem{Karyn}
K. Le Hur, Phys. Rev. B {\bf 85}, 140506(R) (2012).


\bibitem{Camalet}
S. Camalet, P. Degiovanni, J. Schriefl and F. Delduc, Europhysics Letters \textbf{68} 37 (2004).

\bibitem{Dousse}
A. Dousse, L. Lanco, J. Suffczynski, E. Semenova, A. Miard, A. Lemaitre, I. Sagnes, C. Roblin, J. Bloch and P. Senellart, Phys. Rev. Lett. \textbf{101}, 267404 (2008).

\bibitem{Williams}
N. S. Williams, K. Le Hur, and A. N. Jordan, J. Phys. A: Math. Theor. {\bf 44} 385003 (2011).

\bibitem{NegativeT}
S. Braun, P. Ronzheimer, M. Schreiber, S. S. Hodgman, T. Rom, I. Bloch and U. Schneider, Science \textbf{339}, 52-55 (2013).


\bibitem{Lehnert}
M.  Schroer,  M.  Kolodrubetz,  W.  Kindel,  M.  Sandberg, J.  Gao,  M.  Vissers,  D.  Pappas,  A.  Polkovnikov, and K. Lehnert, Phys. Rev. Lett.  {\bf 113}, 050402 (2014).

\bibitem{Roushan}
P. Roushan, C. Neill, Y. Chen, M. Kolodrubetz, C. Quintana,  N.  Leung,  M.  Fang,  R.  Barends,  B.  Campbell, Z.  Chen,  B.  Chiaro,  A.  Dunsworth,  E.  Jeffrey,  J.  Kelly, A.   Megrant,   J.   Mutus,   P.   J.   J.   OMalley,   D.   Sank,
A.  Vainsencher,  J.  Wenner,  T.  White,  A.  Polkovnikov, A. N. Cleland,  and J. M. Martinis, Nature {\bf 515}, 241-244 (2014).

\bibitem{AndreasBerry}
P. J. Leek, J. M. Fink, A. Blais, R. Bianchetti, M. G\" oppl, J. M. Gambetta, D. I. Schuster, L. Frunzio, R. J. Schoelkopf and A. Wallraff, Science \textbf{318}, 1889 (2007).

\bibitem{atomChern}
M. Aidelsburger, M. Lohse, C. Schweizer, M. Atala, J. T. Barreiro, S. Nascimb\` ene, N. R. Cooper, I. Bloch, and N. Goldman, Nature Physics {\bf 11}, 162-166 (2015).

\bibitem{Esslinger}
G. Jotzu, M. Messer, R. Desbuquois, M. Lebrat, T. Uehlinger, D. Greif, and T. Esslinger, Nature {\bf 515},  237-240 (2014).

\bibitem{Guillaume}
N. Doiron-Leyraud, T. Szkopek, T. Pereg-Barnea, C. Proust and G. Gervais, Phys. Rev. B \textbf{91}, 245136 (2015).

\bibitem{Graphene}
K. Bennaceur, J. Guillemette, P. L. L\' evesque, N. Cottenye, F. Mahvash, N. Hemsworth, A. Kumar, Y. Murata, S. Heun, M. O. Goerbig, C. Proust, M. Siaj, R. Martel, G. Gervais and T. Szkopek, 
 Phys. Rev. B \textbf{92}, 125410 (2015).

\bibitem{Kim}
Y. Zhang, Y.-W. Tan, H. L. Stormer and P. Kim, Nature \textbf{438}, 201-204 (2005).

\bibitem{Dicke}
C. Emary and T. Brandes, Phys. Rev. E \textbf{67} 066203 (2003).

\bibitem{Marco}
O. Scarlatella and M. Schir\' o,  arXiv:1611.09378.

\bibitem{Xu}
P. Xu, A. Holm Kiilerich, R. Blattmann, Y. Yu, S.-Liang Zhu and K. Molmer, Phys. Rev. A \textbf{96}, 010101 (2017).

\bibitem{Garry}
G. Goldstein, C. Aron and C. Chamon, Phys. Rev. B {\bf 92}, 174418 (2015).

\bibitem{Nalbach}
P. Nalbach, S. Vishveshwara, and A. A. Clerk, Phys. Rev. B {\bf 92}, 014306 (2015).

\bibitem{Marquardt}
O. Viehmann, J. von Delft and F. Marquardt,  New J. Phys. \textbf{15}, 035013 (2013).

\bibitem{Maryland}
B. Neyenhuis, J. Smith, A. C. Lee, J. Zhang, P. Richerme, P. W. Hess, Z.-X. Gong, A. V. Gorshkov, and C. Monroe,  arXiv:1608.00681.

\bibitem{Rey}
B. Yan, S. A. Moses, B. Gadway, J. P. Covey, K. R. A. Hazzard, A. Maria Rey, D. S. Jin and J. Ye, Nature {\bf 501} 521 (2013).

\bibitem{Martinis}
R. Barends {\it et al.} Nature \textbf{534}, 222 (2016).

\bibitem{AntoineOptique}
H. Labuhn, D. Barredo, S. Ravets, S. de L\' es\' eleuc, T. Macrì, T. Lahaye and A. Browaeys, Nature \textbf{534}, 667 (2016).

\bibitem{Rydberg}
T. L. Nguyen, J.-M. Raimond, C. Sayrin, R. Cortinas, T. Cantat-Moltrecht, F. Assemat, I. Dotsenko,
S. Gleyzes, S. Haroche, G. Roux, T. Jolicoeur, and M. Brune. Towards quantum simulation with circular
Rydberg atoms. ArXiv e-prints, July 2017.


\bibitem{Kuramoto}
J. A. Acebr\' on, L. L. Bonilla, C. J. P. Vicente, F. Ritort, and R. Spigler Rev. Mod. Phys. \textbf{77}, 137 (2005).

\bibitem{NeuroRecent}
V. Flovik, F. Macia and E. Wahlstr\" om, Scientific Reports \textbf{6},  32528 (2016).

\bibitem{Lesage}
F. Lesage and H. Saleur, Phys. Rev. Lett. \textbf{80}, 4370 (1998).

\bibitem{Kayanuma}
K. Saito, M. Wubs, S. Kohler, Y. Kayanuma and P. Hanggi, Phys. Rev. B \textbf{75}, 214308 (2007).

\bibitem{Damski}
B. Damski, Phys.Rev.Lett. \textbf{95} 035701 (2005).

\bibitem{Hershfield}
S. Hershfield, Phys. Rev. Lett. \textbf{70}, 2134 (1993).

\bibitem{Natan}
P. Mehta and N. Andrei, Phys. Rev. Lett. \textbf{96}, 216802 (2006).

\bibitem{Doyon}
B. Doyon, Phys. Rev. Lett. \textbf{99}, 076806 (2007).

\bibitem{Prasenjitcurrent}
P. Dutt, J. Koch, J. E. Han and K. Le Hur,  Annals of Physics \textbf{326} 2963-2999 (2011).

\bibitem{LauraLeticia}
L. Foini, L. F. Cugliandolo and A. Gambassi,  J. Stat. Mech. (2012) P09011.

\bibitem{Laura}
L. Foini, A. Gambassi, R. Konik and L. F. Cugliandolo,  arXiv:1610.00101.

\bibitem{Cecile}
C. Monthus, J. Stat. Mech. (2017) 043302.

\bibitem{HaldaneLi}
H. Li and F. D. M. Haldane, Phys. Rev. Lett. \textbf{101}, 010504 (2008).

\bibitem{RANA}
R. Thomale, A. Sterdyniak, N. Regnault and B. Andrei Bernevig, Phys. Rev. Lett. \textbf{104}, 180502 (2010).

\bibitem{Gabi}
N. Lanat\`a, H. U. R. Strand, Y. Yao and G. Kotliar, Phys. Rev. Lett. \textbf{113}, 036402 (2014).

\bibitem{Pollmann}
R. Singh, J. H. Bardarson and F. Pollmann, New J. Phys. \textbf{18}, 023046 (2016).

\bibitem{Alet}
D. J. Luitz, N. Laflorencie and F. Alet, Phys. Rev. B \textbf{91}, 081103 (2015).

 
 \bibitem{BudichDiehl}
 J. C. Budich and S. Diehl,  Phys. Rev. B \textbf{91}, 165140 (2015).
 
 \bibitem{Uhlmann}
 A. Uhlmann, Rep. Math. Phys. \textbf{24}, 229 (1986).
 
 \bibitem{Shi}
  J. Zhu, M. Shi, V. Vedral, X. Peng, D. Suter and J. Du, EPL, \textbf{94} (2011) 20007.
 
\bibitem{Gefen}
R. S. Whitney and Y. Gefen, Phys. Rev. Lett. \textbf{90}, 190402 (2003).

\bibitem{Hamburg}
 N. Fl\" uschner, D. Vogel, M. Tarnowski, B. S. Rem, D.-S\" oren L\" uhmann, M. Heyl, J. C. Budich, L. Mathey, K. Sengstock and C. Weitenberg,  arXiv:1608.05616.

\bibitem{Igor}
S. F. Caballero-Benitez, G. Mazzucchi, I. B. Mekhov, Phys. Rev. A \textbf{93}, 063632 (2016).

\bibitem{Strack}
P. Strack and S. Sachdev,  Phys. Rev. Lett. \textbf{107}  277202 (2011).

\bibitem{Pierre}
P. Nataf, M. Dogan and K. Le Hur, Phys. Rev. A \textbf{86}, 043807 (2012).
  
 \bibitem{Luca}
L. Perfetti, B. Sciolla, G. Biroli, C. J. van der Beek, C. Piovera, M. Wolf and T. Kampfrath, Phys. Rev. Lett. \textbf{114}, 067003 (2015).

\bibitem{SassettiWeiss}
M. Sassetti and U. Weiss, Phys. Rev. Lett. \textbf{65}, 2262 (1990).

\bibitem{Shiba}
H. Shiba, Prog. Theor. Phys. \textbf{54}, 967 (1975).

 \bibitem{Barankov}
 R. A. Barankov and L. S. Levitov, Phys. Rev. Lett. \textbf{96}, 230403 (2006).
 
 \bibitem{Carleo}
G. Carleo and M. Troyer, Science \textbf{355}, 602 (2017).
  
  \bibitem{mobile1}
T. Fukuhara, A. Kantian, M. Endres, M. Cheneau, P. Schauss, S. Hild, D. Bellem, U. Schollw\" ock, T. Giamarchi, C. Gross, I. Bloch, S. Kuhr Nature Physics \textbf{9}, 235 (2013).

  \bibitem{Fisher}
M. P. A. Fisher, G. Grinstein and D. S. Fisher, Physical Review B. \textbf{40} 546-70 (1989).

\bibitem{Jaksch}
D. Jaksch, C. Bruder, J. I. Cirac, C.W. Gardiner, and P. Zoller, Phys. Rev. Lett. \textbf{81}, 3108 (1998).

\bibitem{Greiner}
M. Greiner, O. Mandel, T. Esslinger, T. W. H\" ansch and I. Bloch, Nature \textbf{415}, 39-44 (2002).

\bibitem{Maurice}
K. Le Hur and T. M. Rice, Annals of Physics \textbf{324}, 1452 (2009).  

\bibitem{Demler}
J. Bauer, C. Salomon and E. Demler, Phys. Rev. Lett. \textbf{111}, 215304 (2013).

\bibitem{GiamarchiMillis}
T.  Giamarchi  and  A.  J.  Millis,  Phys.  Rev.  B \textbf{46},  9325 (1992).

\bibitem{Laurent}
G. Bo\' eris,  L. Gori, M. D. Hoogerland, A. Kumar, E. Lucioni, L. Tanzi, M. Inguscio, T. Giamarchi, C. D'Errico, G. Carleo, G. Modugno and L. Sanchez-Palencia, Phys. Rev. A \textbf{93}, 011601 (Rapid Comm.) (2016).

\bibitem{Stephan}
S. Rachel, N. Laflorencie, H. Francis Song and K. Le Hur,  Phys. Rev. Lett. \textbf{108}, 116401 (2012).

\bibitem{Francis}
H. F. Song, S. Rachel, C. Flindt, I. Klich, N. Laflorencie and K. Le Hur,  Phys. Rev. B \textbf{85}, 035409 (2012), Editors' Suggestion.

\bibitem{AlexReview}
A. Petrescu, H. F. Song, S. Rachel, Z. Ristivojevic, C. Flindt, N. Laflorencie, I. Klich, N. Regnault and K. Le Hur, J. Stat. Mech. (2014) P10005.

\bibitem{Nicolas-PhysRep}
N. Laflorencie, Physics Report \textbf{643}, 1-59 (2016).

\bibitem{Munichmicro}
J. F. Sherson, C. Weitenberg, M. Endres, M. Cheneau, I. Bloch and S. Kuhr Nature \textbf{467}, 68 (2010).

\bibitem{Harvardmicro}
W. S. Bakr, J. I. Gillen, A. Peng, S. F\" olling and M. Greiner, Nature \textbf{462}, 74-77 (2009).

\bibitem{Stony}
B. Gadway, D. Pertot, J. Reeves, M. Vogt and D. Schneble, Phys. Rev. Lett. \textbf{107}, 145306 (2011).

\bibitem{Juliette}
J. Billy, V. Josse, Z. Zuo, A. Bernard, B. Hambrecht, P. Lugan, D. Cl\' ement, L. Sanchez-Palencia, P. Bouyer, and A. Aspect, Nature \textbf{453}, Issue 7197, pp. 891-894 (2008).

\bibitem{ManyBodyLoca}
J.-y. Choi, S. Hild, J. Zeiher, P. Schauss, A. Rubio-Abadal, T. Yefsah, V. Khemani, D. A. Huse, I. Bloch and C. Gross, Science \textbf{352}, 1547 (2016).

\bibitem{Angelakis}
S. Restrepo, J. Cerrillo, V. M. Bastidas, D. G. Angelakis and T. Brandes, Phys. Rev. Lett. \textbf{117}, 250401 (2016).

\bibitem{EdmondThierry}
E. Orignac and T. Giamarchi, Phys. Rev. B \textbf{64} 144515 (2001).

\bibitem{Alex}
A.  Petrescu  and  K.  Le  Hur,  Phys.  Rev.  B \textbf{91},  054520  (2015).

\bibitem{Fallani}
M.  Mancini,   G.  Pagano,   G.  Cappellini,   L.  Livi,   M.  Rider, J.  Catani,  C.  Sias,  P.  Zoller,  M.  Inguscio,  M.  Dalmonte,    and L.  Fallani,  ArXiv  e-prints (2015),  arXiv:1502.02495  [cond-mat.quant-gas].

\bibitem{Atala}
M. Atala, M. Aidelsburger, M. Lohse, J. T. Barreiro, B. Paredes and I. Bloch, Nature Physics \textbf{10}, 588-593 (2014).

\bibitem{Zoller}
M. Lacki, H. Pichler, A. Sterdyniak, A. Lyras, V. E. Lembessis, O.  Al-Dossary,  J.  C.  Budich,   and  P.  Zoller,  Phys.  Rev.  A \textbf{93}, 013604 (2016)

\bibitem{Artificial1}
P. Roushan {\it et al.} Nature Physics  \textbf{13}, 146-151  (2017).

\bibitem{L13}
J. Koch, A. A. Houck, K. Le Hur,  and S. M. Girvin, Phys. Rev. A \textbf{82}, 043811 (2010).

\bibitem{L14}
A. Petrescu, A. A. Houck and K. Le Hur, Phys. Rev. A \textbf{86}, 053804 (2012).

\bibitem{Goldman}
N. Goldman and J. Dalibard, Phys. Rev. X \textbf{4}, 031027 (2014).

\bibitem{Cayssol}
J. Cayssol, B. D\' ora, F. Simon and R. Moessner, Phys. Status Solidi RRL \textbf{7} 101-108 (2013).

\bibitem{Leo}
S. M. Cronenwett, T. H. Oosterkamp and L. P. Kouwenhoven, Science \textbf{281}, 540 (1998).

\bibitem{David} 
D. Goldhaber-Gordon, H. Shtrikman, D. Mahalu, D. Abusch-Magder, U. Meirav and M. A. Kastner, Nature 391, 156 (1998).

\bibitem{Leonid}
L. I.  Glazman and M. Pustilnik, New Directions in Mesoscopic Physics (Towards Nanoscience), eds. R. Fazio, V.F. Gantmakher, and Y. Imry (Kluwer, Dordrecht, 2003), pp. 93-115.

\bibitem{Astafiev}
 O. Astafiev, A. M. Zagoskin, A. A. Abdumalikov, Yu. A. Pashkin, T. Yamamoto, K. Inomata, Y. Nakamura, and J. S. Tsai, Science \textbf{327}, 840-843 (2010).
 
 \bibitem{Chalmers}
 I.-C. Hoi, C.M. Wilson, G. Johansson, J. Lindkvist, B. Peropadre, T. Palomaki and P. Delsing, New J. Phys. \textbf{15} 025011 (2013).

\bibitem{Costi}
T. A. Costi, Phys. Rev. Lett. \textbf{80}, 1038 (1998).

 
  \bibitem{Sbierski}
B. Sbierski, M. Hanl, A. Weichselbaum, H. E. T\" ureci, M. Goldstein, L. I. Glazman, J. von Delft and A. Imamoglu, Phys. Rev. Lett. \textbf{111}, 157402 (2013).
 
 \bibitem{Leclair}
 A. Leclair, F. Lesage, S. Lukyanov and H. Saleur, Phys. Lett. \textbf{A235} 203-208 (1997).
 
 \bibitem{Shen}
  J.-T.  Shen  and  S.  Fan,  Phys.  Rev.  Lett. \textbf{98},  153003 (2007).
 
 \bibitem{Busch}
  P.  Longo,  P.  Schmitteckert  and  K.  Busch,  Phys.  Rev. Lett. \textbf{104}, 023602 (2010).
  
  \bibitem{Yudson}
  V. I. Yudson and P. Reineker, Phys. Rev. A \textbf{78}, 052713 (2008).
  
  \bibitem{HakanAlex}
  M.  Malekakhlagh, A. Petrescu and H. E. T\" ureci, Phys. Rev. A \textbf{94}, 063848 (2016).
       
   \bibitem{Stanford}
  R. M. Potok, I. G. Rau, H. Shtrikman, Y. Oreg and D. Goldhaber-Gordon, Nature, \textbf{446}, 7132, pp. 167-171 (2007); A. J. Keller, L. Peeters, C. P. Moca, I. Weymann, D. Mahalu, V. Umansky, G. Zar\' and, D. Goldhaber-Gordon, 
  Nature \textbf{526}, 237-240 (2015).
  
  \bibitem{LPNMarcoussis}
  Z. Iftikhar, S. Jezouin, A. Anthore, U. Gennser, F.D. Parmentier, A. Cavanna and F. Pierre, Nature \textbf{526}, 233 (2015).

 \bibitem{MarcoKaryn}
   M. Schir\' o and K. Le Hur, Phys. Rev. B \textbf{89}, 195127 (2014).
   
  \bibitem{Olesia}
  O. Dmytruk, M. Trif, C. Mora and P. Simon, Phys. Rev. B \textbf{93}, 075425 (2016).
  
  \bibitem{Audrey}
  A. Cottet, T. Kontos and B. Dou\c{c}ot,  Phys. Rev. B \textbf{91}, 205417 (2015).

 \bibitem{ENSTakis}
  M. R. Delbecq, V. Schmitt, F. D. Parmentier, N. Roch, J. J. Viennot, G. F\`eve, B. Huard, C. Mora, A. Cottet and T. Kontos, Phys. Rev. Lett. \textbf{107}, 256804 (2011).
  
  \bibitem{DengChina}
  G.-W. Deng, L. Henriet, D. Wei, S.-X. Li, H.-O. Li, G. Cao, M. Xiao, G.-C. Guo, M. Schiro, K. Le Hur and G.-P. Guo,  arXiv:1509.06141.

 \bibitem{MarkusRC}
 M. B\" uttiker, A. Pr\^etre and H. Thomas, Phys. Rev. Lett. \textbf{70}, 4114 (1993).

 \bibitem{Gabelli}
 J. Gabelli, G. F\`eve, J.-M. Berroir and B. Pla\c{c}ais, Rep. Prog. Phys. \textbf{75} 126504 (2012).
      
 \bibitem{Nigg}
S. E. Nigg, R. Lopez and M. B\" uttiker, Phys. Rev. Lett. \textbf{97}, 206804 (2006).

\bibitem{ChristopheKaryn}
C. Mora and K. Le Hur, Nature Physics \textbf{6}, 697 (2010).
  
\bibitem{Matveev}
K. A. Matveev, Zh. Eksp. Teor. Fiz. \textbf{98}, 1598 (1990) [Sov. Phys. JETP \textbf{72}, 892 (1991)].

\bibitem{Matveev2}
K. A. Matveev, Phys. Rev. B \textbf{51}, 1743 (1995).

\bibitem{Georg}
K. Le Hur and G. Seelig, Phys. Rev. B \textbf{65}, 165338 (2002).

\bibitem{Hamamoto}
Y. Hamamoto, T. Jonckheere, T. Kato and T. Martin, Phys. Rev. B \textbf{81}, (2010) 153305.

\bibitem{Ledoussal}
Y. Etzioni, B. Horovitz and P. Le Doussal, Phys. Rev. Lett. \textbf{106}, 166803 (2011).

\bibitem{Michele1}
M. Filippone, K. Le Hur and C. Mora, Phys. Rev. Lett. \textbf{107}, 176601 (2011).

\bibitem{Michele2}
M. Filippone, K. Le Hur and C. Mora,  Phys. Rev. B \textbf{88}, 045302 (2013). 

\bibitem{Prasenjit}
P. Dutt, T. L. Schmidt, C. Mora and Karyn Le Hur,  Phys. Rev. B \textbf{87}, 155134 (2013).

\bibitem{MicheleChristophe}
M. Filippone and C. Mora Phys. Rev. B \textbf{86}, 125311 (2012).

\bibitem{Coleman}
P. Coleman, L. Ioffe and A. M. Tsvelik, Phys. Rev. B \textbf{52}, 6611, (1995).

\bibitem{Christophe2}
C. Mora and K. Le Hur, Phys. Rev. B \textbf{88}, 241302 (2013).

\bibitem{Grosfeld}
A. Golub and E. Grosfeld, Phys. Rev. B \textbf{86}, 241105(R) (2012).

\bibitem{Muller}
T. M\" uller, R. Thomale, B. Trauzettel, E. Bocquillon and O. Kashuba,  arXiv:1701.03050.

\bibitem{Michele3}
D. Litinski, P. W. Brouwer and M. Filippone,  arXiv:1612.04822.

\bibitem{YiKane}
H. Yi and C.L. Kane,  Phys. Rev. B \textbf{57}, R5579 (1998); H. Yi, Phys. Rev. B \textbf{65}, 195101 (2002). 

\bibitem{Oreg}
Y.  Oreg,  G.  Refael,   and  F.  von  Oppen,  Phys. Rev. Lett. \textbf{105}, 177002  (2010).

\bibitem{Lutchyn}
R. M.  Lutchyn,   J.D.  Sau and  S.  Das  Sarma,  Phys. Rev. Lett. \textbf{105}, 077001  (2010).
 
 \bibitem{LoicHerviou}
L. Herviou, K. Le Hur, and C. Mora, Phys. Rev. B (2016), Editor's Suggestion.

\bibitem{Beri}
B. B\' eri, Phys. Rev. Lett. \textbf{110}, 216803 (2013).

\bibitem{Michaeli}
K. Michaeli, L. Aviad Landau, E. Sela, L. Fu,  arXiv:1608.00581.

\bibitem{BeriCooper}
B. Beri and N. R. Cooper,  Phys. Rev. Lett. \textbf{109}, 156803 (2012).

\bibitem{EggerAltland}
A.   Altland   and   R.   Egger,  Phys. Rev. Lett. \textbf{110}, 196401 (2013).

\bibitem{Erik}
E. Eriksson, C. Mora, A. Zazunov and R. Egger, Phys. Rev. Lett. \textbf{113}, 076404 (2014).

\bibitem{Artificial2}
J. Dalibard, F. Gerbier, G. Juzeli\" unas and P. \" Ohberg, Rev. Mod. Phys. \textbf{83}, 1523 (2011).

\bibitem{Feng}
X.-Y. Feng, G.-M. Zhang and T. Xiang, Phys. Rev. Lett. \textbf{98}, 087204 (2007).

\bibitem{NewPRBpaper}
K. Le Hur, A. Soret and F. Yang, arXiv:1703.07322, to be published in Phys. Rev. B. 

\bibitem{Sedrakyan}
T. A. Sedrakyan, L. I. Glazman and A. Kamenev, Phys. Rev. Lett. \textbf{114}, 037203 (2015).

\bibitem{MajoranaReview}
S. M. Albrecht, A. P. Higginbotham, M. Madsen, F. Kuemmeth, T. S. Jespersen, J. Nygard, P. Krogstrup, and C. M. Marcus, Nature \textbf{531}, 206 (2016).

\bibitem{Sela}
L. Aviad Landau and E. Sela,  arXiv:1609.09257.

\bibitem{Zazunov}
A. Zazunov, F. Buccheri, P. Sodano and R. Egger,  arXiv:1611.07307.

\bibitem{TakisReview}
J.J. Viennot, M.R. Delbecq, L.E. Bruhat, M.C. Dartiailh, M. Desjardins, M. Baillergeau, A. Cottet and T. Kontos, Comptes Rendus Physique \textbf{17} (7), 705-717 (2016).

\bibitem{AndrewReview}
B. Sothmann, R. S\' anchez and A. N. Jordan, Nanotechnology \textbf{26}, 032001 (2015).

\bibitem{LoicNano}
L. Henriet, A. N. Jordan and K. Le Hur, Phys. Rev. B \textbf{92}, 125306 (2015).

\bibitem{UdsonChristophe}
U. C. Mendes and C. Mora, Phys. Rev. B \textbf{93}, 235450 (2016).

\bibitem{Guo1}
G.-W. Deng {\it et al.} Phys. Rev. Lett. \textbf{115}, 126804 (2015).

\bibitem{Guo2}
G.-W. Deng {\it et al.} Nano Letters \textbf{15}, 6620 (2015).

\bibitem{Halperin}
L. Borda, G. Zarand, W. Hofstetter, B. I. Halperin and J. von Delft, Phys. Rev. Lett. \textbf{90}, 026602 (2003).

\bibitem{Keller}
A. J. Keller {\it et al.}  Nature Phys. \textbf{10} (2014) 145-150.

\bibitem{ChristopheSUN}
C. Mora, P. Vitushinsky, X. Leyronas, A. A. Clerk and K. Le Hur,  Phys. Rev. B \textbf{80}, 155322 (2009) (Editors' Suggestion).

\bibitem{Achim}
A. Rosch, J. Kroha and P. W\" olfle, Phys. Rev. Lett. \textbf{87}, 156802 (2001).

\bibitem{KarynLoss}
K. Le Hur, P. Simon and D. Loss, Phys. Rev. B \textbf{75}, 035332 (2007).

\bibitem{breuer_petruccione_2002}
H.-P. Breuer and F. Petruccione, The theory of open quantum systems, (2002).

\bibitem{DonohueGiamarchi}
P. Donohue and T. Giamarchi, Phys. Rev. B \textbf{63}, 180508 (2001).

 \bibitem{TokunoGeorges}
A. Tokuno and A. Georges, New Journal of Physics \textbf{16}, 073005 (2014).

\bibitem{Crepin}
F. Cr\' epin, N. Laflorencie, G. Roux, and P. Simon, Phys. Rev. B \textbf{84}, 054517 (2011).

\bibitem{piraud_et_al_2014}
M. Piraud, F. Heidrich-Meisner, I. P. McCulloch, S. Greschner, T. Vekua, and U. Schollwock, Phys. Rev. B 91, 14040
(2015).

\bibitem{greschner_et_al_2015}
S. Greschner, M. Piraud, F. Heidrich-Meisner, I. P. McCulloch, U. Schollwock, and T. Vekua, Phys. Rev. Lett. \textbf{115}, 190402 (2015).

\bibitem{greschner_et_al_2016}
S. Greschner, M. Piraud, F. Heidrich-Meisner, I. P. McCulloch, U. Schollwock, and T. Vekua, ArXiv e-prints (2016),

\bibitem{petrescu_et_al_2017}
A. Petrescu, M. Piraud, G. Roux, I. P. McCulloch, and K. Le Hur, Phys. Rev. B \textbf{96}, 014524 (2017).

\bibitem{Laughlin}
R. B. Laughlin, Phys. Rev. Lett. \textbf{50}, 1395 (1983).

\bibitem{Grusdt2014}
F. Grusdt and M. Honing, Phys. Rev. A \textbf{90}, 053623 (2014).

\bibitem{matsubara_matsuda_1956}
Matsubara, Takeo and Matsuda, Hirotsugu, Progress of Theoretical Physics \textbf{16}, 569 (1956); 

\bibitem{batyev_braginskii_1984}
E. G. Batyev and L. S. Braginskii, Sov. Phys. JETP \textbf{60} (1984).

\bibitem{svistunov_kuklov_2003}
A. B. Kuklov and B. V. Svistunov, Phys. Rev. Lett. \textbf{90}, 100401 (2003).

 \bibitem{altman_et_al_2003}
 Ehud Altman and Walter Hofstetter and Eugene
Demler and Mikhail D Lukin, New Journal of Physics \textbf{5}, 113 (2003).
 
\bibitem{duan_et_al_2003}
L.-M. Duan, E. Demler, and M. D. Lukin, Phys.
Rev. Lett. \textbf{91}, 090402 (2003). 

\bibitem{isacsson_et_al_2005}
A. Isacsson, M.-C. Cha, K. Sengupta, and S. M. Girvin, Phys. Rev. B 72, 184507 (2005).

\bibitem{Paramekanti}
A.  Dhar,  M.  Maji,  T.  Mishra,  R.  V.  Pai,  S.  Mukerjee, and A.  Paramekanti,  Phys.  Rev.  A \textbf{85},  041602  (2012); A.  Dhar, T. Mishra, M. Maji, R. V. Pai, S. Mukerjee,  and A. Paramekanti, Phys. Rev. B
\textbf{87}, 174501 (2013).

\bibitem{Leonardo}
M. Calvanese Strinati, E. Cornfeld, D. Rossini, S. Barbarino, M. Dalmonte, R. Fazio, E. Sela and L. Mazza,  arXiv:1612.06682.

\bibitem{L1}
R. N. Palmer and D. Jaksch, Phys. Rev. Lett. \textbf{96}, 180407 (2003).

\bibitem{L2}
A. S. Sorensen, E. Demler,  and M. D. Lukin, Phys. Rev. Lett. \textbf{94}, 086803 (2005).

\bibitem{L3}
M. Hafezi, A. S. Sorensen, E. Demler,   and M. D. Lukin, Phys. Rev. A \textbf{76},  023613 (2007).

\bibitem{L4}
L. Hormozi, G. Moller  and S. H. Simon, Phys. Rev. Lett. \textbf{108}, 256809 (2012).

\bibitem{L5}
N.  R.  Cooper  and  J.  Dalibard,  Phys.  Rev.  Lett. \textbf{110},  185301 (2013).

\bibitem{L6}
N. Y. Yao, A. V. Gorshkov, C. R. Laumann, A. M. L\" auchli, J. Ye and M. D. Lukin, Phys. Rev. Lett. \textbf{110}, 185302 (2013).

\bibitem{L7}
A. Sterdyniak, B. A. Bernevig, N. R. Cooper,   and N. Regnault, Phys. Rev. B \textbf{91}, 035115 (2015).

\bibitem{L8}
M. Hafezi, M. D. Lukin,  and J. M. Taylor, New Journal of Physics, \textbf{15}, 063001 (2013).

\bibitem{L9}
E. Kapit, M. Hafezi,   and S. H. Simon, Phys. Rev. X \textbf{4},  031039 (2014).

\bibitem{L10}
J.  Cho,  D.  G.  Angelakis,    and  S.  Bose,  Phys.  Rev.  Lett. \textbf{101}, 246809 (2008).

\bibitem{L11}
A. L. C. Hayward, A. M. Martin,  and A. D. Greentree, Phys. Rev. Lett. \textbf{108}, 223602 (2012).

\bibitem{L12}
C. Noh and D. G. Angelakis, Reports on Progress in Physics \textbf{80},  016401 (2017), arXiv:1604.04433 [quant-ph].

\bibitem{L15}
B. M. Anderson, R. Ma, C. Owens, D. I. Schuster, and J. Simon Phys. Rev. X \textbf{6}, 041043 (2016).

\bibitem{Kane}
C. L. Kane, R. Mukhopadhyay,   and T. C. Lubensky, Phys. Rev. Lett. \textbf{88}, 036401 (2002).

\bibitem{TeoKane}
J. C. Y. Teo and C. L. Kane, Phys. Rev. B \textbf{89}, 085101 (2014).

\bibitem{Neupert}
T. Neupert, C. Chamon, C. Mudry and R. Thomale, Phys. Rev. B \textbf{90}, 205101 (2014).

\bibitem{Pump1}
M. Lohse, C. Schweizer, O. Zilberberg, M. Aidelsburger and I. Bloch, Nature Phys. \textbf{12}, 350 (2016).

\bibitem{Pump2}
S. Nakajima, T. Tomita, S. Taie, T. Ichinose, H. Ozawa, L. Wang, M. Troyer and Y. Takahashi, Nature Physics \textbf{12}, 296-300 (2016).

\bibitem{Angelakistopo}
J. Tangpanitanon, V. M. Bastidas, S. Al-Assam, P. Roushan, D. Jaksch and D. G. Angelakis, Phys. Rev. Lett. \textbf{117}, 213603 (2016).






\end{thebibliography}

\end{document}